\documentclass[a4paper,11pt]{article}
\usepackage{a4wide,amsmath,amssymb,bbm}
\usepackage[english]{babel}

\usepackage{color}
\usepackage{slashed}
\usepackage[normalem]{ulem}

\usepackage{tikz,tikz-cd} 
\usetikzlibrary{calc}
\usetikzlibrary{arrows}

\usepackage[pdfencoding=auto,psdextra]{hyperref}
\usepackage{enumerate}
\hypersetup{
    colorlinks,
    linkcolor={black},
    citecolor={black},
    urlcolor={black}
}

\makeatletter
 \let\old@startsection=\@startsection
 \let\oldl@section=\l@section
 \renewcommand{\@startsection}[6]{\old@startsection{#1}{#2}{#3}{#4}{#5}{#6\mathversion{bold}}}
 \renewcommand{\l@section}[2]{\oldl@section{\mathversion{bold}#1}{#2}}
\makeatother

\makeatletter
\DeclareFontFamily{OMX}{MnSymbolE}{}
\DeclareSymbolFont{MnLargeSymbols}{OMX}{MnSymbolE}{m}{n}
\SetSymbolFont{MnLargeSymbols}{bold}{OMX}{MnSymbolE}{b}{n}
\DeclareFontShape{OMX}{MnSymbolE}{m}{n}{
    <-6>  MnSymbolE5
   <6-7>  MnSymbolE6
   <7-8>  MnSymbolE7
   <8-9>  MnSymbolE8
   <9-10> MnSymbolE9
  <10-12> MnSymbolE10
  <12->   MnSymbolE12
}{}
\DeclareFontShape{OMX}{MnSymbolE}{b}{n}{
    <-6>  MnSymbolE-Bold5
   <6-7>  MnSymbolE-Bold6
   <7-8>  MnSymbolE-Bold7
   <8-9>  MnSymbolE-Bold8
   <9-10> MnSymbolE-Bold9
  <10-12> MnSymbolE-Bold10
  <12->   MnSymbolE-Bold12
}{}

\let\llangle\@undefined
\let\rrangle\@undefined
\DeclareMathDelimiter{\llangle}{\mathopen}%
                     {MnLargeSymbols}{'164}{MnLargeSymbols}{'164}
\DeclareMathDelimiter{\rrangle}{\mathclose}%
                     {MnLargeSymbols}{'171}{MnLargeSymbols}{'171}
\makeatother

\usepackage{mathrsfs}

\usepackage{bm}
\usepackage{comment}
\usepackage{braket}

\newcommand{\Tr}{\mathrm{Tr}}
\newcommand{\be}{\begin{equation}}
\newcommand{\ee}{\end{equation}}
\newcommand{\alg}{\mathfrak}

\DeclareMathAlphabet{\mathdsl}{U}{bbm}{m}{sl}

\newcommand{\dd}{{\bm{d}}}
\newcommand{\EE}{{\bm{E}}}
\newcommand{\OO}{{\bm{\Omega}}}

\newcommand{\F}{{\mathcal{F}}}
\newcommand{\FF}{{\bm{\F}}}

\newcommand{\Csp}{{\mathfrak{c}}}
\newcommand{\Fsp}{{\mathfrak F}}
\newcommand{\n}[1]{#1'{}}
\newcommand{\dr}{w}
\newcommand{\Ad}{\operatorname{ad}}
\newcommand{\AD}{\operatorname{Ad}}
\newcommand{\Kappa}{\mathcal{K}}
\newcommand{\TT}{\mathbb{T}}

\makeatletter

\@addtoreset{equation}{section} \makeatother

\begin{document}

\phantom{.}

\vspace{30pt}

\begin{center}
{\huge{\bf Supergravity
solution-generating techniques \\
and canonical transformations of $\sigma$-models\\
\vspace{8pt}
from $O(D,D)$}} 

\vspace{80pt}

Riccardo Borsato \ \ and \ \ Sibylle Driezen

\vspace{15pt}

{
\small {\it 
Instituto Galego de F\'isica de Altas Enerx\'ias (IGFAE), Universidade de  Santiago de Compostela, Spain}\\
\vspace{12pt}
\texttt{riccardo.borsato@usc.es, sibylle.driezen@usc.es}}\\

\vspace{100pt}

{\bf Abstract}
\end{center}
\noindent
Within the framework of the flux formulation of Double Field Theory (DFT) we employ a generalised Scherk-Schwarz ansatz and discuss the classification of the twists  that in the presence of the strong constraint give rise to constant generalised fluxes interpreted as gaugings.
We analyse the various possibilities of turning on the fluxes $H_{ijk},F_{ij}{}^k,Q_i{}^{jk}$ and $R^{ijk}$, and the solutions for the twists allowed in each case. 
While we do not impose the DFT (or equivalently supergravity) equations of motion, our results provide solution-generating techniques in supergravity when applied to a background that does solve the DFT equations.
At the same time, our results give rise also to canonical transformations of 2-dimensional $\sigma$-models, a fact
which is interesting especially because these are integrability-preserving transformations on the worldsheet. 
Both the solution-generating techniques of supergravity and the canonical transformations of 2-dimensional $\sigma$-models arise as  maps that leave the generalised fluxes of DFT and their flat derivatives invariant.
These maps  include the known abelian/non-abelian/Poisson-Lie T-duality transformations, Yang-Baxter deformations, as well as  novel generalisations of them.

\pagebreak 
\tableofcontents

\setcounter{page}{1}
\newcounter{nameOfYourChoice}

\section{Introduction}

An important concept in physics is that of symmetry. Symmetries are  powerful tools to constrain the form of solutions of physical theories, and they may be used as a guiding principle to construct new physical models.
Certain symmetries are particularly useful because they connect physical solutions as well as different theories that are at first sight unrelated. A prominent example is T-duality in string theory, which provides such a connection between strings on backgrounds with very different geometry~\cite{Buscher:1987sk,Buscher:1987qj}.\footnote{Recently a notion of ``T-duality'' for point particles was introduced in~\cite{Klimcik:2020rxa}.} 
In this paper we will employ  generalised notions of T-duality that go beyond the  case of reduction on a torus  implemented by  Buscher's rules. When necessary we will refer to the latter case as ``abelian'' T-duality, to distinguish it from the ``non-abelian'' T-duality (NATD) transformation of~\cite{delaOssa:1992vci,Gasperini:1993nz,Giveon:1993ai}. NATD may be viewed as a generalisation where a set of \emph{non-abelian} isometries of the string background are dualised to give rise to a new solution of the low-energy (super)gravity equations of motion, modulo a known anomalous-free condition~\cite{Alvarez:1994np,Elitzur:1994ri}. The transformation can be implemented by a Busher-like gauging procedure of the initial isometries, similar to the abelian case.
Global issues of NATD still need to be understood \cite{Giveon:1993ai} and currently the transformation is viewed as a solution-generating technique rather than a fully-fledged symmetry of string theory~\cite{Sfetsos:2010uq,Lozano:2011kb}. 
Similar comments apply to a  further generalisation of NATD that goes under the name of ``Poisson-Lie'' T-duality (PLTD). In this case there is still a notion of dualisation of a symmetry algebra, but the symmetry does not need to be realised as an isometry of the initial background \cite{Klimcik:1995ux,Klimcik:1995dy}.
This advantage puts the original and dual models on an equal footing 
 at the expense of losing the interpretation of the gauging procedure.
All these generalised notions of T-duality transformations were recently shown to admit $\alpha'$-corrections that promote them to solution-generating techniques in the bosonic and heterotic strings at least to 2 loops (first order in $\alpha'$)~\cite{Borsato:2020wwk,Hassler:2020tvz,Codina:2020yma}.

Double Field Theory (DFT) \cite{Siegel:1993th,Hull:2009mi,Hohm:2010jy,Hohm:2010pp} is an attempt to make (abelian) T-duality manifest at the level of the low-energy action, at the expense of doubling the coordinates $x^m$ of the $D$-dimensional target-space by pairing them with a set of ``dual'' coordinates $\tilde x_m$ in $X^M=(\tilde x_m,x^m)$.
Diffeomorphisms, B-field gauge transformations and T-duality maps are combined into the $O(D,D)$ group, defined by the matrices $O_M{}^N$ that leave the $O(D,D)$ metric $\eta_{MN}$ invariant
\be\label{eq:eta}
\eta_{MN}=\left(
\begin{array}{cc}
0 & \delta^m{}_n\\
\delta_m{}^n & 0
\end{array}
\right),
\ee
meaning that $O_M{}^PO_N{}^Q\eta_{PQ}=\eta_{MN}$.
The action of DFT is manifestly invariant under constant $O(D,D)$ transformations of the coordinates $X^M$ and of the dynamical fields of the theory, i.e.~the generalised metric $\mathcal H_{MN}$ and the generalised dilaton $d$, where the latter is an $O(D,D)$ scalar. The DFT action reduces to the low-energy effective action of the string when constraining   the fields   to depend only on the physical coordinates $x^m$,  also known as the ``strong constraint''.
By relaxing the strong constraint DFT may also  be viewed as a way to go beyond supergravity to provide a description of backgrounds that are locally non-geometric. Nevertheless, in this paper  we will  simply use DFT as a convenient $O(D,D)$-covariant rewriting of (super)gravity, and thus the strong constraint is always assumed. We will use in particular  the so-called ``flux formulation'' of DFT \cite{Geissbuhler:2013uka} where the dynamical fields are the generalised dilaton $d$ and a generalised vielbein ${E_A{}^M}$ for the generalised metric, and where the equations of motion are  written exclusively in terms of the generalised fluxes ${\F_{ABC}}$ and ${\F_A}$ and their (flat) derivatives.

\textbf{(Super)gravity solution-generating techniques ---}
One motivation of our work is the study and  possibly the classification of solution-generating techniques in (super)gravity, where we  focus on those that admit a description in terms of $O(D,D)$. 
In this case  we assume that  the starting point is a (super)gravity solution, say with Neveu-Schwarz-Neveu-Schwarz  (NSNS) fields $G_{mn},B_{mn},\phi$ for the metric, Kalb-Ramond field and dilaton. 
The question  is whether it is possible to construct a map from these fields to a new set of  $G'_{mn},B'_{mn},\phi'$ that  also give rise to a (super)gravity solution.
Our strategy  is to rewrite the $D$-dimensional fields  in terms of the doubled fields of DFT, and demand that $G_{mn},B_{mn},\phi$ and $G'_{mn},B'_{mn},\phi'$  give rise to the same generalised fluxes  and flat derivatives ${\F}={\F'},\partial{\F}=\partial{\F}'$. This is a sufficient condition to have again a (super)gravity solution, and it is the mechanism that applies also for the generalised T-duality transformations.\footnote{Invariance of the generalised fluxes under NATD and PLTD was observed and used already in~\cite{Hassler:2017yza,Demulder:2018lmj,Sakatani:2019jgu,Catal-Ozer:2019hxw,Borsato:2020wwk,Hassler:2020tvz}, and in the case of the Yang-Baxter deformations that we discuss later in~\cite{Catal-Ozer:2019tmm,Borsato:2020bqo}. Ref.~\cite{Codina:2020yma} advocated more general local $O(D,D)$ transformations leaving the generalised fluxes invariant.} 
As remarked, we are interested in classifying solution-generating techniques in (super)gravity rather than the actual (super)gravity solutions, and for this reason at no point we need to impose the DFT equations of motion.

\textbf{Canonical transformations of $\sigma$-models ---}
Another motivation of our work is the classification of canonical transformations of 2-dimensional $\sigma$-models, see~\cite{Osten:2019ayq} for a similar discussion and the recent~\cite{Alfonsi:2021bot} for the relation between the symplectomorphism approach to T-duality and geometric quantisation. Consider a $\sigma$-model $S=\int d^2\sigma\mathcal L$ 
whose Lagrangian can be put in the form\footnote{We are ignoring  the overall string tension. A dot denotes derivatives with respect to the worldsheet time $\tau$, and a prime with respect to the worldsheet spatial coordinate $\sigma$. In this case we limit the discussion to the classical level, and we therefore ignore the dilaton.} $\mathcal L=\tfrac12 G_{mn}\dot x^m\dot x^n-\tfrac12 G_{mn}x'{}^mx'{}^n-B_{mn}\dot x^mx'{}^n$, which is the Polyakov action in conformal gauge. Introducing momenta $p_m$ conjugate to $x^m$ and going to the first-order formalism, the Lagrangian is $\mathcal L=p_m\dot x^m-\mathcal H$ with the Hamiltonian $\mathcal H=\tfrac12 \Psi^M\mathcal H_{MN}\Psi^N$, where one has 
$\Psi^M\equiv\left( p_m, x'{}^m\right)$ and
\be\label{eq:H}
\mathcal H_{MN}\equiv\left(\begin{array}{cc}
G^{mn}& -(G^{-1}B)^m{}_n\\
(BG^{-1})_m{}^n& G_{mn}-(BG^{-1}B)_{mn}
\end{array}
\right),
\ee
which coincides with the known parametrisation commonly used for the generalised metric of DFT.\footnote{If we further identify the momenta with the spatial derivatives of a dual set of coordinates $p_m=\tilde x'_m$ then $X^M=(\tilde x_m,x^m)$ and
we have the action of Tseytlin's double $\sigma$-model~\cite{Tseytlin:1990va} $
S=\int d^2\sigma\left(\tfrac12 \partial_1 X^M(\eta_{MN}+\Omega_{MN})\partial_0X^N
-\tfrac12 \partial_1X^M\mathcal H_{MN}\partial_1X^N\right)$,
where  the matrix $
\Omega_{MN}\equiv\left(\begin{array}{cc}
0& \delta^m_n\\
-\delta_m^n& 0
\end{array}
\right)
$
denotes the topological term used also in~\cite{Copland:2011wx}.
In this discussion we want to look at the standard $\sigma$-model in the Hamiltonian formalism and we therefore refrain from going to the  doubled $\sigma$-model.\label{foot:DSM}}
Canonical Poisson brackets for $x^m$ and $p_m$ translate into Poisson brackets\footnote{
The  Poisson brackets are 
$\{\Psi_M(\sigma_1),\Psi_N(\sigma_2)\}=\tfrac12\eta_{MN}(\partial_1-\partial_2)\delta_{12}-\tfrac12 \Omega_{MN}(\partial_1+\partial_2)\delta_{12}$
where we use the shorthand notation $\partial_{1,2}=\partial_{\sigma_{1,2}}$, $\delta_{12}=\delta(\sigma_1-\sigma_2)$, and $\Omega_{MN}$ was defined in footnote~\ref{foot:DSM}.
With appropriate boundary conditions $(\partial_1+\partial_2)\delta_{12}$ vanishes in the sense of distributions and  the Poisson brackets can be rewritten as $\{\Psi_M(\sigma_1),\Psi_N(\sigma_2)\}=\eta_{MN}\partial_{\sigma_1}\delta(\sigma_1-\sigma_2)$, which is the expression that is commonly used and the one we use in the text. On the line $(\sigma\in \mathbb R)$ test functions should be smooth functions with compact support, and on the circle $(\sigma \in S^1)$ they should be periodic. In the case of the circle one cannot drop terms with $(\partial_1+\partial_2)\delta_{12}$ if the boundary conditions for test functions allow for non-trivial winding. We refer
to~\cite{Osten:2019ayq} for a discussion that uses the original form of the Poisson brackets.} 
 for $\Psi_M$,
  and redefining the phase-space variables as $\Psi_A=E_A{}^M\Psi_M$
via a generalised vielbein $E_A{}^M$ for the generalised metric $\mathcal H_{MN}$, one has
\be
\{\Psi_A(\sigma_1),\Psi_B(\sigma_2)\}=-\F_{AB}{}^C(\sigma_1)\Psi_C(\sigma_1)\delta_{12}+\eta_{AB}\partial_{\sigma_1}\delta(\sigma_1-\sigma_2),
\ee
where the generalised fluxes $\F_{ABC}$ appear  again.
Therefore when two different $\sigma$-models  admit a rewriting of the phase-space variables in terms of $\Psi_A$ and $\Psi'_A$ respectively, and when they give rise to the same generalised fluxes $\F_{ABC}$ when computing the Poisson brackets, the two $\sigma$-models are related by a canonical transformation --- \emph{possibly up to zero modes}.
This last remark is a consequence of the fact that only the \emph{spatial derivative} of the coordinates $x^m$ appear in $\Psi_M$, and therefore with the above argument one is not able to claim that the zero-mode contribution to the Poisson brackets remains invariant under the map.
To use a uniform terminology throughout the paper we will  use  the term ``solution-generating techniques'' notwithstanding that we have in mind   canonical transformations of $\sigma$-models as well.

\textbf{Integrable $\sigma$-models ---} Worldsheet (classical) integrability is a statement about the on-shell 2-dimensional $\sigma$-model, whose equations of motion can be put in the form of a flatness condition for an object known as the ``Lax connection''.\footnote{This is a weak notion of integrability, since in general one should argue that the full tower of conserved charges obtained from the monodromy matrix are in involution.}
Canonical transformations  provide  on-shell identifications of the two $\sigma$-models, so that if one of the two models admits a formulation in terms of a Lax connection, one can use the canonical transformation to construct the Lax connection  for the other $\sigma$-model, and argue integrability also in that case.
Integrability has played an important role in the understanding of the $AdS_5/CFT_4$ correspondence~\cite{Beisert:2010jr}, and much progress was reached also in lower dimensional holographic examples. 
Being able to generate new supergravity solutions (for example starting from $AdS_5\times S^5$) that retain worldsheet integrability gives hope of applying exact methods to non-maximally supersymmetric backgrounds. Additionally, at least when the transformations can be understood as deformations of the original model, this motivates also the search of the corresponding transformation of the  holographically-dual conformal field theory (in this example $\mathcal N=4$ super Yang-Mills).
Yang-Baxter (YB) deformations are integrability-preserving solution-generating techniques that recently have been extensively studied from various points of view, and that started to be relevant for the superstring in~\cite{Delduc:2013qra,Kawaguchi:2014qwa}. At least the so-called ``homogeneous'' YB deformations can be interpreted as solution-generating techniques in the sense of this paper, since they leave the generalised fluxes invariant. The original formulation, the so-called ``$\eta$-model'' or ``inhomogeneous'' YB-model constructed in~\cite{Klimcik:2002zj,Klimcik:2008eq} will also be discussed later.
The homogeneous YB deformations are based on solutions of the classical YB equation on the algebra of isometries of the original background, and they can be applied to generic isometric backgrounds~\cite{Borsato:2018idb} including the integrable $AdS_5\times S^5$ background, giving rise to generalisations of TsT transformations~\cite{Osten:2016dvf}.
Preliminary proposals for the corresponding deformations of $\mathcal N=4$ super Yang-Mills were put forward in~\cite{vanTongeren:2015uha,Araujo:2017jkb}.
As remarked, we are interested in classifying integrability-preserving transformations of 2-dimensional $\sigma$-models, rather than the actual integrable models, and for this reason at no point we need to impose the existence of a Lax connection.

\textbf{$\alpha'$-corrections ---}
The low-energy effective action of the string has higher-derivative $\alpha'$-corrections that admit an $O(D,D)$-covariant formulation at least to 2-loop order.\footnote{Recently a tension was identified at 4 loops for the quartic-Riemann terms multiplied by $\zeta(3)$ that appear for the (super)string~\cite{Hronek:2020xxi}.}
In particular, as at leading order in $\alpha'$, the $\alpha'$-corrected equations of motion of DFT may still be written in the flux formulation  in terms of the generalised fluxes and their flat derivatives exclusively.
This observation can be used to extend the   solution-generating techniques that we classify to higher orders in $\alpha'$.
This strategy was first employed in~\cite{Borsato:2020bqo} to obtain the first $\alpha'$-correction for homogeneous YB deformations, and later for NATD and PLTD\footnote{{For applications of PLTD to obtain string solutions and for different approaches to discuss the $\alpha'$-corrections when applying PLTD see for example~\cite{Eghbali:2020ozg,Eghbali:2018ohx,Eghbali:2017ydo} and references therein.}} in~\cite{Borsato:2020wwk,Hassler:2020tvz,Codina:2020yma}. As argued in~\cite{Codina:2020yma} the same methods can be applied to more general $O(D,D)$-covariant solution-generating techniques, as the ones that we consider in this paper.

\textbf{Summary of the paper ---} 
Starting with section~\ref{sec:ansatz}, we will employ an ansatz for the generalised vielbeins  that is known in the literature as ``generalised Scherk-Schwarz reductions''~\cite{Aldazabal:2011nj,Geissbuhler:2011mx,Grana:2012rr}. Under this ansatz we  identify a $d$-dimensional subspace of the full $D$-dimensional spacetime, so that we can  discuss the more general case of solution-generating techniques acting non-trivially   in $d\leq D$ dimensions. Our discussion  will  be local and we will not discuss global issues of the solutions. We will further restrict ourselves to the case of \emph{constant} generalised fluxes in $d$ dimensions.   This set-up encompasses the generalised T-dualities and the prominent integrable deformations so far considered in the literature.  Subsequently, we  discuss the $O(d,d)$ parametrisation of the ``twist'' used in the reduction, as well as the constraints and the redundancies that arise. We finally explain  methods that will  turn out to be useful to treat in particular cases with non-vanishing $H$-flux. 
In section~\ref{sec:orbits} we  present  the classification of the ``orbits'', namely the possibilities of turning on the different  components of the generalised flux $\F_{IJK}$ (usually denoted as $H_{ijk}, F_{ij}{}^k, Q_i{}^{jk}, R^{ijk}$), and their ``representatives'', i.e.~the solutions for the twists that they allow.
While we initially focus on the fields of the NSNS sector, in section~\ref{sec:RR} we  discuss the Ramond-Ramond (RR) fields of the type II superstring as well.
We finish in section~\ref{sec:concl} with  conclusions and an outlook.
In appendix~\ref{app:notation} we collect our conventions on notation, in appendix~\ref{app:DFT} we give a brief recap on some aspects of DFT and gauged DFT that are relevant for our discussion, in appendix~\ref{app:parU} we discuss how to obtain the parametrisation of the twist that we use, in appendix~\ref{app:RR} we give more details on the formulations to include the RR fields of type II,  in appendix~\ref{app:eom} we review the DFT equations of motion in the flux formulation, and in appendix~\ref{app:PDE} we report on  other attempts we made to treat orbits with non-vanishing $H$-flux.

\section{Reduction ansatz and constant fluxes}\label{sec:ansatz}
We start by discussing a specific ansatz for the generalised vielbein of DFT. In turn this implies a specific ansatz for the NSNS fields (metric, B-field and dilaton) of the class of backgrounds that we consider. We refer to appendix~\ref{app:DFT} for some definitions and details that may be helpful for readers not familiar with DFT.

\subsection{Generalised Scherk-Schwarz ansatz}
To be more general, we assume that our backgrounds are parametrised by coordinates $\bm{x^m}=(\dot x^{\dot\mu},y^\mu)$, where we take $\bm m=0,\ldots,D-1$ and $\mu$ takes $d\leq D$ values. In principle $y^\mu$ may include the time direction.
In this splitting of coordinates $\dot x^{\dot \mu}$ will play the role of ``spectators'' --- we will never specify the $\dot x^{\dot \mu}$ dependence of the solution and $\dot x^{\dot \mu}$ will not participate in the solution-generating technique. The interesting discussion will therefore involve only the coordinates $y^\mu$. From now on, we will use a boldface notation for coordinates and fields of the full $D$-dimensional spacetime, to distinguish them from the coordinates and fields of the $d$-dimensional spaces. We do not make any assumptions  regarding their global properties, in fact our discussion will be valid only in local patches. We refer to~\cite{Inverso:2017lrz} for a discussion on how to construct generalised Leibniz parallelisable spaces from  generalised Scherk–Schwarz uplifts of gauged supergravities.

The ansatz we take for the generalised vielbein and generalised dilaton is~\cite{Aldazabal:2011nj,Geissbuhler:2011mx,Grana:2012rr}
\be\label{eq:twist}
\EE_{\bm A}{}^{\bm M}(\bm  x)=\dot \EE{}_{\bm A}{}^{\bm I}(\dot x)\ \bm U_{\bm I}{}^{\bm M}(y),\qquad\qquad
\bm d(\bm x)=\dot d(\dot x)+\lambda(y).
\ee
The factorisation of the dependence on the coordinates is what  plays a crucial role. We will use a dot for fields depending only on spectator coordinates. In the following we will distinguish between $M,N,\ldots$ indices and $I,J,\ldots$ indices, and similarly for the boldface version, and the reason will be clarified in section~\ref{sec:parU}. See also appendix~\ref{app:notation} for a recap of our conventions on notation. The matrix $\bm U$ is in general an element of $O(D,D)$, and $\bm U$ and $\lambda$ are usually called \emph{twists}.
Because of the role of spectators of the coordinates $\dot x^{\dot \mu}$, it is natural to take $\bm U$ of block form, not mixing $\dot\mu$ and $\mu$ directions, and acting as the identity in the spectator block, so that\footnote{Equation~\eqref{eq:twist} is rather natural and it is hard to imagine a more general ansatz if we want to exploit $O(D,D)$. A possible generalisation of~\eqref{eq:split-U} is to replace the identity matrix in the spectator block with another $O(D-d,D-d)$ element. The simplest example is \be
U_{\dot I}{}^{\dot M}=\left(
\begin{array}{cc}
e^{-\gamma(y)}  \delta^{\dot \mu}{}_{\dot \nu}& 0\\
0 & e^{\gamma(y)} \delta_{\dot \mu}{}^{\dot \nu}
\end{array}
\right),
\ee
where $\gamma$ playes the role of a generically $y$-dependent warping factor in front of $\bm G_{\dot\mu\dot\nu}$ and $\bm B_{\dot\mu\dot\nu}$, see~\cite{Geissbuhler:2011mx,Cagnacci:2018buk}.}
\be\label{eq:split-U}
\bm U=
\left(
\begin{array}{cc}
\bm {\mathsf a} & \bm {\mathsf b}\\
\bm {\mathsf c} & \bm {\mathsf d}
\end{array}
\right),\qquad\qquad
 U=
\left(
\begin{array}{cc}
 {\mathsf a} &  {\mathsf b}\\
 {\mathsf c} &  {\mathsf d}
\end{array}
\right),
\ee 
where we assume that the indices are placed as $\bm U_{\bm I}{}^{\bm M}$ and $U_I{}^M$, and
\be\label{eq:split-U-2}
\bm {\mathsf a}=
\left(
\begin{array}{cc}
 1_{D-d} &  0\\
 0 & {\mathsf a}
\end{array}
\right),\qquad
\bm {\mathsf b}=
\left(
\begin{array}{cc}
 0 &  0\\
 0 & {\mathsf b}
\end{array}
\right),\qquad
\bm {\mathsf c}=
\left(
\begin{array}{cc}
 0 &  0\\
 0 & {\mathsf c}
\end{array}
\right),\qquad
\bm {\mathsf d}=
\left(
\begin{array}{cc}
 1_{D-d} &  0\\
 0 & {\mathsf d}
\end{array}
\right).
\ee 
In the rest of this paper we will therefore work with the twist matrix $U_I{}^M\in O(d,d)\subset O(D,D)$.
This set-up  is in fact known as ``generalised Scherk-Schwarz compactifications'' and used also in gauged DFT, see appendix~\ref{app:DFT} for a short recap of certain aspects. 
Let us also mention that if we define the generalised metric $\bm{\mathcal H}^{\bm{MN}}=\bm{E_A{}^M\mathcal H^{AB}E_B{}^N}$ and  parametrise it as in~\eqref{eq:H} in terms of $\bm{G_{mn}}$ and $\bm{B_{mn}}$ (and similarly for the dotted fields) then equation~\eqref{eq:twist} implies that $\bm{M}\equiv \bm{G-B}$ is of the form
\be\label{eq:ODDM}
\bm M=(\bm {\dot M\mathsf{b+d}})^{-1}(\bm {\dot M\mathsf{a+c}}),
\ee
where $\bm{\dot M}\equiv \bm{\dot G}-\bm{\dot{B}}$.
This is the known transformation of $\bm M$ under $O(D,D)$ transformations.
In other words the generalised Scherk-Schwarz ansatz is selecting a class of backgrounds with metric, B-field (and dilaton) of a specific form.

Because of the above ansatz, the Weitzenb\"ock connection constructed out of $\bm{U_I{}^M}$ is non-vanishing only when the $\bm{IJK}$ indices are of the type $IJK$, i.e.~they are vector indices of the $O(d,d)\subset O(D,D)$ subgroup.
The generalised fluxes then become
\be
\begin{aligned}
\FF_{\bm {ABC}}&=\dot \FF_{\bm{ABC}}+\dot \EE_{\bm A}{}^I\dot \EE_{\bm B}{}^J\dot \EE_{\bm C}{}^K\F_{IJK},\\
\FF_{\bm A}&=\dot \FF_{\bm A} +\dot \EE_{\bm A}{}^I\F_I,
\end{aligned}
\ee
where $\dot \FF_{\bm{ABC}}=3\dot \OO_{\bm {[ABC]}}$, $\dot \OO_{\bm {ABC}}=\dot \EE_{\bm A}{}^{\dot\mu}\partial_{\dot\mu}\dot \EE_{\bm B}{}^{\bm J}\dot \EE_{\bm{ CJ}}$, $\dot \FF_{\bm A}=\dot\OO^{\bm B}{}_{\bm{BA}}+2\dot \EE_{\bm A}{}^{\dot \mu}\partial_{\dot\mu}\dot d$, and 
\be
\F_{IJK}=3\Omega_{[IJK]},\qquad\qquad
\F_I=\Omega^J{}_{JI}+2U_I{}^\mu\partial_\mu \lambda,\qquad\qquad
\Omega_{IJK}=U_I{}^{\mu}\partial_{\mu} U_J{}^{ N}U_{K N}.
\ee
After defining the flat derivative $\partial_{\bm A}\equiv \EE_{\bm A}{}^{\bm M}\partial_{\bm M}$ one finds that 
\be\label{eq:df}
\begin{aligned}
\partial_{\bm A}\FF_{\bm {BCD}}&=\dot \EE_{\bm A}{}^{\dot \mu}\partial_{\dot \mu}\FF_{\bm {BCD}}
+\dot \EE_{\bm A}{}^I\dot \EE_{\bm B}{}^J\dot \EE_{\bm C}{}^K\dot \EE_{\bm D}{}^LU_{I}{}^\mu\partial_\mu\F_{JKL},\\
\partial_{\bm A}\FF_{\bm B}&=\dot \EE_{\bm A}{}^{\dot \mu}\partial_{\dot \mu}\FF_{\bm B}+\dot \EE_{\bm A}{}^I\dot \EE_{\bm B}{}^JU_{I}{}^\mu\partial_\mu\F_{J}.
\end{aligned}
\ee
We will now further restrict the assumptions behind our calculations, and present the constraints coming from Bianchi identities.

\subsection{Constant fluxes and pre-Roytenberg algebra}
In general $\F_{IJK},\F_I$ may be $y$-dependent but in this paper we assume them to be constant. One motivation comes from interpreting the components of the 3-form generalised flux as the coefficients defining the brackets of a Courant algebroid~\cite{Grana:2008yw,Halmagyi:2009te,Blumenhagen:2012pc} on the Roytenberg algebra~\cite{Roytenberg:1999,Roytenberg:2001am}.
In general the Jacobiator of these brackets  is non-vanishing, as  can be checked in an $O(d,d)$-covariant way by calculating the Jacobiator of the C-bracket (see appendix~\ref{app:DFT} for its definition) of the generalised vielbeins (here the twists $U$)~\cite{Geissbuhler:2013uka}. In fact
\be
[[U_{ I},U_{ J}]_{(\text C)},U_{ K}]_{(\text C)}^{ M}=[\F_{ IJ}{}^{ L}U_{ L},U_{ K}]_{(\text C)}^{ M}
=(\F_{ IJ}{}^{ H}\F_{ KLH}-\partial_{ K}\F_{ IJL}+\tfrac12\partial_{ L}\F_{ IJK})U^{ LM},
\ee
and the Jacobiator is
\be
\text{Jac}(U_{ I},U_{ J},U_{ K})^{ M}=(-\mathcal Z_{ IJKL}+\tfrac12 \partial_{ L}\F_{ IJK})U^{ LM}.
\ee
On the strong constraint the Bianchi identities for the generalised fluxes imply $\mathcal Z_{ IJKL}=0$ (see appendix~\ref{app:DFT}). If we in addition want to deal with the structure  of a Lie algebra rather than the one of a Courant algebroid, it is natural to take the generalised fluxes constant, $\partial_{ L}\F_{ IJK}=0$, such that   the Jacobiator vanishes.
The assumption of constant fluxes (both  $\F_{IJK}$ and  $\F_I$) will simplify considerably our calculations to identify the solution-generating techniques, and ultimately this is the main motivation for this assumption.
In particular it will be easier to find different twists $(U,\lambda)$ and $(U',\lambda')$ that give rise to the same generalised fluxes $\F_{IJK}=\F'_{IJK},\F_I=\F'_I$, and their constancy immediately implies that not only  $\FF_{\bm{ABC}},\FF_{\bm A}$ but also their flat derivatives remain the same, see~\eqref{eq:df}.\footnote{In the context of gauged supergravities the constancy of the generalised fluxes is imposed to make sure that the $(D-d)$-dimensional theory does not depend on the coordinates $y$. Our motivation is different and at least in principle there would be nothing wrong with taking $\F_{IJK},\F_I$ to be $y$-dependent.}
Let us remark that demanding the invariance of $\F_{IJK},\F_I$ is a stronger condition compared to the invariance of $\FF_{\bm{ABC}},\FF_{\bm A}$, and relaxing this assumption may lead to interesting generalisations on which we will comment in the conclusions.

Let us now discuss the constraints coming from Bianchi identities for the generalised fluxes $\F_{IJK}$ and  $\F_I$. When these are constant the Bianchi identities (on the strong constraint) reduce to
\be
\F_{ [IJ}{}^{ H}\F_{ K]LH}=0, \qquad\quad
\F^{ K}\F_{ IJK}=0, \qquad\quad  
\F^{ I}\F_{ I}-\tfrac{1}{6}\F^{ IJK}\F_{ IJK}= 0 . \label{eq:BianchiFconst}
\ee
The first can be understood as the Jacobi identity for a Lie algebra, while the rest are constraints involving also $\F_I$. 
We introduce Lie algebra generators $\TT_I=(\tilde T^i,T_i)$ so that $[\TT_I,\TT_J]=\F_{IJ}{}^K\TT_K$, where indices are raised and lowered with the ad-invarant metric,
\begin{equation}
\eta_{IJ}=\left(\begin{array}{cc}0&\delta^i_j\\\delta_i^j&0\end{array}\right) ,
\end{equation}
that corresponds to the pairing $\llangle \TT_I,\TT_J\rrangle=\eta_{IJ}$. We adopt the usual notation for the components of the 3-form flux $\F_{IJK}$ when choosing upper or lower indices  $\F_{ijk}=H_{ijk},\F_{ij}{}^k=F_{ij}{}^k,\F_i{}^{jk}=Q_i{}^{jk},\F^{ijk}=R^{ijk}$. Notice that such an algebra is unimodular $\F_{IJ}{}^J=0$. The commutation relations of the ``pre-Roytenberg algebra''\footnote{We follow the terminology usually employed in the literature, so that ``Roytenberg algebra'' is used in the case of the Courant algebroid where $\F_{IJK}$ are generically non-constant, while ``pre-Roytenberg algebra'' is just a Lie algebra with a pairing $\eta_{IJ}$ that splits generators as $\TT_I=(\tilde T^i,T_i)$.} spanned by $\TT_I$, and that we denote by $\alg r$, read
\be\label{eq:preR}
[T_i,T_j]=F_{ij}{}^kT_k+H_{ijk}\tilde T^k,\qquad
[T_i,\tilde T^j]=Q_i{}^{jk}T_k-F_{ik}{}^j\tilde T^k,\qquad
[\tilde T^i,\tilde T^j]=Q_k{}^{ij}\tilde T^k+R^{ijk}T_k.
\ee
The Lie algebra of the  Drinfel'd double is obtained by setting $H=R= 0$ and keeping  $F$ and $Q$. When only $R=0$ the above  Lie algebra is known as \emph{quasi-Manin} triple.
In terms of all the fluxes the Bianchi identities read
\begin{align}
F_{[ij}{}^kH_{lm]k}=0,\label{eq:Jacobi1}\\
F_{[ij}{}^kF_{l]k}{}^m-H_{k[ij}Q_{l]}{}^{mk}=0,\label{eq:Jacobi2}\\
F_{ij}{}^kQ_k{}^{lm}+H_{ijk}R^{klm}-4Q_{[i}{}^{k[l}F_{j]k}{}^{m]}=0,\label{eq:Jacobi3}\\
Q_k{}^{[jl}Q_i{}^{m]k}-F_{ik}{}^{[j}R^{lm]k}=0,\label{eq:Jacobi4}\\
Q_k{}^{[ij}R^{lm]k}=0.\label{eq:Jacobi5}
\end{align}
\be \label{eq:BianchiF1}
H_{ijk}\F^k+F_{ij}{}^k\F_k=0,\qquad
Q_i{}^{jk}\F_k-F_{ik}{}^j\F^k=0,\qquad
R^{ijk}\F_k+Q_k{}^{ij}\F^k=0 , 
\ee
\be \label{eq:BianchiContr}
H_{ijk}R^{ijk} + 3 F_{ij}{}^k Q_k{}^{ij} = 6 \F_i \F^i .
\ee
The more generic case is the one when all fluxes $F_{ij}{}^k,H_{ijk},Q_i{}^{jk},R^{ijk}$ are non-zero. Nevertheless it is interesting to consider simpler  cases in which only some of the fluxes are turned on. We will call ``orbits'' the classes that have a definite set of fluxes turned on. The elements of each orbit (namely the possible solutions for the twist $U$) will be called ``representatives'' of the orbits. Turning on fewer fluxes essentially corresponds to studying sub-orbits. The possible cases are organised in the structure of a diamond, see Figure~\ref{fig:orbits}, where  each node of the diamond corresponds to a possible orbit and  its dual orbit under a rigid $T$-transformation, connected by an arrow. In particular under a rigid $T$-transformation $F\leftrightarrow Q$ and $H\leftrightarrow R$, see also the next section. The four corners of the diamond correspond to orbits that are self-dual under rigid $T$-transformations.\footnote{Self-duality is in general only at the level of the orbit, but it may be that even some representatives are self-dual.}
\begin{figure}
\centering
\begin{tikzpicture}
  \begin{scope}
\node (emp) at (0,0) {$(\emptyset)$};
\node (F) at (-3,-1) {$(F)\leftrightarrow (Q)$};
\node (H) at (3,-1) {$(H)\leftrightarrow (R)$};
\node (FQ) at (-6,-2) {$(F,Q)$};
\node (FH) at (-2,-2) {$(F,H)\leftrightarrow (Q,R)$};
\node (FR) at (2,-2) {$(F,R)\leftrightarrow (H,Q)$};
\node (HR) at (6,-2) {$(H,R)$};
\node (FQR) at (-3,-3) {$(F,Q,R)\leftrightarrow (F,H,Q)$};
\node (FHR) at (3,-3) {$(F,H,R)\leftrightarrow (H,Q,R)$};
\node (FHQR) at (0,-4) {$(F,H,Q,R)$};
  \end{scope}
\end{tikzpicture}
\caption{Diamond representing the possible (sub)orbits. Arrows relate orbits connected by rigid $T$-transformations. The orbits at the four corners of the diamond are self-dual under rigid $T$-transformations.}
\label{fig:orbits}
\end{figure}
The ``empty'' orbit $(\emptyset)$ with all fluxes vanishing contains as representative the $d$-dimensional torus with no flux turned on. One possible representative of the $(F)$-orbit is any background with isometries whose algebra has structure constants $F_{ij}{}^k$, so that in the $(Q)$-orbit one finds for example their non-abelian T-dual backgrounds. Other examples worth mentioning are the torus with $H$-flux in the $(H)$-orbit, Wess-Zumino-Witten (WZW) models in the $(F,H)$ orbit, and Poisson-Lie (PL) symmetric models in the $(F,Q)$ orbit.
In section~\ref{sec:orbits} we will focus on each orbit and discuss the representatives that appear in each of them.

\subsection{$O(d,d)$ parametrisation of the twist $U$}\label{sec:parU}
Under mild assumptions (see~\cite{Dibitetto:2012rk} and appendix~\ref{app:parU}) the most general $O(d,d)$ parametrisation of the twist $U$ is
\be\label{eq:parU}
U_I{}^M=
\left(
\begin{array}{cc}
\delta^i_j & \beta^{ij}\\
0 & \delta_i^j
\end{array}
\right)
\left(
\begin{array}{cc}
\delta^j_k & 0\\
b_{jk} & \delta_j^k
\end{array}
\right)
\left(
\begin{array}{cc}
(\rho^t)^k{}_\mu & 0\\
0 & (\rho^{-1})_k{}^\mu
\end{array}
\right),
\ee 
where $\beta^{ij}$ and $b_{ij}$ are antisymmetric ($\beta^{ij}=-\beta^{ji}$, $b_{ij}=-b_{ji}$) and $\rho_\mu{}^j$ are components of the matrix $\rho\in GL(d)$. From now on we will use the shorthand notation $\rho_i{}^\mu\equiv (\rho^{-1})_i{}^\mu$, so that the placing of indices will indicate if we use $\rho$ or its inverse.
Our motivation is to find \emph{different} $U, \n U$ and $\lambda, \n \lambda$ that give rise to the same generalised fluxes $\F_{IJK},\F_I$. Notice however that twists that are related by the group  of allowed gauge transformations --- i.e.~$GL(d)$ diffeomorphisms and gauge transformations of the B-field in $d$ dimensions --- should not be considered as genuinely different. We will call $H_{\text{geom}}$ this group. Therefore, rather than $O(d,d)$, we should take $U$ to belong to the coset $O(d,d)\setminus H_{\text{geom}}$. Under a $GL(d)$ diffeomorphism  one has the transformation $U_I{}^M\to U_I{}^NR_N{}^M$ where~\cite{Hohm:2012gk}
\be
R_M{}^N=\left(
\begin{array}{cc}
(r^t)^\mu{}_\nu & 0\\
0 & (r^{-1})_\mu{}^\nu
\end{array}
\right)\qquad\text{with}\qquad (r^{-1})_\mu{}^\nu=\partial_\mu V^\nu.
\ee
Therefore, we see that the action of $GL(d)$ diffeomorphisms is all reabsorbed in the transformation of $\rho$ in~\eqref{eq:parU}, meaning $\rho_i{}^\mu\to \rho_i{}^\nu\partial_\nu V^\mu$. This also implies that $\beta$ and $b$ in~\eqref{eq:parU}, while they may be $y$-dependent, they must transform as scalars under $GL(d)$ diffeomorphisms.
This is the reason why we prefer to distinguish between $i,j$ and $\mu,\nu$ indices, as well as $I,J$ and $M,N$ for the double indices.
Gauge transformations of the B-field in $d$ dimensions are obtained by the transformations  $U_I{}^M\to U_I{}^NJ_N{}^M$ where~\cite{Hohm:2012gk}
\be
J_M{}^N=\left(
\begin{array}{cc}
\delta^\mu{}_\nu & 0\\
x_{\mu\nu} & \delta_\mu{}^\nu
\end{array}
\right),\qquad\text{with}\qquad x_{\mu\nu}=\partial_{[\mu}\xi_{\nu]}.
\ee
It is easy to see that this transformation is completely reabsorbed just by the redefinition of $b$ in~\eqref{eq:parU} as $b_{ij}\to b_{ij}+\rho_i{}^\mu\rho_j{}^\nu\partial_{[\mu}\xi_{\nu]}$.
To conclude, when we say that $U$ and $\n U$ must be different we mean that they must be different elements of the coset $O(d,d)\setminus H_{\text{geom}}$, or in other words that there exists no element $h\in H_{\text{geom}}$ such that $U=  \n Uh$.

An important feature in our discussion is that the ansatz~\eqref{eq:twist} is invariant under the redefinition
\be \label{eq:twistRedundancy}
\bm\dot\EE_{\bm A}{}^{\bm I}\to\bm\dot \EE_{\bm A}{}^{\bm J}\bm V_{\bm J}{}^{\bm I},
\qquad
\bm U_{\bm I}{}^{\bm M}\to \bm V^{\bm J}{}_{\bm I}\bm U_{\bm J}{}^{\bm M}.
\ee
with $\bm V_{\bm I}{}^{\bm J}\in O(D,D)$ constant and decomposed in terms of a constant matrix $V_{I}{}^{J}\in O(d,d)$ similar as $\bm U$  in equations~\eqref{eq:split-U} and~\eqref{eq:split-U-2}.
 This redefinition by $V_{ J}{}^{ I}$ does not give rise to new backgrounds, because it signals only the redundancy of the freedom in the decomposition between $\dot\EE$ and $U$ in~\eqref{eq:twist}. However, under the rotation by $V$  the generalised fluxes will be rotated as $\n\F_{IJK}=V_I{}^UV_J{}^VV_K{}^W\F_{UVW}$ and $\n \F_I=V_I{}^U\F_U$. This  shows that different forms of the commutation relations of the pre-Roytenberg algebra $\alg r$ may be in fact related by rigid $O(d,d)$ transformations, and  therefore they need to be considered physically equivalent.
Hence, orbits are  classified by  \emph{equivalent classes}, where the equivalence is given by a rigid $O(d,d)$ change of basis. Nevertheless, in section~\ref{sec:orbits} we will  list them in the intuitive way of Figure~\ref{fig:orbits}, since the grouping into equivalent classes may require additional knowledge on the pre-Roytenberg algebra.
On the other hand, given a background with $\dot\EE_{\bm A}{}^{\bm I}$ and $\bm U_{\bm I}{}^{\bm M}$, if we are able to find a $\bm U'_{\bm I}{}^{\bm M}$ that yields the same generalised fluxes \emph{up to} a rigid $O(d,d)$ transformation, then we can apply a compensating transformation on $\dot\EE_{\bm A}{}^{\bm I}$ to make sure that $\FF_{\bm{ABC}},\FF_{\bm A}$ remain invariant.

Another interesting rigid $O(d,d)$ transformation, {that now can give rise to {new} inequivalent representatives} for the twist $U$, is
\begin{equation} \label{eq:UtransAuto}
\bm U_{\bm I}{}^{\bm M}\to \bm{ \tilde{U}}_{\bm I}{}^{\bm M} = \bm W_{\bm I}{}^{\bm J}\bm U_{\bm J}{}^{\bm M},
\end{equation}
where again $\bm W_{\bm I}{}^{\bm J}$ is decomposed in terms of a costant matrix $W_I{}^J\in O(d,d)$ similar as $\bm U$  in equations~\eqref{eq:split-U} and~\eqref{eq:split-U-2}, and $W_I{}^J$  implements an automorphism of the pre-Roytenberg algebra $\mathfrak{r}$. Notice that here we are not transforming $\dot\EE_{\bm A}{}^{\bm I}$. In particular, the fact that $W$ is an automorphism, i.e.~it satisfies  
 \begin{equation}
 W_I{}^{I'} W_J{}^{J'} W_{K}{}^{K'} \F_{I'J'K'} = \F_{IJK} , 
 \end{equation}
guarantees that the transformation \eqref{eq:UtransAuto} leaves the generalised fluxes $\F_{IJK}$ invariant. When studying (super)gravity solution-generating techniques we additionally require the automorphism to satisfy $W_I{}^{J} \F_J = \F_I$. In this case, 
 if $U_I{}^M$ is a representative, than the transformation \eqref{eq:UtransAuto} generates a new representative $\tilde{U}_I{}^M = W_I{}^J U_J{}^M$. Let us point out that --- since $W\in O(d,d)$ --- after this transformation the twist can still be parametrized as in \eqref{eq:parU}. 
When viewing the transformation as a solution-generating technique, it will be important to understand in which cases $W$ gives rise to gauge transformations in $H_{\text{geom}}$. This may be complicated to discuss in general, but in certain explicit examples  particular conclusions can be made. For instance in the $(F)$-orbit, modding out by $GL(d)$ transformations requires the automorphism $W$ to be \textit{outer}. Interestingly, when  $\mathfrak{r}$ is not semisimple  outer automorphisms may also involve continuous parameters. For example, given a 2-cocyle $\omega:\alg r\to\alg r$ (see section~\ref{s:Forbit} for  the definition), then $W=\exp(\zeta\omega)$ is an automorphism of the algebra with $\zeta\in\mathbb R$ a continuous parameter. In order to avoid having $W$ inner we have to impose that $\omega$ is not coboundary. Notice finally that using \eqref{eq:twistRedundancy}, the transformation by $W$ in \eqref{eq:UtransAuto} may equally  be seen as leaving the twist $U$ invariant but transforming the spectator contribution $\dot\EE_{\bm A}{}^{\bm I}$ of the background as $\dot\EE_{\bm A}{}^{\bm I} \rightarrow \dot\EE_{\bm A}{}^{\bm J}\bm{ W^I{}_J}$.

In general $O(d,d)$ transformations with ``off-diagonal'' components will  reshuffle the types of fluxes in  complicated ways.
Given the commutation relations~\eqref{eq:preR},
 an $O(d,d)$ redefinition $\TT_I'=h_I{}^J\TT_J$ with\footnote{This is~\eqref{eq:parU} where we take $\rho=1$ since it does not rotate the fluxes. However here $b$ and $\beta$ are \textit{constant} antisymmetric matrices.}
\be
V_I{}^J=\left(
\begin{array}{cc}
\delta^i{}_j +(\beta b)^i{}_j & \beta^{ij}\\
b_{ij} & \delta_i{}^j
\end{array}
\right),
\ee
gives rise to the rotated structure constants
$[T'_i,T'_j]=F'_{ij}{}^kT'_k+H'_{ijk}\tilde T'^k$, $[T'_i,\tilde T'^j]=Q'_i{}^{jk}T'_k-F'_{ik}{}^j\tilde T'^k$,
$[\tilde T'^i,\tilde T'^j]=Q'_k{}^{ij}\tilde T'^k+R'^{ijk}T'_k$,
where
\be \label{eq:rigidOdd}
\begin{aligned} 
H'_{ijk}={}&H_{ijk}+3F_{[ij}{}^lb_{k]l}-3b_{l[i}b_{j|m}Q_{|k]}{}^{lm}+b_{il}b_{jm}b_{kn}R^{lmn},\\
F'_{ij}{}^k={}&F_{ij}{}^k-3F_{[ij}{}^lb_{m]l}\beta^{mk}-H_{ijl}\beta^{lk}+2b_{l[i}Q_{j]}{}^{lk}+3b_{l[i}b_{j|m}Q_{|n]}{}^{lm}\beta^{nk}-b_{il}b_{jm}b_{pn}R^{lmn}\beta^{pk} \\
&+ b_{il} b_{jm} R^{lmk},\  \\
Q'_i{}^{jk} ={}& {Q_i{}^{jk} + 4 b_{l[i} Q_{m]}{}^{l[k} \beta^{j]m} + 3 b_{l[i} b_{m|p} Q_{|n]}{}^{lp} \beta^{nj} \beta^{mk}  - 2 F_{il}{}^{[j} \beta^{k]l} + 3 F_{[in}{}^l b_{m]l} \beta^{nj} \beta^{mk} } \\
& + {  H_{imn} \beta^{jm}\beta^{kn} + b_{il} R^{ljk} - 2 b_{il} b_{mn} R^{ln[j} \beta^{k]m} + \beta^{l[j} \beta^{k]n} b_{lm} b_{np} R^{mpq} b_{qi} , } \\
R'^{ijk}={}&R^{ijk}-3b_{lm}\beta^{l[i}R^{jk]m}-3\beta^{l[i}R^{j|mp}\beta^{|k]q}b_{lm}b_{pq}-\beta^{il}\beta^{jn}\beta^{kr}b_{lm}b_{np}b_{qr}R^{mpq}\\
&+3Q_n{}^{[ij}\beta^{k]n}-6Q_n{}^{l[i}\beta^{j|n}\beta^{|k]m}b_{lm}-3\beta^{l[i}\beta^{j|n}\beta^{|k]q}b_{lm}b_{pq}Q_n{}^{mp}\\
&+3\beta^{l[i}F_{lm}{}^j\beta^{k]m}+3\beta^{l[i}\beta^{j|m}\beta^{|k]p}b_{np}F_{lm}{}^n+\beta^{il}\beta^{jm}\beta^{kn}H_{lmn}.
\end{aligned}
\ee
Given a certain representative of a specific orbit, one may want to apply such an $O(d,d)$ transformation to see if it can be rotated to a different orbit with  less (or just different) types of fluxes  turned on.
In section~\ref{sec:orbits} we will often use this possibility to go from one orbit to another by means of rigid $O(d,d)$ transformations.
Alternatively, the above expressions may be used to classify the automorphisms of $\alg r$ implemented by $b$- and $\beta$-shifts.
Another important case is to take $U'=TU$ related by the rigid $T$-transformation\footnote{The matrix form of $T_I{}^J$ and $\eta_{IJ}$ coincide, but notice the different position of indices in the two definitions.}
\be\label{eq:T}
T_I{}^J=
\left(
\begin{array}{cc}
0 & \delta^{ij}\\
\delta_{ij} & 0
\end{array}
\right),
\ee 
which leads to the usual T-duality-like relations among the fluxes
\be\label{eq:dualF}
\begin{aligned}
&\n\F^i=\delta^{ij}\F_j,\qquad
&&\n H_{ijk}=\delta_{il}\delta_{jm}\delta_{kn}R^{lmn},\qquad
&&&\n Q_i{}^{jk}=\delta_{il}\delta^{jm}\delta^{kn}F_{mn}{}^l,\\
&\n\F_i=\delta_{ij}\F^j,\qquad
&&\n R^{ijk}=\delta^{il}\delta^{jm}\delta^{kn}H_{lmn},\qquad
&&&\n F_{ij}{}^{k}=\delta_{il}\delta_{jm}\delta^{kn}Q_{n}{}^{lm}.
\end{aligned}
\ee
This mechanism is what allows us to describe solution-generating techniques that involve also rigid $T$-transformations, like abelian/non-abelian/Poisson-Lie T-duality.

Let us remark that the concept of ``Poisson-Lie plurality'' --- namely the possibility of decomposing the same Drinfel'd double $\alg d$ in terms of different choices of subalgebras $\alg d=\alg g\oplus\tilde{\alg g}=\alg g'\oplus\tilde{\alg g}'$ --- can be understood as the transformation of $U$ by a constant $O(d,d)$ matrix $V$ as in~\eqref{eq:twistRedundancy}. What we have here is a generalisation of the traditional definition of PL plurality in the sense that we do not require $\alg r$ to have the structure of a Drinfel'd double, and any $O(d,d)$ transformation may in principle be considered, even those relating different orbits.\footnote{For explicit examples on how to relate representatives of the $(F,H)$ and of the $(F,Q)$ orbits see e.g.~\cite{Sakatani:2021eqt}.}

Let us now look at the explicit expressions for the components of the fluxes in the parametrisation~\eqref{eq:parU} used for the twist $U$.
One has
\be
\begin{aligned}
\F_{IJK}&=\delta_{[I}{}^i\delta_J{}^j\delta_{K]}{}^k\ {H_{ijk}}
+3\delta_{[I}{}^i\delta_J{}^j\delta_{K]k}\ F_{ij}{}^k
+3\delta_{[I}{}^i\delta_{Jj}\delta_{K]k}\ Q_i{}^{jk}
+\delta_{[Ii}\delta_{Jj}\delta_{K]k}\ {R^{ijk}},\\
\F_{I}&=\delta_{Ii}\ \F^i+\delta_{I}{}^i\ \F_i,
\end{aligned}
\ee
where 
\be
H_{ijk}\equiv\F_{ijk}=\F_{[ijk]},\qquad 
F_{ij}{}^k\equiv\F_{ij}{}^k=\F_{[ij]}{}^k,\qquad
Q_i{}^{jk}\equiv\F_i{}^{jk}=\F_i{}^{[jk]},\qquad
R^{ijk}\equiv\F^{ijk}=\F^{[ijk]},
\ee
are\footnote{See for example~\cite{Geissbuhler:2013uka} for similar expressions. The last two equations could be rewritten also as
\be\label{eq:FF3alt}
\begin{aligned}
Q_i{}^{jk}&=\partial_i\beta^{jk}
-2\beta^{l[j}\dr_{il}{}^{k]}
-\beta^{m[j}\beta^{k]n}H_{imn}\\
R^{ijk}&=-3\beta^{l[i}\partial_l\beta^{jk]}
-3\beta^{l[i}\beta^{j|m}\dr_{lm}{}^{|k]}
-\beta^{l[i}\beta^{j|m}\beta^{|k]n}H_{lmn}.
\end{aligned}
\ee}
\be\label{eq:FF3}
\begin{aligned}
H_{ijk}&= 3(\partial_{[i}b_{jk]}+\dr_{[ij}{}^lb_{k]l}),\\
F_{ij}{}^{k}&=\dr_{ij}{}^k+\beta^{kl}H_{ijl}\\
Q_i{}^{jk}&=\partial_i\beta^{jk}
-2\beta^{l[j}F_{il}{}^{k]}+\beta^{l[j}\beta^{k]m}H_{ilm}\\
R^{ijk}&=-3\beta^{l[i}Q_l{}^{jk]}
+3\beta^{l[i}\beta^{j|m}F_{lm}{}^{|k]}-\beta^{l[i}\beta^{j|m}\beta^{|k]n}H_{lmn}.
\end{aligned}
\ee
Here we defined
\be\label{eq:def-dr}
\partial_i\equiv \rho_i{}^\mu\partial_\mu,
\qquad\qquad
\dr_{ij}{}^k\equiv -2\rho_{[i}{}^\mu\rho_{j]}{}^\nu\partial_\mu\rho_\nu{}^k.
\ee
To make $\dr_{ij}{}^k$ appear also in the generalised flux with one index, we rewrite $\lambda=\bar \lambda-\tfrac12 \log\det \rho$ so that using $d\log\det\rho=\Tr(d\rho\rho^{-1})$ we find
\be\label{eq:FF1}
\F_i=\dr_{ij}{}^j+2\partial_i\bar\lambda,\qquad\qquad
\F^i=
\partial_j\beta^{ji}
+\beta^{ik}(\dr_{kj}{}^j+2\partial_k\bar\lambda)
=\partial_j\beta^{ji}
+\beta^{ik}\F_k.
\ee
Notice that \emph{if} $\dr_{ij}{}^k$ can be interpreted as structure constants of a Lie algebra, then its definition is simply the Maurer-Cartan (MC) identity for $\rho$, which can then be taken of MC form $\rho=g^{-1}dg=dy^\mu \rho_\mu{}^it_i$, where $t_i\in Lie(G)$ are the generators of the Lie algebra with structure constants $\dr_{ij}{}^k$ and $g\in G$.
This happens for example for all orbits that have $H_{ijk}=0$, since in this case the Bianchi identities imply that $F_{ij}{}^k$ solves the Jacobi identity and the above equations give $F_{ij}{}^k=\dr_{ij}{}^k$.
We will see that the discussion for orbits with non-vanishing $H$-flux is more complicated, and in the next section we present some methods that we will use in this case.

\subsection{Twist ansatz for orbits with $H$-flux} \label{s:AnsatzOrbitsH}
The methods explained in this section may be useful in general, but later we will actually use them  only for orbits which have non-vanishing $H_{ijk}$, so this section is not necessary for reading the discussion of the other orbits.\footnote{For a discussion on dealing with orbits with $H\neq 0$ without appealing to the methodology of this section, see appendix \ref{app:PDE}.} The particular issue with non-vanishing  $H$-flux is that $\dr_{ij}{}^k$ may not be interpreted as structure constants of a Lie algebra (it may not even be constant) and therefore $\rho$ can not be taken of MC form. Nevertheless, we can maintain a geometrical interpretation  by exploiting the fact that the \emph{doubled} manifold --- spanned locally by the coordinates $y^\mu$ and their T-duals $\tilde{y}_{\tilde \mu}$ --- can be interpreted as a group manifold $\mathdsl{R}$ associated to the pre-Roytenberg algebra $\mathfrak{r}$, equipped with the  bilinear ad-invariant form $\eta$ of split signature. This is in fact the set-up  also known as DFT$_{\text{WZW}}$ of \cite{Blumenhagen:2015zma} and employed also in~\cite{Demulder:2018lmj}.  We decompose the group elements $\mathdsl{g}(Y^M) \in \mathdsl{R}$ specifically as $\mathdsl{g}(Y) = \tilde{m}(\tilde{y}) m (y)$, where $\tilde{m} = e^{\tilde{m}_i (\tilde{y}) \tilde T^i}$ and $m = e^{m^i (y) T_i}$, which  is always possible locally. Notice that the subspaces spanned by $T_i$ and $\tilde T^i$ are not necessarily Lie subalgebras of $\mathfrak{r}$.\footnote{The interpretation given in~\cite{Demulder:2018lmj} is that the physical manifold is understood as a coset $\mathdsl{R}\setminus \tilde {\mathcal M}$, where $\tilde {\mathcal M}$ is parametrised by the $\tilde y_\mu$. This is possible only when the $R$-flux vanishes, since in this case the generators $\tilde{T}^i$ span a subalgebra of $\mathfrak r$. We will not need to appeal to this coset interpretation.  The parametrisations for the twists that we use manifestly satisfy the strong constraint.} Let us introduce in $\mathfrak{r}$ the adjoint action by $m$ and the right-invariant one-form $dm m^{-1}$ as
\begin{equation}
 m \TT_I m^{-1} = M_I{}^J \TT_J = e^{\text{ad}_{m^i T_i}} \TT_I , \qquad dm m^{-1} = dy^MV_M{}^I \TT_I = dy^\mu V_\mu{}^i T_i + dy^\mu V_{\mu i} \tilde T^i .
\end{equation}
We will denote the inverse of $V_\mu{}^i$ as $V_i{}^\mu$, i.e.~$V_\mu{}^i V_i{}^\nu = \delta_\mu^\nu$ and $V_i{}^\mu V_\mu{}^j = \delta_i^j$.
The MC identity for the one-form $dm m^{-1}$ in $\mathfrak{r}$ projected to the subspaces is
 \begin{equation} \label{eq:MCidforV}
 \begin{aligned}
 2 \partial_{[\mu} V_{\nu]}{}^i &= V_\mu{}^j V_\nu{}^k F_{jk}{}^i + 2 V_{[\mu}{}^j V_{\nu]k} Q_j{}^{ki} + V_{\mu j} V_{\nu k} R^{ijk} , \\
  2 \partial_{[\mu} V_{\nu]i} &= V_{\mu j} V_{\nu k} Q_i{}^{jk}  + 2 V_{[\mu}{}^j V_{\nu] k}  F_{ij}{}^k + V_\mu{}^j{} V_\nu{}^k H_{ijk} .
 \end{aligned}
 \end{equation}
In addition, to first order in the expansion $m = e^{m^i T_i}$, we have $M_i{}^j = \delta_i{}^j + m^l F_{li}{}^j + {\cal O}(m^2)$, $M^i{}_j = \delta^i_j - m^l F_{lj}{}^i + {\cal O}(m^2)$, $M_{ij} = m^l H_{lij} + {\cal O}(m^2)$ and $M^{ij} = m^l Q_l{}^{ij} + {\cal O}(m^2)$, so that we can take $M_i{}^j$ and $M^i{}_j$ invertible at least in a certain neighborhood around the identity, while $M_{ij}$ and $M^{ij}$ may not be invertible.
 
When the $H$-flux is non-vanishing,  depending on the orbit in consideration, we will consider two possible parametrisations for the twist $U_I{}^M$ that are equivalent to further specifications of the more general parametrisation given in \eqref{eq:parU}. Essentially they encompass an ansatz for the $\rho$-twist and a rewriting of the $\beta$- and $b$-twist of \eqref{eq:parU}. In particular, we will consider\footnote{If we relaxed the strong constraint we could simply take for the twist  the components of $\mathdsl g^{-1}d\mathdsl g$, since the MC identity would ensure that they satisfy the correct algebra relations. When imposing the strong constraint this is not possible because of the non-invertibility of the would-be frame field. Therefore one has to look for other solutions for the twists. Notice that $U^{(2)}$ is a generalisation of the twist used in~\cite{Demulder:2018lmj}, which is necessary  if one wants to be able to describe, for example, the so-called Yang-Baxter deformations.}
\begin{equation} \label{eq:parU1}
U^{(1)}_I{}^M =  \begin{pmatrix}
\delta^i_j + \xi^{il} \omega_{lj} & \xi^{ij} \\
\omega_{ij} & \delta_i{}^j
\end{pmatrix}_I{}^J M_J{}^K \begin{pmatrix}
(V^t)^k{}_\mu & 0 \\
0 & V_k{}^\mu
\end{pmatrix}_K{}^M \ ,
\end{equation}
or
\begin{equation} \label{eq:parU2}
U^{(2)}_I{}^M =M_I{}^J  \begin{pmatrix}
\delta^j_k + \xi^{jl} \omega_{lk} & \xi^{jk} \\
\omega_{jk} & \delta_j{}^k
\end{pmatrix}_J{}^K  \begin{pmatrix}
(V^t)^k{}_\mu & 0 \\
0 & V_k{}^\mu
\end{pmatrix}_K{}^M \ ,
\end{equation}
where the $\xi^{ij},\omega_{ij}$ that we use in $U^{(1)}$ and $U^{(2)}$ are different because of the different position of the adjoint action $M_I{}^J$ in the above expressions.  In general a twist given by $U^{(1)}$ may not be rewritten in the form of $U^{(2)}$, and viceversa, so generically speaking the two twists should not be viewed as  equivalent.\footnote{By equating the four block components of $U^{(1)}(\tilde{\xi}, \tilde{\omega})$ and $U^{(2)}(\xi, \omega)$ we will have for $M = \begin{pmatrix}
M_1 & M_2 \\ M_3 & M_4
\end{pmatrix}$ the relations $\tilde{\omega} = (M_4+M_3 \xi) \omega M_1^{-1}$ and $\tilde{\xi} = (M_4 + M_3 \xi)^{-1} M_1 \xi$  constraint with the consistency relations $\omega M_1^{-1} M_2 = (M_4+M_3 \xi )^{-1} M_3 \xi$ and $ M_1 \xi (M_4+M_3 \xi)^{-1} M_3 M_1^{-1} M_2 = M_2  (M_4+M_3 \xi)^{-1} M_3 \xi$. As we will see these are consistent when $M_2$ or $M_3$ are vanishing. In general it may not be possible to solve these constraints since $M_2$ or $M_3$ may not be invertible.}
When, respectively, $M^{ij} = 0$  ($M_{ij} = 0$) we will consider $U^{(1)}$ ($U^{(2)}$) since the inclusion of the adjoint action realises a simple shift\footnote{If we considered instead  $U^{(2)}$ ($U^{(1)}$), the $\rho$-twist would receive a contribution from the (unknown) $\omega_{ij}$ ($\xi^{ij}$) which would further complicate the discussion.} of $\omega_{ij}$  ($\xi^{ij}$) in terms of $M_{ij}$ ($M^{ij}$).  
Notice that both twists are automatically elements of $O(d,d)$ since $\xi^{ij}$ and $\omega_{ij}$ are antisymmetric matrices and  the bilinear form $\eta$ is ad-invariant. Furthermore $\xi^{ij}$ and $\omega_{ij}$ depend only on the coordinates $y^\mu$ such that the twists satisfy manifestly the strong constraint.

In the case that the $R$-flux $R^{ijk}$ is vanishing and $\xi^{ij} = 0$ the twist $U^{(2)}_I{}^M$ coincides with the generalised frame fields of \cite{Demulder:2018lmj} in which a particular solution for $\omega$ was found for general $(F,H,Q)$-flux. In fact, in the parametrisation $U^{(2)}_I{}^M$ we can generalise this solution to include non-vanishing $R^{ijk}$ and $\xi^{ij}$ as well.  In order to do so we follow    \cite{Demulder:2018lmj} and observe that  the flux equations can be written as 
\begin{equation} \label{eq:fluxTS}
\F_{IJK} = M_I{}^{I'} M_J{}^{J'} M_K{}^{K'} (T_{I'J'K'} + S_{I'J'K'}) =  T_{IJK} + S_{IJK},
\end{equation}
where,
\begin{equation} \label{eq:ExprTS}
\begin{alignedat}{3}
T_{IJK} &= 3 \hat{U}_{[I}{}^M \partial_M (\hat{U}_J{}^N) \hat{U}_{K]N}  , &\qquad  &  \hat{U}_I{}^M &= \begin{pmatrix}
(\delta^i_j + \xi^{il} \omega_{lj}) (V^t)^j{}_\mu & \xi^{ij} V_j{}^\mu \\
\omega_{ij} (V^t)^j{}_\mu & V_i{}^\mu
\end{pmatrix}, \\
S_{IJK} &= 3 \Lambda_{[I}{}^L \F_{JK]L}, &\qquad  &
 \Lambda_I{}^J &= \hat{U}_I{}^M V_M{}^J = \begin{pmatrix}
\xi^{ik} V_k{}^\mu V_{\mu j} & \xi^{ij} \\
V_i{}^\mu V_{\mu j} & \delta_i{}^j
\end{pmatrix} ,
\end{alignedat}
\end{equation}
and we have used  the identity 
\begin{equation} \label{eq:DerAdjoint}
\partial_M M_I{}^J = V_M{}^K M_I{}^L \F_{KL}{}^J .
\end{equation}
as well as the fact that the 
 adjoint action is an automorphism of $\mathfrak{r}$, 
 \begin{equation} \label{eq:MAuto}
 \F_{IJK} = M_I{}^{I'} M_J{}^{J'} M_{K}{}^{K'} \F_{I'J'K'} .
 \end{equation}
 While the equations for $F_{ij}{}^k$, $Q_i{}^{jk}$ and $R^{ijk}$ become involved due to the presence of $\xi^{ij}$, the equation for $H_{ijk}$ is independent of $\xi^{ij}$ (and the inclusion of $R^{ijk}$). In particular it reads
\begin{equation}  \label{eq:EqForOmegaparU2}
3 V_i{}^\mu V_j{}^\nu V_k{}^\rho \partial_{[\mu} \omega_{\nu\rho]} = - 2 H_{ijk} - 3 V_{[i}{}^\mu F_{jk]}{}^l V_{\mu l} ,
\end{equation}
where we have defined $\omega_{\mu\nu} = V_\mu{}^i \omega_{ij} V_\nu{}^j$. One can verify using \eqref{eq:MCidforV} that a particular solution for this (inhomogeneous) differential equation is given by $\omega^{\text{\tiny inhom.}} = \bar\omega^{(2)} - \Omega^{(2)}$ where, 
\begin{equation} \label{eq:defomega2}
\bar\omega^{(2)} = \frac{1}{2} V_\mu{}^i V_{\nu i} dy^\mu \wedge dy^\nu , 
\end{equation}
which satisfies
\begin{equation}
\partial_{[\mu}\bar\omega^{(2)}_{\nu\rho]} = \frac{1}{2} \left( -  V_\mu{}^i V_\nu{}^j V_\rho{}^k H_{ijk} - V_{[\mu}{}^i V_{\nu}{}^j V_{\rho ] k}  F_{ij}{}^k  +  V_{[\mu}{}^i  V_{\nu j} V_{\rho]k} Q_i{}^{jk} + V_{\mu i} V_{\nu j} V_{\rho k} R^{ijk} \right)
\end{equation}
and $\Omega^{(2)}$ chosen such that\footnote{When $R^{ijk}=0$, the three-form $d\Omega^{(2)}$  takes a nice expression in terms of the group elements $m$, in particular \cite{Demulder:2018lmj}
\begin{equation}
d\Omega^{(2)} = \frac{1}{12} \llangle dm m^{-1} , [dmm^{-1},dmm^{-1}] \rrangle .
\end{equation}}
\begin{equation} \label{eq:defOmega2unexpl}
d\Omega^{(2)} = \frac{1}{12} \F_{IJK} \  V^I \wedge V^J \wedge V^K  + \frac{1}{6} R^{ijk} \ V_i \wedge V_j \wedge V_k .
\end{equation}
Notice that $\Omega^{(2)}$ may not exist globally and an explicit expression can only be found by choosing particular coordinates in a local patch. The most general solution to \eqref{eq:EqForOmegaparU2} is then    given by $\omega_{ij} =\omega^{\text{\tiny inhom.}}_{ij} + \omega^{\text{\tiny hom.}}_{ij}$ where $\omega^{\text{hom.}}_{ij}$ is a closed two-form. 
 Let us point out that this is a solution for the  $\omega$ form of  $U^{(2)}$, and it is not a solution for the $\omega$  of $U^{(1)}$ in  \eqref{eq:parU1}.  Furthermore, within parametrisation $U^{(2)}$, it is still necessary to solve  the other flux equations for $F_{ij}{}^k$, $Q_i{}^{jk}$ and $R^{ijk}$ in terms of the unknown $\xi^{ij}$. 
Using the equations \eqref{eq:fluxTS} and \eqref{eq:ExprTS} we find in particular
\begin{equation}
\begin{aligned}
S_{ij}{}^k &= 2 F_{ij}{}^k - 2 Q_{[i}{}^{kl} \Lambda_{j]l} + \xi^{kl} H_{ijl} + \xi^{kn} \Lambda_{nl} F_{ij}{}^l , \\
S_i{}^{jk} &= Q_i{}^{jk} + \Lambda_{il} R^{ljk} - 2 F_{il}{}^{[j} \xi^{k]l} - 2  \Lambda_{ml} Q_i{}^{l[j} \xi^{k]m} , \\
S^{ijk} &= 3 Q_l{}^{[ij} \xi^{k]l} + 3 \Lambda_{ml} R^{l[ij} \xi^{k]m} ,
\end{aligned}
\end{equation}
with $\Lambda_{ij}  = V_i{}^\mu V_{\mu j}$, and
\begin{equation}
\begin{aligned}
T_{ij}{}^k &= \hat{\dr}_{ij}{}^k + \xi^{kl} H_{ijl} , \\
T_i{}^{jk} &= \hat{\partial}_i\xi^{jk}
-2\xi^{l[j}\hat{\dr}_{il}{}^{k]}
-\xi^{m[j}\xi^{k]n}H_{imn} , \\
T^{ijk}&=-3\xi^{l[i}\hat{\partial}_l\xi^{jk]}
-3\xi^{l[i}\xi^{j|m}\hat{\dr}_{lm}{}^{|k]}
-\xi^{l[i}\xi^{j|m}\xi^{|k]n}H_{lmn} ,
\end{aligned}
\end{equation}
where we have used eqs.~\eqref{eq:FF3} (by replacing $\beta^{ij}$ with $\xi^{ij}$ and $\rho_i{}^\mu$ with $V_i{}^\mu$) and defined $\hat{\partial}_i = V_i{}^\mu \partial_\mu$ and $\hat{\dr}_{ij}{}^k = - 2 V_{[i}{}^\mu V_{j]}{}^\nu \partial_\mu V_\nu{}^k$. Using the MC identity \eqref{eq:MCidforV} we have
\begin{equation}
\hat{\dr}_{ij}{}^k = - F_{ij}{}^k + 2 Q_{[i}{}^{kl} \Lambda_{j]l} - \Lambda_{il} \Lambda_{jm} R^{lmk} .
\end{equation}
The equations for $\xi^{ij}$ describing the $(F,H,Q,R)$-orbit can then be rewritten as
\begin{align}
& 2 \xi^{kl} H_{ijl} + \xi^{km} \Lambda_{ml} F_{ij}{}^l - \Lambda_{il} \Lambda_{jm} R^{lmk} = 0 , \label{eq:Xi1} \\
& \hat{\partial}_i \xi^{jk} + 2  \Lambda_{il} \xi^{m[j} Q_m{}^{k]l} + \Lambda_{il} R^{ljk} + 2 \Lambda_{il} \Lambda_{mn} \xi^{m[j} R^{k]ln} - \xi^{m[j} \xi^{k]n} H_{imn} = 0 , \label{eq:Xi2} \\
&3 Q_l{}^{[ij} \xi^{k]l} + 3 \xi^{l[i} \xi^{j|m} F_{lm}{}^{|k]} - 3 \xi^{l[i} \xi^{j|m} R^{|k]pq} \Lambda_{lp} \Lambda_{mq} + 2 \xi^{l[i} \xi^{j|m} \xi^{|k]n} H_{lmn} = R^{ijk} , \label{eq:Xi3}
\end{align}
where in the latter equation we have used \eqref{eq:Xi2}. Using \eqref{eq:Xi1} we can rewrite \eqref{eq:Xi3} also as
\begin{equation}\label{eq:Xi4}
3 Q_l{}^{[ij} \xi^{k]l} + 3 \xi^{l[i} \xi^{j|m} F_{lm}{}^{|k]} - 2 \Lambda_{lp} \Lambda_{mq} \xi^{l[i} \xi^{j|m} R^{|k]pq}  - \xi^{l[i} \xi^{j|m} \xi^{|k]n} \Lambda_{np} F_{lm}{}^p = R^{ijk} .
\end{equation} 
We will not solve these equations in general, but rather use them when studying   sub-orbits with $H\neq 0$.
Notice however that for a non-vanishing $R$-flux they do not allow a trivial solution $\xi = 0$.

While we will use these methods to look at orbits with non-vanishing $H$-flux, let us point out that the framework described here can cover the most general representatives of the $(F)$, $(Q)$, $(F,Q)$ and $(F,R)$ orbits derived later.

{\textbf{Comments on other possible ansatze} --- The methodology described above to find a suitable twist when $H\neq 0$  is far from the most general parametrisation considered in \eqref{eq:parU}: there are still several possibilities to generalise the two ansatze considered in \eqref{eq:parU1} and \eqref{eq:parU2}. 
\begin{enumerate}[(i)]
\item The most obvious generalisation is to replace  $V_\mu{}^i$ in \eqref{eq:parU1} and \eqref{eq:parU2} as $V_\mu{}^i \rightarrow A_\mu{}^\nu V_\nu{}^i$  for some matrix $A_\mu{}^\nu (y)$  which, to exclude diffeomorphisms, is not of the form $\partial_\mu a^\nu$ for some (vector) $a^\nu (y)$. Up to a rewriting, this is now equivalent to the general parametrisation \eqref{eq:parU}, which will be more convenient to use in practice.
\item  Another possibility is to consider a different parametrisation of the group elements $\mathdsl{g} \in \mathdsl{R}$. In particular  we can take instead $\mathdsl{g}(Y) = m (\tilde y) \tilde{m} (y)$, with $\tilde{m} = e^{\tilde{m}_i (y) \tilde{T}^i}$ and $m = e^{m^i (\tilde{y}) T_i}$, and define all the quantities of the twist $U$ in terms of the group elements $\tilde{m}$ (so that, again, $U$ manifestly satisfies the strong constraint). 
We define $\tilde{M}_I{}^J \mathdsl{T}_J = \tilde{m} \mathdsl{T}_I \tilde{m}^{-1}$ and $\tilde{V} = d\tilde{m} \tilde{m}^{-1} = dy^\mu \tilde{V}_{\mu i}\tilde T^i + dy^\mu \tilde{V}_\mu{}^i T_i$.  When trying to find non-trivial solutions for the equivalent of $\xi^{ij}$,  this parametrisation could be convenient when the generators $\tilde{T}^i$ span a subalgebra of $\mathfrak{r}$ such that $\tilde{V}_\mu{}^i = 0$ and $\tilde{M}^{ij} = 0$. In that case, we can consider the following twists
\begin{align}
\tilde{U}^{(3)}_I{}^M &=  \begin{pmatrix}
\delta^j_k + \xi^{jl} \omega_{lk} & \xi^{jk} \\
\omega_{jk} & \delta_j{}^k
\end{pmatrix}_I{}^J  \tilde{M}_J{}^K  \begin{pmatrix}
\delta^{kl} (\tilde{V}^t)_{l  \mu}  & 0 \\
0 & \delta_{kl} \tilde{V}^{l \mu}
\end{pmatrix}_K{}^M ,  \label{eq:parU3}\\
\tilde{U}^{(4)}_I{}^M &= \tilde{M}_I{}^J  \begin{pmatrix}
\delta^j_k + \xi^{jl} \omega_{lk} & \xi^{jk} \\
\omega_{jk} & \delta_j{}^k
\end{pmatrix}_J{}^K  \begin{pmatrix}
\delta^{kl} (\tilde{V}^t)_{l  \mu}  & 0 \\
0 & \delta_{kl} \tilde{V}^{l \mu}
\end{pmatrix}_K{}^M , \label{eq:parU4}
\end{align}
where we have included a rigid $T$-transformation, and where we have denoted the inverse of $\tilde{V}_{\mu i}$ as $\tilde{V}^{i\mu}$. Because $\tilde{M}^{ij}=0$ these two parametrisations are actually equivalent. When $\text{span}(\tilde{T}^i)$ does not form a subalgebra, however, we expect that not much is gained compared to the parametrisations of \eqref{eq:parU1} and \eqref{eq:parU2}.
\end{enumerate}

\noindent Although we will not consider these generalisations further, it would certainly be interesting to  systematically understand whether  non-trivial and inequivalent solutions for $\xi$ can be found in specific orbits within these other ansatze. See also appendix \ref{app:PDE} for some further comments.

 }

\section{Classification of orbits}\label{sec:orbits}
In the following we will discuss the orbits found by turning on all combinations of the fluxes $F,H,Q$ and $R$ as in Figure~\ref{fig:orbits}.
Let us remark that when looking at solution-generating techniques in supergravity we want to impose that both $\F_{IJK}=\F'_{IJK}$ and $\F_I=\F'_I$, while when looking at canonical transformations of $\sigma$-models it is enough to impose $\F_{IJK}=\F'_{IJK}$. That means that there may be certain (classical) canonical transformations of the $\sigma$-model that cannot be interpreted as supergravity solution-generating techniques. Examples of this kind are  non-unimodular homogeneous Yang-Baxter deformations, that will be discussed in the $(F)$-orbit (see~\cite{Hronek:2020skb} for some exceptions in this class).

\subsection{$\emptyset$}

The simplest orbit is the one where all fluxes are zero. From the equation for $F_{ij}{}^k$ one gets $\dr_{ij}{}^k=0$, which can be solved  up to $GL(d)$ transformations by 
 $\rho_\mu{}^i=\delta_\mu^i$. The equation for $H_{ijk}$ can be solved at least locally by taking $b_{ij}$  constant up to gauge transformations. Finally, the equation for $Q_i{}^{jk}$ is solved by taking also $\beta^{ij}$  constant, and the equation for $R^{ijk}$ is automatically satisfied. 
If the space is compact,\footnote{Remember that for $\mathbb R^d$ the second de Rham cohomology is trivial, and the exact $b_{ij}$ can be gauged away.} this is the example of the $d$-dimensional torus with no flux, decorated with constant $b$- and $\beta$-shifts.

\subsection{$F$} \label{s:Forbit}
In this orbit we assume that the only non-vanishing flux is the $F$-flux. First let us discuss the restrictions imposed by the Bianchi identities.
They imply that $F_{ij}{}^k$ satisfies the Jacobi identity. We will call $\alg g$ the corresponding Lie algebra with generators $T_i\in \alg g$. The remaining generators $\tilde T^i$ of the pre-Roytenberg algebra $\alg r$ form an abelian algebra $\tilde {\alg g}$, so that  we have $\alg r=\alg g\oplus\tilde{\alg g}$. Moreover $F_{ij}{}^k\F_k=0$ and $F_{ik}{}^j\F^k=0$ imply that
\be \label{eq:ForbitBianchiF1res}
\F_k=0 \text{ if }T_k\in [\alg g,\alg g],\qquad\qquad
\F^k=0 \text{ if }T_k\notin Z(\alg g),
\ee
where $[\alg g,\alg g]$ is the derived algebra and $Z(\alg g)$ is the center of the algebra. Notice that for \emph{semisimple} algebras this means that $\F_i=0$ and $\F^i=0$. Finally we also have the orthogonality condition $\F^i \F_i = 0$.

Let us now turn to solving the equations~\eqref{eq:FF3} for the fluxes in terms of the functions $\rho,b,\beta$ appearing in the parametrisation of the twist $U$.
First we have $F_{ij}{}^k=\dr_{ij}{}^k$ and, therefore,  the definition of $\dr_{ij}{}^k$ in~\eqref{eq:def-dr}  reads like the (left) MC identity. Although we can take $\rho$ of MC form, this solution can be generalised by taking $\rho_i{}^\mu=W_i{}^j\bar\rho_j{}^\mu$ with  $\bar\rho=\bar g^{-1}d\bar g$ of MC form for a $\bar g\in G$ and $W$ an automorphism of the Lie algebra, $W_i{}^lW_j{}^m F_{lm}{}^n=F_{ij}{}^kW_k{}^n$.
To mod out by $GL(d)$ diffeomorphisms, we have to take $W$ to be an \emph{outer} automorphism. The dressing by the automorphism $W$ will burden  the following expressions, but we prefer to keep $W$ explicitly because different outer automorphisms $W$ will correspond to inequivalent representatives.

Solving for $H_{ijk}=0$ globally is a question about the second de Rham cohomology of the  manifold. 
It is well known that if $\alg g$ is the Lie algebra of the compact and connected Lie group $G$, then the $n$-th cohomology group $H^n(G)$ with
real coefficients
is isomorphic to  
$H^n(\alg g,\mathbb R)$, the $n$-th Chevalley-Eilenberg Lie algebra cohomology with coefficients in $\mathbb R$~\cite{Chevalley:1948zz}.
In other words, one has to impose $H_{ijk}=0$  with $b_{ij}$ constant, which leads to the equation $F_{[ij}{}^lb_{k]l}=0$ implying that $b_{ij}$ is a constant    2-cocycle. It is useful to rewrite this in operatorial form. Given a 2-cocycle $\omega_{ij}$ solving $F_{[ij}{}^l\omega_{k]l}=0$ one can construct a linear operator $\omega:\alg g\to \alg g$ such that $\omega T_i=\omega_{ij}T^j$, where algebra indices are raised\footnote{Notice that when we write $T^i$ we mean $T^i=\kappa^{ij}T_j$, which is different from the other generators $\tilde T^i$ of $\alg r$.} and lowered with a symmetric invariant bilinear form on the algebra,  $\kappa_{ij}=\braket{T_i,T_j}$. Then the 2-cocycle condition in operatorial form reads
\be
\omega[x,y]=[\omega x,y]+[x,\omega y],\qquad\qquad x,y\in \alg g.
\ee
Because we want to mod out by gauge transformations of the B-field, we have to impose that $b_{ij}$ is not coboundary, in other words $b_{ij}\neq F_{ij}{}^kx_k$ for some constants $x_k$, or equivalently in operatorial form $b\neq \Ad_x$ for some constant $x\in \alg g$.
If $\alg g$ is semisimple, all 2-cocycles are 2-coboundaries --- the second Lie algebra cohomology is trivial. Therefore, in the compact case interesting solutions to $H_{ijk}=0$ are possible only for $\alg g$ non-semisimple.
In the non-compact case there is no general theorem relating  the Lie algebra and de Rham cohomologies.\footnote{For example $\mathbb R^2$ is de Rham trivial, but the 2-dimensional abelian algebra has non-trivial Lie algebra cohomology.
In fact it admits a non-trivial 2-cocycle giving rise to the three-dimensional Heisenberg-Weyl algebra.}
Putting global issues aside, the equation $H_{ijk}=0$ may be solved for example by $b_{ij}=\hat b_{ij}+\hat {\hat b}_{ij}$ with $\hat b_{ij}=\braket{\hat \omega T_i,T_j}$ and $\hat {\hat b}_{ij}=\braket{\hat {\hat \omega}_{ g}T_i,T_j}$, where $\hat {\hat \omega}_{ g}=W^{-1} \circ \hat {\hat \omega}_{\bar g} \circ W$, $ \hat {\hat \omega}_{\bar g}= \AD_{\bar g}^{-1}\circ\hat {\hat \omega}\circ\AD_{\bar g}$  and both $\hat \omega$ and $\hat {\hat \omega}$ are constant 2-cocycles that are not 2-coboundaries.

The equation for $\beta^{ij}$ coming from $Q_i{}^{jk}=0$ is
\be 
\partial_i  \beta^{jk}
-2 \beta^{l[j}F_{il}{}^{k]}=0,
\ee
and its most generic solution is
\be \label{eq:betaYBdef}
\beta^{ij} =  \braket{ r_{{g}} (T^i) ,  (T^j) },
\ee
with $r_g=W^{-1}\circ r_{\bar g}\circ W$,  $r_{\bar g}=\AD_{\bar g}^{-1}\circ r\circ\AD_{\bar g}$ and $r$  a constant and antisymmetric $r^t=-r$ linear operator on the algebra. Transposition is understood with respect to the ad-invariant symmetric bilinear form of $\mathfrak{g}$, i.e.~$\braket{x,ry}=\braket{r^t x,y}$. 
It is easy to argue that the above is the most general solution.
In fact, we may view $r\equiv  \AD_{\bar{g}} W \circ \beta\circ W^{-1} \AD_{\bar{g}}^{-1}$  just as a redefinition of the variables for which we want to solve the equation. But then, by using $d\AD_g=\AD_g\Ad_{g^{-1}dg}$, one finds that the   differential equation for $\beta$ is equivalent to $\partial_\mu r^{ij}=0$, and therefore $r$ must be constant.

The equation $R^{ijk}=0$ reads
\be \label{eq:betaYB3}
\beta^{l[i} \beta^{j|m}F_{lm}{}^{|k]}=0,
\ee
which is known as the classical Yang-Baxter equation (CYBE) on $\alg g$. Therefore both $ \beta^{ij}$ and $r^{ij}$ must solve the CYBE.
To rewrite it in operatorial form we may introduce the linear operator $r:\alg g\to \alg g$ such that $rT^i=r^{ij}T_j$. Then the CYBE is
\be \label{eq:CYBE}
[rx,ry]-r([rx,y]+[x,ry])=0,\qquad\qquad\forall x,y\in\alg g.
\ee
Notice that $r$ may come with an overall continuous parameter --- in fact its entries may depend on several independent parameters --- so that we can interpret it as a deformation of a representative with $\beta=0$.

It is simple to discuss the rigid $O(d,d)$ automorphisms that leave this orbit invariant. 
From~\eqref{eq:rigidOdd} one sees that the algebra is invariant if and only if $F_{[ij}{}^lb_{k]l}=0$ and $F_{il}{}^{[j}\beta^{k]l}=0$. The former is a constant $b$-shift by a 2-cocycle, while the latter a constant $\beta$-shift where $\beta$ commutes with the adjoint action on the algebra. Both are included in the above solutions.\footnote{
Notice that for such $\beta$ one has $r_{\bar g}=r$ because the adjoint action commutes with $\beta$.}

One also finds
\be
\F_i=F_{ij}{}^j+2\partial_i\bar\lambda,\qquad\qquad
\F^i=-\beta^{jk}F_{jk}{}^i+2\beta^{ij}\partial_j\bar\lambda.
\ee
Notice that for consistency $\partial_i\bar\lambda$ must be constant which means $\partial_\mu(\rho_i{}^\nu\partial_\nu\bar\lambda)=0$. This in turn implies
$
\partial_\mu\partial_\nu\bar\lambda = \partial_\mu\rho_\nu{}^i\rho_i{}^\epsilon\partial_\epsilon\bar\lambda.
$
Antisymmetrising in $\mu,\nu$ and using the invertibility of $\rho$ one gets the condition
\be\label{eq:fdl}
F_{ij}{}^k\rho_k{}^\mu\partial_\mu\bar\lambda=0.
\ee
If an algebra $\alg g$ is such that its derived algebra is strictly a subalgebra $[\alg g,\alg g]\subset \alg g$, i.e.~there are certain generators of $\alg g$ not contained in $[\alg g,\alg g]$, then for those generators $\partial_i\bar\lambda$ is not set to zero by~\eqref{eq:fdl}. Solvable algebras are examples of this kind of algebras, and in this case one may have $\partial_i\bar\lambda=constant$ but $\bar\lambda$ not constant.
If the algebra is semisimple, then the above equation (and the invertibility of $\rho$)  imply $\bar\lambda$ constant. 
Notice in addition that $\beta$ must be such that $\F^i$ is constant. 

To summarise, the $(F)$-orbit contains representatives that are invariant under a group $G$ of isometries. They are found by taking $\rho$ of MC form and setting $\beta=0$. 
From these representatives one may construct the so-called Yang-Baxter (YB) deformations. They are found by switching on $\beta$ of the above form, in general multiplied by a deformation parameter. We refer to the section on the $(F,R)$-orbit for a discussion on the relation of YB-deformations to PL-plurality. Finally we have also included the possibility of outer automorphisms in this $(F)$-orbit.

\textbf{Comments on Yang-Baxter deformations ---}
Notice that if we start from an isometric background ($\beta=0$) then $\F^i=0$, and if we turn on a YB deformation ($\beta\neq 0$) then to have invariance of $\F$ we must have $\beta^{jk}F_{jk}{}^i=2\beta^{ij}\partial_j\bar\lambda$. If the original dilaton is isometric ($\bar\lambda$ is constant),  then this implies the unimodularity condition for $\beta$ and $r$~\cite{Borsato:2016ose}. Otherwise it looks like a generalisation of this condition. 
We can use YB deformations to give an example of the discussion on the geometric interpretation of the backgrounds as in appendix~\ref{app:geom}, and see how the fluxes in \emph{curved} indices change under a YB deformation of an isometric background. We refer to~\cite{Plauschinn:2018wbo} for a comprehensive review on non-geometric backgrounds in string theory. As the starting point ($\beta=b=0$) we take a background that has only non-vanishing $F_{\mu\nu}{}^\rho$. From~\eqref{eq:new-geom-F} with 
\be
h_M{}^N=\left(
\begin{array}{cc}
\delta^\mu{}_\nu & -\n\beta^{\mu\nu}\\
-\n b_{\mu\nu} & \delta_\mu{}^\nu+\n b_{\mu\rho}\n\beta^{\rho\nu}
\end{array}
\right),\qquad\text{where}\quad \n\beta^{\mu\nu}=\rho_i{}^\mu\rho_j{}^\nu\n\beta^{ij},\quad \n b_{\mu\nu}=\rho_\mu{}^i\rho_\nu{}^j\n b_{ij},
\ee
one finds that the fluxes in curved indices after the transformation are
\be
\begin{aligned}
\n H_{\mu\nu\rho}
&=6\n b_{\delta[\mu|}\n \beta^{\delta\alpha}F_{\alpha|\nu}{}^\gamma \n b_{\rho]\gamma},\\
\n F_{\mu\nu}{}^\rho
&=F_{\mu\nu}{}^\rho+4 \n b_{\delta[\mu}\n \beta^{\alpha[\rho}F_{\nu]\alpha}{}^{\delta]},\\
\n Q_\mu{}^{\nu\rho}
&=2\n \beta^{\beta[\nu}F_{\mu\beta}{}^{\rho]},\\
\n R^{\mu\nu\rho}
&=0,
\end{aligned}
\ee
where we used  the 2-cocycle condition for $b$ and the  CYBE for $\beta$.
Notice that we have the relations
\be
\n F_{\mu\nu}{}^\rho=F_{\mu\nu}{}^\rho+2\n b_{\delta[\mu}\n Q_{\nu]}{}^{\rho\delta},
\qquad
\n H_{\mu\nu\rho}=\tfrac32 \n b_{\alpha[\rho}\n F_{\mu\nu]}{}^\alpha .
\ee
To conclude, $R^{\mu\nu\rho}$ remains vanishing, but it is possible to shift $F_{\mu\nu}{}^\rho$ and generate $H_{\mu\nu\rho}$ and $Q_\mu{}^{\nu\rho}$. In general YB backgrounds have therefore an interpretation as T-folds, see also~\cite{Fernandez-Melgarejo:2017oyu}.

\subsection{$Q$}\label{s:Qorbit}
The $(Q)$-orbit is related to the previous one by a rigid $T$-transformation. Because the $(F)$-orbit contains isometric backgrounds, the $(Q)$-orbit will contain their non-abelian T-duals.
Similarly to the previous case, Bianchi identities imply the Jacobi identity for $ Q_i{}^{jk}$. Now the generators $\tilde T^i$ span a non-abelian  algebra $\tilde{\mathfrak{g}}$, while the generators $T_i$ span  an abelian algebra $\alg g$ so that $\alg r=\alg g\oplus\tilde{\alg g}$.

The conditions $ F_{ij}{}^k=\dr_{ij}{}^k=0$ and $ H_{ijk}=0$  may be solved as in the $(\emptyset)$-orbit by taking $ \rho_i{}^\mu=\delta_i{}^\mu$ and, at least locally, $ b$  constant up to gauge transformations.
The equation $ Q_i{}^{jk}=\partial_i\beta^{jk}$ is the interesting one in this orbit and it is easily integrated to 
\be \label{eq:Qorbitbetadef}
 \beta^{ij}= y^\mu\delta_\mu{}^k Q_k{}^{ij}+ \tilde\omega^{ij},
\ee
with $\tilde\omega^{ij}$ the constants of integration.
Finally, imposing $ R^{ijk}=0$ at each order in $y$ and  using the Jacobi identity for $Q_i{}^{jk}$, we find  that $ \tilde\omega$ must be a 2-cocycle of $\tilde{\alg g}$
\begin{equation}
Q_l{}^{[ij}\tilde\omega^{k]l}=0 .
\end{equation}
 Notice that in this case the role of upper and lower indices is exchanged compared to the previous $(F)$-orbit.
The 2-cocycle $\tilde\omega$ may be multiplied by an overall continuous parameter --- in fact its entries may depend on several independent parameters --- so that we can think of it as a deformation of the $\tilde\omega=0$ case.
Also in this discussion the above solutions are already including the rigid $O(d,d)$ transformations of the pre-Roytenberg algebra $\alg r$ under consideration.

For the generalised flux with one index we have
\be
\F_i=2\partial_i{\bar\lambda},\qquad
\F^i= Q_j{}^{ji}+2\tilde\omega^{ik}\partial_k{\bar\lambda}+2{y}^j Q_j{}^{ik}\partial_k{\bar{\lambda}}.
\ee
Constancy of the  fluxes implies $ {\bar\lambda}=\alpha_i y^i+{\bar\lambda}_0$  and $ Q_j{}^{ik}\alpha_k=0$, for all $i,j$, for some constants $\alpha_i$ and $\bar\lambda_0$. Therefore the fluxes become $\F_i=2\alpha_i$, and $\F^i= Q_j{}^{ji}+2\tilde\omega^{ik}\alpha_k$.
Bianchi identities imply
\be
 Q_i{}^{jk}{\F}_k=0,\qquad
 Q_k{}^{ij}{\F}^k=0, \qquad {{\F}^i {\F}_i = 0},
\ee
which again imply that ${\F}_k=0$ if $\tilde T^k\notin Z(\tilde{\alg g})$ and ${\F}^k=0$ if $\tilde T^k\in [\tilde{\alg g},\tilde{\alg g}]$. { Notice that the first and third conditions above hold thanks to $ Q_j{}^{ik}\alpha_k=0$ and antisymmetry of $\tilde{\omega}$. The second condition reads $
{Q}_k{}^{ij} {Q}_l{}^{lk} + 2 {Q}_k{}^{ij} \tilde{\omega}^{kl} \alpha_l = 0 $.
Recall that in the case of $\tilde{\alg g}$  semisimple we have the stronger conditions $\F_i=\F^i=0$. }

The fact that the $(F)$- and $(Q)$-orbits are related by a rigid $T$-transformation means that starting with an $F_{ij}{}^k$ we can  relate it to a $ Q'_i{}^{jk}$ as $F_{ij}{}^k=\delta_{il}\delta_{jm}\delta^{kn}  Q'_n{}^{lm}$. Similarly, $ \F_i=\delta_{ij}\F^{'j}$ and $\F^i=\delta_{ij}\F'_j$, and  the generators are identified as  $T_i=\delta_{ij}\tilde T^{'j}, T^i=\delta^{ij}\tilde T'_{j}$ so that the roles of $\alg g$ and $\tilde{\alg g}$ are exchanged.

\textbf{Comments on non-abelian T-duality ---}
If the starting point is an isometric background with only $F_{ij}{}^k\neq 0$, the above $\beta$ with $\tilde\omega^{ij}=0$ is in fact the $\beta$-twist of non-abelian T-duality (NATD), as can be easily seen by comparing to~\eqref{eq:ODDM}. 
Starting from $\F_i=F_{ij}{}^j+2\partial_i\bar\lambda$ and $\F^i=0$ in the $(F)$-orbit, then invariance of the fluxes  implies $ \F'_i=2\alpha_i=0$ in the $Q'$-orbit, and also $ Q'_j{}^{ji}=\delta^{ik}(F_{kj}{}^j+2\partial_k\bar\lambda)$ which means $ \partial_i\bar\lambda=-F_{ij}{}^j$. For an isometric dilaton $\bar\lambda$ is constant, and one gets also in this case a  unimodularity condition $F_{ij}{}^j=0$ which is the anomaly-free condition of \cite{Alvarez:1994np,Elitzur:1994ri}.
Let us use the example of NATD to see how the fluxes in \emph{curved} indices can transform, following the discussion of appendix~\ref{app:geom}. We start again with an isometric background where only $F_{\mu\nu}{}^\rho\neq 0$.
From~\eqref{eq:new-geom-F} we have
\be
h_M{}^N=\left(
\begin{array}{cc}
-\delta^{\tilde \mu}{}_i\tilde \beta^{ij}\delta_{jk}\rho_\nu{}^k & \delta^{\tilde \mu}{}_i\delta^{ij}\rho_j{}^\nu \\
\delta_{\tilde \mu}{}^i(1+\tilde b\tilde \beta)_i{}^j\delta_{jk}\rho_\nu{}^k & -\delta_{\tilde \mu}{}^i\tilde b_{ij}\delta^{jk}\rho_k{}^\nu
\end{array}
\right)
=\left(
\begin{array}{cc}
-\tilde \beta^{\tilde\mu}{}_\nu & \rho^{\tilde \mu\nu} \\
{\rho^t_{\tilde{\mu}\nu}}+(\tilde b\tilde \beta)_{\tilde \mu\nu} & -\tilde b_{\tilde \mu}{}^\nu
\end{array}
\right),
\ee
{where $\tilde \beta^{\tilde\mu}{}_\nu = \delta^{\tilde \mu}{}_i\tilde \beta^{ij}\delta_{jk}\rho_\nu{}^k$ and $\tilde b_{\tilde \mu}{}^\nu = \delta_{\tilde \mu}{}^i\tilde b_{ij}\delta^{jk}\rho_k{}^\nu$.} We are being explicit in writing all tensors and indices, and we prefer to use a tilde (rather than a prime) both on the functions and on the coordinate indices when they refer to the NATD representative of the $(Q)$-orbit. 
After the NATD transformation the fluxes are
\be
\begin{aligned} \label{eq:curvedIsoTGen}
\tilde H_{\tilde \mu\tilde \nu\tilde \rho}
&={3\delta_{[\tilde\mu}{}^i \delta_{\tilde\nu}{}^j\delta_{\tilde\rho] k}\tilde b_{i}{}^l \tilde b_{j}{}^m F_{lm}{}^k}\\
\tilde F_{\tilde \mu\tilde \nu}{}^{\tilde \rho}
&=
2\delta_{[\tilde\mu }{}^i\delta_{\tilde\nu] j}\delta^{\tilde\rho k}\tilde b_{i}{}^lF_{lk}{}^j\\
\tilde Q_{\tilde \mu}{}^{\tilde \nu\tilde \rho}
&=\delta_{\tilde\mu i}\delta^{\tilde\nu j}\delta^{\tilde\rho k}F_{jk}{}^i,\\
\tilde R^{\tilde \mu\tilde \nu\tilde \rho}
&=0.
\end{aligned}
\ee
where we used that $\tilde \beta$ is a 2-cocycle for the  Lie algebra with structure constants $Q_i{}^{jk}$.
Notice that $\tilde Q_{\tilde \mu}{}^{\tilde \nu\tilde \rho}$ is always non-zero when we dualise a non-abelian algebra, and the background has the interpretation of a T-fold. If we turn on $\tilde b$ we can also generate geometric fluxes. Finally we have the relations
$
\tilde F_{ \mu\nu}{}^{  \rho}=2\tilde b_{\delta[\mu}\tilde Q_{\nu]}{}^{\rho\delta}$, 
$\tilde H_{\mu\nu\rho}=-\tfrac32 \tilde b_{\alpha[\rho}\tilde F_{\mu\nu]}{}^\alpha$, notice the different sign  in the last equation compared to the YB case.

\textbf{Comments on ``deformed T-duals'' ---}
When in the $(Q)$-orbit we turn on $\tilde\omega^{ij}$ (which may include an overall deformation parameter) we generate representatives that can be understood as \emph{deformations} of NATD, and that in~\cite{Borsato:2016pas} were called ``deformed T-duals'' (DTD). See also~\cite{Borsato:2017qsx,Borsato:2018idb}. 
It turns our that YB-deformations are actually related to  DTD models~\cite{Hoare:2016wsk,Borsato:2016pas}, and here we rephrase this fact in  $O(d,d)$ language. In order to do that we will take the point of view of the YB-deformation, and for this reason we will have structure constants $F_{ij}{}^k$ for the non-abelian Lie algebra $\alg g$. This also means that we will need to lower the indices of $\tilde \omega$ with deltas $\tilde\omega_{ij}=\delta_{ik}\delta_{jl}\tilde\omega^{kl}$ to take care of the duality relation. 
First let us point out  that in the above discussion ${\alg g}$ does not need to be semisimple. Actually, if we want $\tilde\omega_{ij}$ to generate a non-trivial deformation from the point of view of DTD, then we must take $\alg g$ non semisimple.\footnote{In fact, given that the second Lie algebra cohomology of semisimple algebras is trivial,  any $\tilde\omega_{ij}$ would be a coboundary (i.e.~$\tilde\omega_{ij}=F_{ij}{}^kc_k$ for some $c_k$). Therefore we could remove $\tilde\omega$ by redefining the coordinates. It may still be interesting to consider $\tilde\omega$ coboundary because we can still relate to it YB deformations, as we are about to see. These will be equivalent to NATD only when the deformation parameter takes \emph{finite} values, the limit $\eta\to 0$ being degenerate. Prominent examples are Jordanian deformations, see~\cite{Kawaguchi:2014qwa,vanTongeren:2019dlq}.}  Therefore  in the following we will not assume  semisimplicity. Nevertheless, we will assume that $\alg g$ is a subalgebra of an algebra $ {\alg f}$ that admits a non-degenerate symmetric invariant bilinear form $\kappa$. If $T_i$ are generators of $\alg g$, we generate the subalgebra $\alg g^*$ of $ {\alg f}$ by $T^i$, where we use $\kappa^{-1}$ to raise the indices of $T_i$.
We can think of $\tilde\omega$ as a linear operator in $ {\alg f}$ if we restrict it to the subalgebras $\tilde\omega:{\alg g}\to {\alg g}^*$.   
Let us now assume that $\tilde\omega$ is invertible in this restriction and let us call $r: {\alg g}^*\to{\alg g}$ the inverse of $\tilde\omega$. This  means that $r\tilde\omega =  P$ and $\tilde\omega r= P^T$, where $ P, P^T$ project on ${\alg g},{\alg g}^*$ respectively.  Because $\tilde\omega$ is a 2-cocycle in $\alg g$, $r$ satisfies the CYBE (on the whole $ {\alg f}$). 
Now, given the coordinates $ y$ used in the DTD model, let us consider the following change of coordinates
\be
 y = \zeta  P^T\frac{1-{\AD_{\bar g}^{-1}}}{\log \AD_{\bar g}}\tilde\omega \log\bar g,\qquad \quad \bar g\in G.
\ee
Here $ y= y^\mu\delta_{\mu i}T^i\in {\alg g}^*$.
One can prove that this change of coordinates implies~\cite{Borsato:2017qsx}
\be
P^T(\Ad_{ y}+\zeta\tilde\omega)P=\zeta P^T(\tilde\omega_{\bar g})P,\qquad
d y=\zeta P^T(\tilde\omega_{\bar g}\bar g^{-1} d\bar g),
\ee
where $\tilde\omega_{\bar g}=\AD_{\bar g}^{-1}\tilde\omega \AD_{\bar g}$, which in components read\footnote{When writing the equations in components we are automatically implementing the projectors, because the indices $i,j$ were restricted to $\alg g$ from the beginning.}
\be
 y^\mu\delta_{\mu k}f_{ij}{}^k+\zeta\tilde\omega_{ij}=\zeta(\tilde\omega_{\bar g})_{ij},\qquad\qquad
d y^\mu=\zeta d\bar y^\nu\bar \rho_\nu{}^i(\tilde\omega_{\bar g})_{ij}\delta^{j\mu},
\ee
where we define $\bar \rho$ such that $ d\bar y^\nu\bar \rho_\nu{}^iT_i=\bar g^{-1}d\bar g$. 
We can take into account the above change of coordinates in the twist $\tilde U$ under consideration by using the above substitution and Jacobian, so that
\be
\tilde U= 
\left(\begin{array}{cc}
0 & \delta \\
\delta & 0
\end{array}\right)
\left(\begin{array}{cc}
1 & \tilde \beta \\
0 & 1
\end{array}\right)
\to
\bar U=
\left(\begin{array}{cc}
0 & \delta \\
\delta & 0
\end{array}\right)
\left(\begin{array}{cc}
1 & \zeta \delta\omega_{\bar g}\delta \\
0 & 1
\end{array}\right)
{
\left(\begin{array}{cc}
\zeta (\bar \rho\omega_{\bar g}\delta)^t & 0 \\
0 & \zeta^{-1} (\bar \rho\omega_{\bar g}\delta)^{-1}
\end{array}\right)
}
\ee
where the last matrix implements the Jacobian, and we are writing explicitly the $\delta^{ij}$ or $\delta_{ij}$ while omitting the indices.
A straightforward calculation gives
\be
\bar U=
\left(\begin{array}{cc}
0 & \eta r_{\bar g}\bar \rho^{-1}\\
-\zeta\omega_{\bar g}\bar \rho^t & \bar\rho^{-1}
\end{array}\right)=
\left(\begin{array}{cc}
1 & \eta r_{\bar g}\\
0 & 1
\end{array}\right)
\left(\begin{array}{cc}
1 & 0\\
-\zeta\omega_{\bar g} & 1
\end{array}\right)
\left(\begin{array}{cc}
\bar \rho^t & 0\\
0 & \bar\rho^{-1}
\end{array}\right)
\ee
where we used that $r$ is the inverse of $\omega$ and $\eta=\zeta^{-1}$. Thanks to a diffeomorphism, we were therefore able to rewrite the twist of DTD in terms of the one of a YB-deformation, plus a shift of $b$ so that $H_{ijk}=0$.
Interestingly, this $\bar U$ is not included in the discussion of appendix~\ref{app:parU}. 

\subsection{$R$}
Having $H=F=Q=0$  immediately implies that $R=0$. Indeed, it is well-known that the $(R)$-orbit is not realisable when imposing the strong constraint. Notice that therefore we cannot include the rigid $T$-transformation of the $(H)$-orbit as a solution-generating technique.

\subsection{$F,Q$} \label{s:FQorbit}

In the $(F,Q)$-orbit the Bianchi identities imply that both the $F$- and $Q$-flux satisfy the Jacobi identity on their own, and in addition there is an identity mixing them
\begin{equation} \label{eq:mixedJac}
F_{ij}{}^k Q_k{}^{lm}  - 4 Q_{[i}{}^{k[l} F_{j]k}{}^{m]}= 0 .
\end{equation}
Hence, $F$ and $Q$ can be interpreted as the structure constants of  Lie algebras $\mathfrak{g}$ and $\tilde{\mathfrak{g}}$ respectively, with generators $T_i$ and $\tilde T^i$.
Together with the ad-invariant pairing $\eta$ the structure $\mathfrak{r} = \mathfrak{g} \oplus \tilde{\mathfrak{g}}$ is known as a Drinfel'd double. Furthermore the Bianchi's for $\F_I$ read
\begin{equation}
\begin{aligned}
F_{ij}{}^k \F_k  = 0 , \qquad  Q_k{}^{ij} \F^k = 0 , \qquad  Q_i{}^{jk} \F_k -  F_{ik}{}^j  \F^k   = 0,  \qquad \F_k \F^k = \frac{1}{2} F_{ij}{}^k Q_k{}^{ij} .
\end{aligned}
\end{equation}
The first two identities imply
\begin{equation}
\F_i = 0~\text{if}~T_i \in [ \mathfrak{g}, \mathfrak{g}], \qquad \F^i =0~\text{if}~ T^i \in [\tilde{\mathfrak{g}} , \tilde{\mathfrak{g}}].  
\end{equation}
Finally notice that by tracing the mixed Jacobi identity \eqref{eq:mixedJac} over $i = l$ and $j = m$ we can also write the last identity as $\F_i \F^i = Q_i{}^{ji} F_{jk}{}^k$.

Let us now  solve the  flux equations~\eqref{eq:FF3} for the twist $U(\rho, b , \beta)$. The solution for the $\rho$- and $b$-twist found from the $F$- and $H$-flux equation, respectively, are equivalent to the solutions derived and explained in the $(F)$-orbit, see section \ref{s:Forbit}. Recall that, up to diffeomorphisms, we have $\rho_i{}^\mu = W_i{}^j \bar{\rho}_j{}^\mu$ with $W$ an outer automorphism of $\mathfrak{g}$ and $\bar{\rho} = \bar{g}^{-1} d \bar{g}$ of (left-invariant) MC form.  The $\beta$-twist, however, must now solve
\begin{equation} \label{eq:PDEbeta}
Q_i{}^{jk} = \partial_i \beta^{jk} + 2 \beta^{[j|l} F_{il}{}^{k]}    .
\end{equation}
Notice that this is a linear inhomogeneous partial differential equations (PDE). The most general solution to such equations is found by adding the most general homogeneous solution to a particular solution $\beta_{\text{\tiny inhom.}}$ of \eqref{eq:PDEbeta}. The $\beta$-twist of Poisson-Lie symmetric backgrounds \cite{Klimcik:1995ux,Klimcik:1995dy,Sfetsos:1997pi}, appropriately dressed by the automorphism $W$, is an example of such a particular solution. It is given by
\begin{equation} \label{eq:FQbeta}
\beta_{\text{\tiny inhom.}}^{ij} = \llangle  \AD_{\bar{g}^{-1}} \cdot P \cdot \AD_{\bar{g}} \widetilde{W} (\tilde{T}^i) , \widetilde{W} (\tilde{T}^j ) \rrangle ,
\end{equation}
where $\widetilde{W} = W^{-t}$ is an automorphism of $\tilde{\mathfrak{g}}$, $P$ projects on $\mathfrak{g}$, and recall that we have defined the bracket $\llangle \mathdsl{T}_I , \mathdsl{T}_J \rrangle = \eta_{IJ}$.\footnote{The property $\widetilde{W} = W^{-t}$ follows from requiring that $W$ and $\widetilde{W}$ form an automorphism in the Drinfel'd double as $W(\mathfrak{r}) =   \widetilde{W} (\tilde{\mathfrak{g}}) \oplus W (\mathfrak{g})$ which preserves the bilinear form $\eta_{IJ}$.} The most general homogeneous solution is known from the discussion in the $(F)$-orbit, i.e.~given by \eqref{eq:betaYBdef}, which we denote here as
\begin{equation}
\beta^{ij}_{\text{\tiny hom.}} = \eta  \langle  r_{\bar{g}} ( T^i ) , T^j \rangle ,
\end{equation}
where, recall, $r_{\bar{g}} \equiv W^{-1} \circ \AD^{-1}_{\bar{g}} \circ r  \circ \AD_{\bar{g}} \circ W$,  $r^t=-r$ and $T^i = \kappa^{ij} T_j$ with $\kappa_{ij} = \langle T_i, T_j \rangle$. Notice also that here we explicitly introduce a \textit{deformation} parameter $\eta$. At this point this is not necessary, but we can do it because we are solving a homogeneous equation, and it will be useful to solve for the $R$-flux equation. Therefore, turning on $\beta_{\text{\tiny hom.}}$ can be seen as a deformation of the PL-symmetric background at $\eta=0$.
Concluding, the most general solution to \eqref{eq:PDEbeta} is $\beta^{ij} = \beta_{\text{\tiny inhom.}}^{ij} + \beta^{ij}_{\text{\tiny hom.}}$.
Solving for $R^{ijk} = 0$  gives additional algebraic conditions on the operator $r$. In particular, by expanding order by order in $\eta$,  it must satisfy the equations\footnote{This is where the explicit parameter $\eta$ becomes useful. Because of the presence of $\eta$, the $R$-flux equation gives two  equations for $\beta_{\text{\tiny inhom.}}$ and $\beta_{\text{\tiny hom.}}$ which have a nice interpretation. However this splitting is not necessary, and in general one gets one single algebraic constraint on $\beta$.}
\begin{align}
2 \beta^{l[i}_{\text{\tiny inhom.}} \beta^{j|m}_{\text{\tiny hom.}} F_{lm}{}^{|k]} - \beta^{l[i}_{\text{\tiny hom.}} Q_l{}^{jk]}  = 0 , \label{eq:hYBPLCondition}\\ 
 \beta^{l[i}_{\text{\tiny hom.}} \beta^{j|m}_{\text{\tiny hom.}} F_{lm}{}^{|k]} = 0 ,
\end{align}
where we have used the identity $\beta_{\text{\tiny inhom.}}^{l[i}\beta_{\text{\tiny inhom.}}^{j|m} F_{lm}{}^{|k]} -\beta_{\text{\tiny inhom.}}^{l[i}Q_l{}^{jk]}=0$ (see e.g.~\cite{Sfetsos:1997pi}). Therefore, from the second condition we find again the requirement that $r$ must satisfy the CYBE \eqref{eq:CYBE}.\footnote{Let us point out that when $\tilde{\mathfrak{g}}$ is abelian, and thus $Q_i{}^{jk}$ is vanishing, then $\beta_{\text{\tiny inhom.}}^{ij} = 0$, and we safely reduce to the most general representative of the $(F)$-orbit in which $r$ must satisfy the CYBE.} The first condition, on the other hand, should be viewed as a compatibility condition between $\beta_{\text{\tiny inhom.}}$ and $\beta_{\text{\tiny hom.}}$. In order to interpret it in a field-independent way it will be convenient to write out the components of the adjoint action by $\bar{g}$ in the Drinfel'd double $\mathfrak{r}$,
\begin{equation}
\bar{g} T_i \bar{g}^{-1} = (\AD_{\bar{g}})_i{}^j T_j , \qquad \bar{g} \tilde{T}^i \bar{g}^{-1} = (\AD_{\bar{g}})^{ij} T_j + (\AD_{\bar{g}})^i{}_j \tilde{T}^j .
\end{equation}
From $\AD_{\bar{g}} \in O(d,d)$ and $\AD_{\bar{g}} \circ \AD_{{\bar{g}}^{-1}} = 1$ we can derive the relations $(\AD_{\bar{g}})^i{}_j = (\AD_{{\bar{g}}^{-1}})_j{}^i$, $(\AD_{\bar{g}^{-1}})^{ij} = (\AD_{\bar{g}})^{ji}$ and  $(\AD_{\bar{g}})^{ij} = - (\AD_{{\bar{g}}^{-1}})_l{}^i (\AD_{\bar{g}})^{ml} (\AD_{\bar{g}})_m{}^j$.
The  solution for $\beta^{ij}$ can now be written as 
\begin{equation} 
\beta^{ij}=(W^{-1})_l{}^i \bar{\beta}^{lm} (W^{-1})_m{}^j , \qquad \bar{\beta}^{ij} = (\AD_{\bar{g}})^{il} (\AD_{\bar{g}^{-1}})_l{}^j + \eta (\AD_{\bar{g}^{-1}})_l{}^i r^{lm} (\AD_{\bar{g}^{-1}})_m{}^j .
\end{equation}
Using the previous relations,  the automorphism properties for $W$ and those  for $\AD_{\bar{g}}$ in $\mathfrak{r}$, in particular using
\begin{equation} \label{eq:FQAutoQ}
Q_i{}^{jk} = (\AD_{\bar{g}})_i{}^l (\AD_{\bar{g}^{-1}})_m{}^j (\AD_{\bar{g}^{-1}})_n{}^k Q_l{}^{mn} - 2 (\AD_{\bar{g}})_i{}^l (\AD_{\bar{g}^{-1}})_m{}^{[j} (\AD_{\bar{g}})^{k]n} F_{ln}{}^m ,
\end{equation}
we find that the compatibility condition \eqref{eq:hYBPLCondition} takes the following simple form
\begin{equation}\label{eq:r2cQ}
r^{l[i} Q_l{}^{jk]} = 0 ,
\end{equation}
which is the condition of $r^{ij}$ being a 2-cocycle of $\alg{\tilde g}$.

It is interesting to see in which cases we are genuinely in the $(F,Q)$-orbit modulo rigid $O(d,d)$ transformations. From~\eqref{eq:rigidOdd} we find that  we can turn off the $Q$-flux --- and thus describe only the $(F)$-orbit --- 
when it is of the form $Q_i{}^{jk} = 2 F_{il}{}^{[j} \beta^{k]l}$ for some constant antisymmetric $\beta^{ij}$ which satisfies the CYBE $\beta^{l[i} F_{lm}{}^j \beta^{k]m} = 0$ on $\mathfrak{g}$. 
Equivalently, the $F$-flux can be turned off by a rigid $O(d,d)$ when it is of the form $F_{ij}^k = -2 b_{l[i} Q_{j]}{}^{lk}$ for some constant antisymmetric $b$ that satisfies the CYBE $b_{l[i} b_{jm} Q_{k]}{}^{lm} = 0$ on $\tilde{\mathfrak{g}}$.

For the generalised flux $\F_I$ we have
\begin{equation}
\F_i = F_{ij}{}^j + 2\partial_i \bar{\lambda} , \qquad \F^i = Q_j{}^{ji} - \beta^{jl} F_{jl}{}^i + 2 \beta^{ij} \partial_j \bar{\lambda} .
\end{equation}
Similarly to the $(F)$-orbit  we must have that $F_{ij}{}^k \partial_k \bar{\lambda} = 0$ such that $\partial_i \bar{\lambda} = \alpha_i$, and thus $\F_i$, is constant. Additionally, notice that the $\beta$-twist must be such that $\F^i$ is constant. One can verify that this is already implied by using  the ordinary Jacobi identities,  the trace of the mixed Jacobi identity $\eqref{eq:mixedJac}$ over $j,m$ and the mixed Bianchi identity for $\F^i$ and $\F_i$.

Summarising, in this orbit we identified the possibility of a generalisation of a homogeneous YB-deformation --- compatible as a 2-cocycle with the $Q$-flux --- which deforms the ordinary PL symmetric backgrounds \cite{Klimcik:1995ux,Klimcik:1995dy}
 described previously in the literature.\footnote{As in the case of the standard homogeneous YB-deformation of the $(F)$-orbit, it should be possible to understand also this case as a version of PL plurality, and it would be interesting to see this explicitly.}  Interestingly these deformations can be implemented even when the initial background has no isometry, in contrast to the usual homogeneous YB-deformations described in the $(F)$-orbit.

\subsection{$F,R$} \label{s:FRorbit}

In the $(F,R)$-orbit the Bianchi identities imply that $F_{ij}{}^k$ are the structure constants of the Lie algebra $\mathfrak{g}$ generated by $T_i$ and that the $R$-flux satisfies $F_{nm}{}^{[k} R^{ij]n} = 0$. In turn the latter implies the (weaker) condition that $R_{ijk} = \kappa_{il} \kappa_{jm} \kappa_{kn} R^{lmn}$ with $\kappa_{ij} = \langle T_i , T_j \rangle$ is a 3-cocycle of $\mathfrak{g}$. For the fluxes with one-index we must impose that $F_{ij}{}^k \F_k = F_{ik}{}^j \F^k = R^{ijk} \F_k = 0$ as well as the orthogonality $\F_i \F^i =0$. Hence we must have the same  conditions as given in \eqref{eq:ForbitBianchiF1res} together with $R^{ijk} \F_k = 0$. 

The solutions for $(\rho, b , \beta)$ to the flux equations \eqref{eq:FF3}  are equivalent to the solutions of the $(F)$-orbit apart from the condition that follows from the $R$-flux equation. Hence, the expression for the $\beta$-twist is given in \eqref{eq:betaYBdef}  and the solution for $\rho$ is given by $\rho_i{}^\mu = W_i{}^j \bar{\rho}_j{}^\mu$, where $W$ is an outer automorphism  and  $\bar{\rho} = \bar{g}^{-1} d \bar{g}$ is a left-invariant MC of $\mathfrak{g}$.  The $R$-flux equation now reads  by using $W, \AD_{\bar{g}} \in \text{Aut}(\mathfrak{g})$
\begin{equation} \label{eq:FRRflux1}
( W_{\bar g})_l{}^i (W_{\bar g})_m{}^j (W_{\bar g})_n{}^k R^{lmn} = 3  {r}^{l[i} {r}^{j|m} F_{lm}{}^{|k]} , \qquad W_{\bar g} \equiv W \cdot \AD_{\bar{g}} .
\end{equation}
Using the fact that in $\mathfrak{r}$ the adjoint action by $\bar{g}$ is of block-diagonal form, i.e.~$(\AD_{\bar{g}})_{ij} = (\AD_{\bar{g}})^{ij} = 0$, and that $\AD_{\bar{g}}$ is an automorphism of $\mathfrak{r}$, one can derive the following identity
$
( \AD_{\bar g})_l{}^i (\AD_{\bar g})_m{}^j (\AD_{\bar g})_n{}^k R^{lmn} = R^{ijk} ,
$
so that the condition \eqref{eq:FRRflux1} is in reality field-independent and can be written as
\begin{equation} \label{eq:FRRflux2}
R^{ijk} = 3  \bar{r}^{l[i} \bar{r}^{j|m} F_{lm}{}^{|k]}  , \qquad \bar{r}^{ij} \equiv (W^{-1})_l{}^i r^{lm} (W^{-1})_m{}^j .
\end{equation}
A natural choice is to take the $R$-flux of the form
\begin{equation} \label{eq:FRRfluxassumption}
R^{ijk} = \alpha \kappa^{il} \kappa^{jm} F_{lm}{}^k, 
\end{equation}
which is a 3-cocycle of $\mathfrak{g}$, and where $\alpha$ is a real constant. This is of course always a possible choice.  Assuming it from now on, \eqref{eq:FRRflux2} requires that ${r}$ solves the modified CYBE (mCYBE), that is
\begin{equation} \label{eq:mCYBE}
[r x, ry] - r([rx,y]+[x,ry]) = - c^2 [x,y] , \qquad \forall x, y \in \mathfrak{g} ,
\end{equation}
with, up to redefinitions of the $r$ operator, $c^2 = \alpha =  \{-1, +1\}$. Here  $c^2=1$ is known as a split $r$-matrix while $c^2 =-1$ as a non-split $r$-matrix.
Notice that when ${r}$ satisfies the mCYBE then so does $\bar{r}$. Turning on an outer automorphism $W$ can therefore be seen as mapping different solutions of the mCYBE to each other. When $\mathfrak{g}$ is a semisimple (and bosonic) Lie algebra, the canonical solution of the mCYBE  is known as the Drinfel'd-Jimbo $r$-matrix \cite{Drinfeld:1985rx,Jimbo:1985zk} which is unique \cite{Levendorskii1991} up to a $GL(l, \mathbb{C})$ freedom on the Cartan subalgebra (CSA) directions of $\mathfrak{g}$  with $l$ the rank of $\mathfrak{g}$. In particular the canonical  Drinfel'd-Jimbo $r$-matrix is given in a Cartan-Weyl basis of the complexified algebra $\mathfrak{g}_\mathbb{C}$ and annihilates the Cartan generators while multiplying positive and negative roots with $\mp c$ respectively. When the real form $\mathfrak{g}_\mathbb{C}$ is compact, no split solutions exist  and one can only consider a non-split $r$-matrix. When the real form is non-compact, however, both possibilities can exist.   Since an outer automorphism of semisimple algebras maps the CSA  and the set of simple roots to itself,  the inclusion of $W$ will not affect the canonical $r$-matrix.

To truly sit in the $(F,R)$-orbit we must take $c^2 \neq 0$  such that the $R$-flux does not vanishes. This means that ``turning on" $r$ (equivalently, $\beta$), e.g.~by means of a deformation parameter,  can not be seen as a solution-generating technique in the $(F,R)$-orbit.
Hence, in the $(F,R)$-orbit, and assuming the flux configuration of eq.~\eqref{eq:FRRfluxassumption},  we can describe only the split ($c^2=1$) and the non-split ($c^2 = -1$) inhomogeneous Yang-Baxter models (see e.g.~\cite{Klimcik:2002zj,Klimcik:2008eq,Hoare:2016ibq}). One can verify that in these cases we are genuinely in $(F,R)$ under  rigid $O(d,d)$ equivalence relations. In other words, we cannot turn off the $R$-flux and $F$-flux as can be seen by rewriting \eqref{eq:rigidOdd} for {$H=Q=H'=Q' =0$} such that
\begin{equation} \label{eq:FRrigidOddFR}
R'^{ijk} = 3   \bar{r}^{l[i} \bar{r}^{j|m} F_{lm}{}^{|k]}  + 3  \beta^{l[i} {\beta}^{j|m} F_{lm}{}^{|k]} 
\end{equation}
for some constant antisymmetric matrix $\beta^{ij}$. In general there may be no real solution for $\beta^{ij}$ that sets $R'^{ijk}$ to zero.\footnote{Interestingly, it would be possible to remove the $R$-flux if the algebra $\alg g$ admitted both split and non-split solutions of the mCYBE. It would also be interesting to understand the consequences of removing $R$ by relaxing the reality of $\beta$.} When we cannot turn off  $R^{ijk}$, we can also conclude that we cannot turn off $F_{ij}{}^k$. Otherwise, we would be in the $(R)$-orbit and recall that this orbit cannot be realised under the strong constraint.

However, an interesting observation is that these $(F,R)$-representatives can be mapped, using a particular rigid $O(d,d)$ transformation, to the self-dual $(F,Q)$-orbit of Poisson-Lie symmetric backgrounds. From~\eqref{eq:rigidOdd} and taking $b_{ij}= 0$ and $\beta^{ij} =  \langle \bar{r} T^i , T^j \rangle$ we find that\footnote{{Notice that here we do not require $Q'=0$ so that we do not have the expression \eqref{eq:FRrigidOddFR} for $R'$.}}
\begin{equation} \label{eq:FQfluxesEta}
H'_{ijk}=0, \quad F'_{ij}{}^k = F_{ij}{}^k, \quad Q'{}_i{}^{jk} = - 2 F_{il}{}^{[j} \beta^{k]l}, \quad R'{}^{ijk}  = 0,
\end{equation}
for generic $c^2$, so that this discussion applies also to the homogeneous YB-deformations of the $(F)$-orbit.  Here $Q'{}_i{}^{jk}$ are the structure constants of $\mathfrak{g}_r$ whose Lie bracket is defined as
\begin{equation}  \label{eq:gRLiebracket}
[x,y]_r = [rx , y] + [x, ry], \qquad \forall x,y \in \mathfrak{g} ,
\end{equation}
and which precisely underlies the bi-algebra structure corresponding to the Drinfel'd double $\mathfrak{r} = \mathfrak{g} \oplus \mathfrak{g}_r$ that in the $c^2=-1$ case gives rise to the PL-duality between the inhomogeneous YB-model and the $\lambda^\star$-deformation \cite{Hoare:2015gda,Sfetsos:2015nya,Klimcik:2015gba}. Therefore, when assuming the expression for $R^{ijk}$ given in \eqref{eq:FRRfluxassumption}, the discussion of the solution-generating techniques in this orbit (as well as the discussion on the $\F_I$-flux equations) is implicitly captured by the discussion given in section \ref{s:FQorbit}.
As already remarked, the above rigid $O(d,d)$ transformation that sends us to the $(F,Q)$-orbit can be understood as a notion of a PL-plurality transformation. 

\subsection{$Q,R$}
In the $(Q,R)$-orbit the Bianchi identities imply that $Q_i{}^{jk}$ can be interpreted as structure constants of a Lie algebra and that $R^{ijk}$ is a  3-cocycle. Solving the flux equations  \eqref{eq:FF3}  gives, for the same reasons as in section~\ref{s:Qorbit} for the $(Q)$-orbit, that $\rho_\mu{}^i = \delta_\mu^i$ and that $\beta^{ij}$ is given by \eqref{eq:Qorbitbetadef}. Using the Bianchi identity the $R$-flux equation becomes
\begin{equation}
R^{ijk} = - 3 \tilde{\omega}^{l[i} Q_l{}^{jk]} ,
\end{equation}
where recall $\tilde{\omega}^{ij}$ is an antisymmetric constant matrix. Importantly, a rigid $O(d,d)$ transformation as in~\eqref{eq:rigidOdd} can undo this $R$-flux contribution at no other expense, and therefore we are effectively describing the $(Q)$-orbit.\footnote{{In section~\ref{s:Qorbit} for the $(Q)$-orbit, $\tilde{\omega}^{ij}$ was constrained to be a 2-cocycle in order to have vanishing $R$-flux. In that discussion it was assumed that the symmetry in~\eqref{eq:twistRedundancy} had been fixed, or in other words that $\dot{\bm E}$ is not allowed to transform. Here we do not need to impose any constraint on $\tilde{\omega}^{ij}$ because the $R$-flux is not required to vanish at the start, it is rather removed by the transformation~\eqref{eq:twistRedundancy} under which $\dot{\bm E}$ is allowed to transform.} } Hence, the genuine $(Q,R)$-orbit with non-vanishing $R$-flux can not be realised when imposing  the strong constraint, 
and therefore we do not expect to be able to
apply a rigid T -transformation from the $(F, H)$-orbit when the $H$-flux is non-vanishing.
 This is reminiscent of the known anomaly obstructions to gauging global symmetries in WZW models~\cite{Witten:1991mm}, which are part of the $(F,H)$-orbit.\footnote{Notice that in our framework we can only deform/dualise one copy of the symmetry group of the WZW model, e.g.~the left one.}

\subsection{$F,Q,R$}

In the $(F,Q,R)$-orbit the $F_{ij}{}^k$ flux can still be interpreted as the structure constants of a Lie algebra $\mathfrak{g}$ {spanned by  the generators $T_i$},  while in general the $Q_i{}^{jk}$ flux satisfies  Bianchi identities mixed with $R^{ijk}$ and $F_i{}^{jk}$, see \eqref{eq:Jacobi3}---\eqref{eq:Jacobi5}. 

To solve for the twist functions $(\rho, b , \beta)$ of $U$ parametrised as in \eqref{eq:parU} we can observe that the flux equations for $H_{ijk}, F_{ij}{}^k$ and $Q_i{}^{jk}$ are identical to the equations in the $(F,Q)$-orbit of section \ref{s:FQorbit}. Therefore, the $b$-twist is found from the second de Rham cohomology of the manifold while the $\rho$- and $\beta$-twist take the form\footnote{See  section \ref{s:FQorbit} for the definitions of the various objects, which  are not affected by the presence of the $R$-flux.} 
\begin{equation}
\rho = g^{-1} d g , \qquad \beta^{ij} =  \llangle  \AD_{{g}^{-1}} \cdot P \cdot \AD_{{g}} \tilde{T}^i , \tilde{T}^j  \rrangle +  \eta  \langle  r_g ( T^i ) , T^j \rangle .
\end{equation}
with $g \in \exp \mathfrak{g}$ and $r_g = \AD_{g^{-1}} \circ r \circ \AD_g$. For simplicity we have dropped here the possibility of automorphisms of $\mathfrak{r}$ but recall from the general discussion around eq.~\eqref{eq:UtransAuto} that this is always possible.\footnote{For example, recall that $\rho_i{}^\mu = W_i{}^j \bar{\rho}_j{}^\mu $ with $W_i{}^j \in \text{Out}(\mathfrak{g})$ and $\bar{\rho} = \bar{g}^{-1} d\bar{g}$, $\bar{g} \in \exp \mathfrak{g}$, is also a solution for $\rho$. When we take for instance the automorphism $W_I{}^L \in \text{Aut} (\mathfrak{r})$ to be block-diagonal, i.e.~$W_{ij} = W^{ij} =  0$ then we will have $ {\widetilde{W}^i{}_j} \equiv W^i{}_j = (W^{-1})_j{}^i$ as well as the relations $Q_i{}^{jk} = W_i{}^l W^j{}_m W^k{}_n Q_l{}^{mn}$ and $R^{ijk} = W^i{}_l W^j{}_m W^k{}_n R^{lmn}$, so that the solution of the $Q$-flux equation  for $\beta^{ij}$ given in \eqref{eq:FQbeta} still holds.}
What does change in this orbit, however,  is the algebraic $R$-flux equation. We find
\begin{equation}
R^{ijk} = 3   \beta^{l[i}_{\text{\tiny hom.}} \beta^{j|m}_{\text{\tiny hom.}} F_{lm}{}^{|k]} - 3 \beta^{l[i}_{\text{\tiny hom.}} Q_l{}^{jk]} +
6 \beta^{l[i}_{\text{\tiny inhom.}} \beta^{j|m}_{\text{\tiny hom.}} F_{lm}{}^{|k]}  ,
\end{equation}
or in terms of the adjoint action by $g$ in $\mathfrak{r}$,
\begin{equation}
(\AD_g)_l{}^i (\AD_g)_m{}^j (\AD_g)_n{}^k R^{lmn} = 3 \eta^2 r^{l[i} r^{j|m} F_{lm}{}^{k]} - 3\eta r^{l[i} Q_l{}^{jk]}
\end{equation}
where we have used the identity \eqref{eq:FQAutoQ} (which we note  is not affected by the presence of $R$-flux). In the $(F,Q)$ orbit, in order to have vanishing $R$-flux, $r$ was required to be a solution of the CYBE for $F_{ij}{}^k$ and to satisfy the 2-cocycle condition with $Q_i{}^{jk}$. These conditions are now relaxed by the presence of the non-trivial $R$-flux. Notice that the right-hand-side of the above equation is constant (i.e.~$g$-independent), which puts a strong constraint on the left-hand-side. Taking $g = e^x$ with $x\in \mathfrak{g}$ and expanding order by order in $x$ we have respectively at leading and first order
\begin{equation} \label{eq:RfluxFQR}
\begin{aligned}
 R^{ijk} = 3 \eta^2 r^{l[i} r^{j|m} F_{lm}{}^{k]} - 3\eta r^{l[i} Q_l{}^{jk]} , \qquad
 F_{lm}{}^{[i} R^{jk]m} =0 .
 \end{aligned}
\end{equation}
The first condition then implies that we can remove the $R$-flux using a rigid $O(d,d)$ transformation. Indeed taking in  \eqref{eq:rigidOdd}   $b = 0$ and $\beta = -\eta r$ we  can turn off the $R$-flux at the expense of  shifting the $Q$-flux and, therefore,  we are effectively describing here the $(F,Q)$-orbit. Therefore we do not expect to be able to apply a rigid $T$-transformation from the $(F,H,Q)$-orbit of quasi-Manin triples to the $(F,Q,R)$-orbit.

\subsection{$H$} \label{s:Horbit}
We now start the discussion of orbits with non-vanishing $H$-flux, following the framework explained in section \ref{s:AnsatzOrbitsH}. 
The first example is the $(H)$-orbit. Notice that the Bianchi identities for $\F_{IJK}$ are trivially satisfied while those for $\F_I$ require $H_{ijk} \F^k =0$ and the orthogonality $\F_i \F^i = 0$.

Using the notation of  section \ref{s:AnsatzOrbitsH}, for the adjoint action by $m$ and the one-form $V^I$ we simply find in this orbit that
\begin{equation} \label{eq:adjointoneformH}
M_I{}^J = \begin{pmatrix}
\delta^i{}_j & 0 \\
m^k H_{ijk} & \delta_i{}^j
\end{pmatrix} , \qquad V_\mu{}^i = \partial_\mu m^i, \qquad V_{\mu i} = - \frac{1}{2} V_\mu{}^j m^k H_{ijk} .
\end{equation}
As argued, since $M^{ij} = 0$, we continue with the twist $U^{(1)}_I{}^M$ to solve the flux equations \eqref{eq:FF3}. In terms of the functions $\rho, b, \beta$ of \eqref{eq:parU} we have $\rho_\mu{}^i = V_\mu{}^i$, $\beta^{ij} = \xi^{ij}$ and $b_{ij} = \omega_{ij} + 2 V_{[i}{}^\mu V_{\mu j]}$. In fact this $\rho_\mu{}^i$ can be gauged away by a diffeomorphism to get $\rho_\mu{}^i=\delta_\mu^i$, which we understand as a consequence of the ansatz $U^{(1)}_I{}^M$, and thus we have $w_{ij}{}^k = 0$. Consequently, solving for $F_{ij}{}^k = 0$ constrains $\beta^{ij}$ to satisfy
\begin{equation}\label{eq:betaH0}
\beta^{kl} H_{ijl} = 0 .
\end{equation}
Solving for $Q_i{}^{jk} = 0$ gives that $\beta^{ij} $ must be constant, and $R^{ijk} = 0$ is solved automatically.  From~\eqref{eq:rigidOdd} one sees that this constant $\beta$-transformation satisfying \eqref{eq:betaH0} is precisely an automorphism of the pre-Roytenberg algebra, and therefore it can be removed by a rigid $O(d,d)$ transformation as in~\eqref{eq:twistRedundancy}.  Then  we are describing the example of the torus with $H$-flux, see e.g.~\cite{Shelton:2005cf}. To complete the discussion we can also solve for the $b$-twist, or equivalently $\omega_{ij}$. By defining $b_{\mu\nu} = \rho_\mu{}^i b_{ij} \rho_\nu{}^j$ and observing that $b_{\mu\nu} = \omega_{\mu\nu} - 2 \bar\omega_{\mu\nu}$, with $\bar\omega$ defined as in \eqref{eq:defomega2}, we find that
\begin{equation}
 3 \partial_{[\mu} \omega_{\nu\rho]} = -2 V_\mu{}^i V_\nu{}^j V_\rho{}^k H_{ijk}
 \end{equation} 
 which has a particular solution $\omega^{\text{\tiny inhom.}}_{\mu\nu} = \frac{4}{3} V_\mu{}^i V_{\nu i}$, so that we have an explicit expression at least locally. The most general solution is $\omega_{\mu\nu} = \omega^{\text{\tiny inhom.}}_{\mu\nu} + \omega^{\text{\tiny hom.}}_{\mu\nu} $ with $\omega^{\text{\tiny hom.}}$ a closed two-form.  
 
 For  the generalised flux $\F_I$ we have $\F_i = 2 \partial_i \bar{\lambda}$ and $\F^i = 0$ (after setting $\beta=0$ by the rigid $O(d,d)$ transformation) so that the above Bianchi identities are automatically satisfied.

\textbf{Another class of representatives.}
Let us now present another class of representatives within the $(H)$-orbit that are not captured by the above discussion and the methods of section \ref{s:AnsatzOrbitsH}.\footnote{For more comments on solving the $(H)$-orbit without relying on section \ref{s:AnsatzOrbitsH} see appendix \ref{app:PDE}.}
We will try to look for representatives with non-vanishing $w_{ij}{}^k$. In particular we will  restrict to the case of \emph{constant} $w_{ij}{}^k$ and we will want to interpret them as structure constants of a Lie algebra $\alg g$, so that we will prefer to use the notation $f_{ij}{}^k=w_{ij}{}^k$. This can be achieved simply by taking $\rho=g^{-1}dg$ of MC form, with $g$ a group element of $G$ such that $\alg g=Lie(G)$.
Then from $F_{ij}{}^k=0$ we have
\be
f_{ij}{}^k=-\beta^{kl}H_{ijl}.
\ee
For $\alg g$ non-abelian this is possible only if we  turn on a certain $\beta$, and in general it is consistent only if $\beta^{kl}H_{ijl}$ is constant.  The Jacobi identity  for the structure constants $f_{ij}{}^k$ now implies 
\be
\beta^{nl}H_{n[ij}H_{k]lp}\beta^{mp}=0.
\ee
While a non-constant $\beta$ may be still possible, let us take a  constant $\beta$  to simplify  the discussion further. Then from $Q_i{}^{jk}=0$ it follows
\be\label{eq:condbetabetaH}
\beta^{l[j}f_{il}{}^{k]}=0,\qquad\implies\qquad
\beta^{jl}\beta^{km}H_{ilm}=0,
\ee
which is a weaker condition compared to~\eqref{eq:betaH0}. Notice that when this condition holds, both 
 $R^{ijk}=0$  and  the Jacobi identity for $f_{ij}{}^k$ are automatically satisfied. 
For the fluxes $\F_I$ we have that $\F_i=-H_{ijl} \beta^{jl} +  2 \partial_i \bar{\lambda}$ and $\F^i =  2 \beta^{ik} \partial_k\bar\lambda $, where~\eqref{eq:condbetabetaH} was used, and the Bianchi identities involving $\F_I$ are automatically satisfied. Constancy of the fluxes imposes also the condition $f_{ij}{}^k \partial_k \bar{\lambda} = 0$, which  is equivalent to the Bianchi identity $H_{ijk} \F^k = 0$.

One can check that in $d=3$  there is no solution to~\eqref{eq:condbetabetaH} but already in $d=4$ there are several. Given a basis $e_i$ we first write the 3-form $H$-flux as
\be
H=4h_{[l}e_i\wedge e_j\wedge e_{k]}=
h_1 e_2\wedge e_3\wedge e_{4}
-h_2 e_1\wedge e_3\wedge e_{4}
+h_3 e_1\wedge e_2\wedge e_{4}
-h_4 e_1\wedge e_2\wedge e_{3}.
\ee
Then  one possible solution to~\eqref{eq:condbetabetaH} is found by setting for example all $\beta^{ij}=0$ except $\beta^{23}$, and $h_1=h_4=0$. The only non-vanishing components of the structure constants are then
\be
f_{14}{}^2=-h_2\beta^{23},\qquad\qquad f_{14}{}^3=-h_3\beta^{23},
\ee
which define a Heisenberg algebra
\be
[t_1,t_4]=-\beta^{23}(h_2t_2+h_3t_3).
\ee
Interestingly we may view this class of representatives as a deformation of the representatives described previously.
It would be nice to extend the methods of section \ref{s:AnsatzOrbitsH} to include these also.

\subsection{$H,R$}
In the $(H,R)$-orbit the Bianchi identities read 
\begin{equation}
H_{ijk} R^{klm} = 0 , \qquad H_{ijk} \F^k = 0, \qquad R^{ijk} \F_k = 0 , \qquad \F_i \F^i = 0 .
\end{equation}
Following section \ref{s:AnsatzOrbitsH}, one can verify that, compared to the $(H)$-orbit, the adjoint action by $m$ and the one-form $dm m^{-1}$ do not change with the presence of $R$-flux and thus are given in \eqref{eq:adjointoneformH}. Using the Bianchi identities notice that we therefore have that $V_{\mu l} R^{ijl} = 0$.\footnote{This is consistent with the MC identity \eqref{eq:MCidforV} for $V_\mu{}^i$.}
Continuing with the parametrisation $U^{(1)}_I{}^M$ and solving for the flux equations with $\rho_\mu{}^i = V_\mu{}^i$, $\beta^{ij} = \xi^{ij}$ and $b_{ij} = \omega_{ij} + 2 V_i{}^\mu V_{\mu j}$ we ultimately find that $R^{ijk}$ must be vanishing and, therefore, that we reduce to the $(H)$-orbit. Hence,  we must conclude that the $(H,R)$-orbit (with non-vanishing $R$-flux) can not be realised within the framework of section \ref{s:AnsatzOrbitsH}.

\subsection{$F,H$} \label{s:FHorbit}

In the $(F,H)$-orbit the Bianchi identities imply that $F_{ij}{}^k$ satisfies the Jacobi identity and can be interpreted as the structure constants of a Lie algebra that we will call $\mathfrak{f}$. We should however point out that the generators are not $T_i$ (since their commutation relations in $\alg r$ are of the form $[T,T]\sim FT+H\tilde T$), and we will denote them by $t_i\in \alg f$. Moreover, the Bianchi identities imply that $H_{ijk}$ is a 3-cocycle of $\mathfrak{f}$ and that $H_{ijk} \F^k + F_{ij}{}^k \F_k =0$, $F_{ik}{}^j \F^k = 0$ and $\F_i \F^i = 0$, and in particular $\F^k = 0$ if ${t_k} \notin Z(\alg f)$.  When $\mathfrak{f}$ is semisimple the Bianchi identities imply $\F_i = \F^i = 0$.

Let us now ask when we are genuinely inside this orbit, and not in simpler sub-orbits, or in other words when there exists a transformation of the pre-Roytenberg algebra that sets either $F$ or $H$ (or both) to zero. From~\eqref{eq:rigidOdd} one sees that $H$ can be removed by a rigid $O(d,d)$  if it is of the form $H_{ijk}=-3F_{[ij}{}^lb_{k]l}$ for some constant $b$, i.e. if $H$ is a coboundary for $\alg f$. We therefore want to restrict ourselves to $H$ in the third Lie algebra cohomology of $\mathfrak{f}$.  For  $\mathfrak{f}$ semisimple $H^3(\mathfrak{f}, \mathbb{R}) = \mathbb{R}^n$ \cite{BSMF_1950__78__65_0}, where $n$ labels the number of simple factors in $\mathfrak{f}$, such that one can take $H_{ijk} = \sum_{a=1}^n \alpha_a F_{ij}{}^l \kappa_{lk}$ with $\kappa_{ij} =\langle t_i , t_j \rangle$, and $\alpha_a$, $a = 1, \ldots, n$ real constants. Viceversa, starting from $(F,H)$ and using~\eqref{eq:rigidOdd} one sees that $F$ can be removed if it is of the form $F_{ij}{}^k=H_{ijl}\beta^{lk}$ for some constant $\beta$ which  additionally satisfies $H_{ilm}\beta^{jl}\beta^{km}=0$ such that the $Q$ and $R$-fluxes are to remain vanishing. 

Following section \ref{s:AnsatzOrbitsH}, the adjoint action by $m$ now takes the form
\begin{equation} \label{eq:FHadjoint}
M_I{}^J = \begin{pmatrix} 
M^i{}_j & 0 \\
M_{ij} & M_i{}^j
\end{pmatrix} ,
\end{equation}
where, using ad-invariance of $\eta_{IJ}$, we have $M^i{}_j = (M^{-1})_j{}^i$ and $M_{ij} =   - M_i{}^l M_{kl } (M^{-1})_j{}^k$. Furthermore, using that  $M \in \text{Aut}(\mathfrak{r})$ we have the following identities
\begin{equation} \label{eq:AutoFH}
\begin{aligned}
& F_{ij}{}^l M_l{}^k = M_i{}^l M_j{}^m F_{lm}{}^k, \\
& F_{ij}{}^l M_{lk} + H_{ijl} (M^{-1})_k{}^l = 2 M_{[i |l} M_{|j]}{}^m F_{mk}{}^l  +  M_i{}^l M_{j}{}^m H_{lmk} ,
\end{aligned}
\end{equation}
of which the last identity can be rewritten as
\begin{equation} \label{eq:AutoFH2}
M_i{}^l M_j{}^m M_k{}^n H_{lmn} = H_{ijk} + 3 F_{[ij}{}^l M_{k]}{}^n M_{ln} . 
\end{equation}
Notice in addition that  the components $M_i{}^j$ give an automorphism of $\mathfrak{f}$.

Since $M^{ij} = 0$ we prefer the parametrisation $U^{(1)}_I{}^M$ for the twist, that corresponds to
\begin{equation}
\rho_i{}^\mu = M_i{}^j V_j{}^\mu , \qquad \beta^{ij} = \xi^{ij} , \qquad b_{ij} = \omega_{ij} + M_{ik} M_j{}^k \ .
\end{equation}
Using the MC identity \eqref{eq:MCidforV}, the expression for the derivatives of $M_I{}^J$ \eqref{eq:DerAdjoint} and the automorphism properties, we find $\dr_{ij}{}^k = F_{ij}{}^k$. Hence we might as well take the $\rho$-twist as $\rho_i{}^\mu = W_i{}^j \bar{\rho}_j{}^\mu$ with $\bar{\rho} = \bar{g}^{-1} d \bar{g}$, $\bar{g}\in \exp\mathfrak{f}$ and $W\in \text{Out}(\mathfrak{f})$.
Solving for the $F$-flux equation now implies that
\begin{equation} \label{eq:betaFHcond}
\beta^{kl} H_{ijl} = 0 .
\end{equation}
Solving for $Q_i{}^{jk} = 0$ and $R^{ijk} = 0$ then simply gives the homogeneous YB solution for $\beta^{ij} = \langle r_g (t^i) , t^j \rangle $ ---  as in \eqref{eq:betaYBdef} but defined with different generators  --- in which the constant antisymmetric operator $r$ solves the CYBE {on $\mathfrak{f}$}. Notice that $\beta$ is, however, constrained by the condition \eqref{eq:betaFHcond}.  Finally, to solve for the $b$-twist  we rewrite the $H$-flux equation as $3 \partial_{[\mu} b_{\nu\rho]} = \rho_\mu{}^i \rho_\nu{}^j \rho_\rho{}^k H_{ijk}$ with $b_{\mu\nu} = \rho_\mu{}^i (\omega_{ij} + M_{ik} M_j{}^k) \rho_\nu{}^j$, and observe that for $\tilde{b}_{\mu\nu} \equiv \rho_\mu{}^i M_{ik} M_j{}^k \rho_\nu{}^j $ we find
\begin{equation}
3 \partial_{[\mu} \tilde{b}_{\nu\rho]} = \rho_{\mu}{}^i  \rho_\nu{}^j \rho_\rho{}^k H_{ijk} + 2 V_\mu{}^i V_\nu{}^j V_\rho{}^k H_{ijk} + 3 V_{[\mu}{}^i V_\nu{}^j V_{\rho]k} F_{ij}{}^k ,
\end{equation}
where we have used \eqref{eq:MCidforV}, \eqref{eq:DerAdjoint}, \eqref{eq:AutoFH} and \eqref{eq:AutoFH2}.
From the definitions for $\bar\omega^{(2)}$ and $\Omega^{(2)}$ given in \eqref{eq:defomega2} and \eqref{eq:defOmega2unexpl} we can solve the $H$-flux equation for $b_{\mu\nu}$ by
\begin{equation} \label{eq:bTwistFH}
b_{\mu\nu} = \bar\omega^{(2)}_{\mu\nu} - \Omega^{(2)}_{\mu\nu} +  \tilde{b}_{\mu\nu} =   \bar\omega^{(2)}_{\mu\nu} - \Omega^{(2)}_{\mu\nu} +  \rho_\mu{}^i M_{ik} M_j{}^k \rho_\nu{}^j .
\end{equation}
Notice that the most general solution adds  closed two-forms admitted by the manifold to $b_{\mu\nu}$ but also  that this solution might not exist globally. 

Finally for  the equations for the generalised flux $\F_I$ in this orbit we have
\begin{equation}
\F_i = F_{ij}{}^j + 2 \alpha_i , \qquad \F^i = - \beta^{jl} F_{jl}{}^i + 2 \beta^{il} \alpha_l
\end{equation}
{where we have used $\beta^{kl} H_{lij} = 0$} and where, as before, $\alpha_i = \partial_i \bar{\lambda}$ is a constant, which implies $F_{ij}{}^k\alpha_k=0$.  The Bianchi identities for $\F_I$ therefore read
\begin{equation}
 \begin{aligned}
F_{ij}{}^l F_{lm}{}^m - H_{ijl} \beta^{mn} F_{mn}{}^l =0, \qquad 2 F_{il}{}^j \beta^{lm} \alpha_m - F_{il}{}^j \beta^{mn} F_{mn}{}^l = 0 . 
\end{aligned}
\end{equation}
One can verify that the second identity guarantees also the orthogonality condition $\F^i \F_i = 0$.   Together with  the second identity and the Jacobi identity for $F_{ij}{}^k$ one can verify that all these conditions guarantee also the constancy of $\F^i$. Notice that if the algebra $\mathfrak{f}$ is semisimple and the dilaton is isometric ($\alpha_i = 0$) these condition reduce to the requirement that $\alg f$ as well as  $\beta$ and the $r$-operator are unimodular~\cite{Borsato:2016ose}.

An example of a $\sigma$-model in the $(F,H)$-orbit is the Principal Chiral Model (PCM) with the addition of a Wess-Zumino (WZ) term (in the absence of spectators in fact we describe the whole $\sigma$-model). When fixing the correct normalisation between the coupling in front of the PCM and the WZ actions, this becomes the WZW model \cite{Witten:1983ar}, which is in addition a conformal field theory. The pure NSNS bosonic-string background $AdS_3\times S^3\times T^4$ can be realised as an $SL(2,\mathbb R)\times SU(2)$ WZW model  plus four free bosons parametrising the torus.

Inhomogeneous and homogeneous YB-deformations of PCM + WZ models were discussed in~\cite{Hoare:2020mpv}. The homogenous-type deformations leave the generalised fluxes invariant and therefore can be included in our discussion. However we would like to point out that the homogeneous YB-deformations of~\cite{Hoare:2020mpv} are always such that the image of the $r$-matrix (which is a subalgebra $\alg f$ of the full Lie algebra of the group used to construct the models) is a \emph{solvable} algebra. As a consequence of Cartan's criterion, the components of the $H$-flux computed from the WZ term (e.g. $H_{ijk}=\alpha f_{ij}{}^l\kappa_{kl}$ for some constant $\alpha$) are zero when all three legs are along $\alg f$. Selecting a solvable subalgebra is equivalent to splitting the coordinates $\bm{x^m}$ of the full target-space into spectator coordinates $\dot x^{\dot \mu}$ and coordinates $y^\mu$ such that $H_{\mu\nu\rho}=0$. Therefore in our language the homogeneous deformations of~\cite{Hoare:2020mpv} should be viewed as a deformation of the simpler $(F)$-orbit.  Notice that the solvability condition can be seen as a possible solution to~\eqref{eq:betaFHcond}.
New deformations of the PCM + WZ model correspond to solving~\eqref{eq:betaFHcond} whilst relaxing the solvability condition.

Let us now turn to the generically inhomogeneous YB-deformation of the PCM + WZ model of~\cite{Klimcik:2019kkf,Klimcik:2020fhs,Hoare:2020mpv}. To compare to those results let us consider the $O(d,d)$ matrix
\be
\bar U_I{}^J=
\left(
\begin{array}{cc}
\mathsf{a}& \mathsf{b}\\
\mathsf{c} & \mathsf{d}
\end{array}
\right)
=
\left(
\begin{array}{cc}
\gamma^{-1} \cosh (\zeta r\kappa) & \gamma^{-1}\kappa^{-1}\sinh( \zeta\kappa r)\\
\gamma\kappa\sinh (\zeta r\kappa) & \gamma\cosh( \zeta\kappa r)
\end{array}
\right),
\ee
where $\gamma,\zeta\in \mathbb R$ are parameters, $\kappa_{ij}$ is a symmetric matrix that we will interpret as the Killing metric of an algebra $\alg f$, and $r^{ij}$ is an antisymmetric matrix. No further restriction on $r$ is required for $\bar U$ to be in $O(d,d)$.
However, to have an integrable  $\sigma$-model  further conditions are necessary, see~\cite{Klimcik:2019kkf,Klimcik:2020fhs,Hoare:2020mpv}.
If we take $\bar{\mathcal H}^{IJ}=\bar U_K{}^I\dot{\mathcal H}^{KL}\bar U_L{}^J$, where simply
\be
\dot{\mathcal H}^{IJ}=
\left(
\begin{array}{cc}
\kappa_{ij} & 0\\
0 & \kappa^{ij}
\end{array}
\right),
\ee
and in the absence of spectators, then using~\eqref{eq:ODDM} it is easy to check that the combination of metric and B-field $\bar M=\bar G-\bar B$ parametrising $\bar {\mathcal H}$ will be
\be
\bar M=\left(\frac{1+\gamma^2}{1-\gamma^2}-e^{-2\zeta\kappa r}\right)^{-1}
\left(\frac{1+\gamma^2}{1-\gamma^2}+e^{-2\zeta\kappa r}\right)\kappa.
\ee
Comparing to equation (1.2) of~\cite{Hoare:2020mpv} we see that in order to rewrite the $\sigma$-model action in an $O(d,d)$ covariant form it is sufficient to take for example the twist\footnote{To relate to the parameters in~\cite{Hoare:2020mpv} we have to identify $\frac{1+\gamma^2}{1-\gamma^2}=e^{\chi}$ and $\zeta=-\rho/2$.}
\be
U=\left(
\begin{array}{cc}
1 & 0\\
\hat b & 1
\end{array}
\right)
\bar U
\left(
\begin{array}{cc}
v^t & 0\\
0 & v^{-1}
\end{array}
\right),
\ee
with $v=dgg^{-1}$ and $g\in \exp\alg f$, and where the $\hat b$-shift of the B-field  takes care of the WZ term.
Let us mention that $\bar U$ is of the form of~\eqref{eq:parU} with 
\be
\bar \rho=\gamma^{-1}(\cosh(\zeta\kappa r))^{-1},\quad
\bar \beta=\gamma^{-2}\kappa^{-1}\sinh(\zeta\kappa r)(\cosh(\zeta\kappa r))^{-1},\quad
\bar b=\gamma^2\kappa \sinh(\zeta r\kappa)\cosh(\zeta r\kappa ).
\ee
Let us now ignore the WZ term for a moment ($\hat b=0$). The $v$-twist is very simple and when taken alone it gives rise to $F$-flux only, in particular  $F_{ij}{}^k=-f_{ij}{}^k$ in terms of the structure constants of $\alg f$.
The multiplication by $\bar U$ from the left can be seen as a rigid $O(d,d)$ rotation, and for this reason (when $\hat b=0$) we can still interpret the model as being in the $(F)$-orbit. 
On the other hand, if we include $\bar U$ when computing the fluxes we obtain
\be
\begin{aligned}
H_{ijk}&=-3\gamma^3(\cosh \zeta\kappa r)_{[i}{}^l (\cosh \zeta\kappa r)_{j}{}^m (\sinh \zeta\kappa r)_{k]}{}^n f_{lmn},\\
F_{ij}{}^k&=-\gamma[(\cosh \zeta\kappa r)_{i}{}^l (\cosh \zeta\kappa r)_{j}{}^m (\cosh \zeta r\kappa )^{k}{}_n
+2(\cosh \zeta\kappa r)_{[i}{}^l (\sinh \zeta\kappa r)_{j]}{}^m (\sinh \zeta r\kappa )^{k}{}_n]f_{lm}{}^n,\\
Q_{i}{}^{jk}&=-\gamma^{-1}[(\sinh \zeta\kappa r)_{i}{}^l (\sinh \zeta r\kappa )^{j}{}_m (\sinh \zeta r\kappa )^{k}{}_n
+2(\cosh \zeta\kappa r)_{i}{}^l (\sinh \zeta r\kappa r)^{[j}{}_m (\cosh \zeta r\kappa )^{k]}{}_n]f_{l}{}^{mn},\\
R^{ijk}&=-3\gamma^{-3}(\sinh \zeta r\kappa )^{[i}{}_l (\sinh \zeta r\kappa )^{j}{}_m (\cosh \zeta r\kappa )^{k]}{}_n
f^{lmn},
\end{aligned}
\ee
where the indices of $f_{ij}{}^k$ are raised and lowered with $\kappa^{-1}$ and $\kappa$. Notice that  all the fluxes above are non-zero. While, as already mentioned, at $\hat b=0$ this is just a rigid $O(d,d)$ rotation of a representative of the $(F)$-orbit, when  including now $\hat b$ the new fluxes written in terms of the above ones are
\be 
\begin{aligned} 
H'_{ijk}={}&H_{ijk}+3F_{[ij}{}^l\hat b_{k]l}-3\hat b_{l[i}\hat b_{j|m}Q_{|k]}{}^{lm}+\hat b_{il}\hat b_{jm}\hat b_{kn}R^{lmn}+\hat H_{ijk},\\
F'_{ij}{}^k={}&F_{ij}{}^k+2\hat b_{l[i}Q_{j]}{}^{lk}+ \hat b_{il} \hat b_{jm} R^{lmk},\  \\
Q'_i{}^{jk} ={}& Q_i{}^{jk}  + \hat b_{il} R^{ljk} ,\\
R'^{ijk}={}&R^{ijk},
\end{aligned}
\ee
where $\hat H_{ijk}$ is the shift of the $H$-flux produced by $\hat b$. The requirement is that in the undeformed case $(\zeta=0,\gamma=1)$ the $H$-flux is proportional to the structure constants of $\alg f$, namely
$-3f_{[ij}{}^l\hat b_{k]l}+\hat H_{ijk}=\alpha f_{ijk}$ for some coefficient $\alpha$.
It would be interesting to see if there is a different parametrisation of the twist of the inhomogeneous YB-deformation of PCM + WZ that gives rise to simpler expressions for the fluxes, and if there is a parametrisation in which the fluxes are manifestly constant. This would help understand what is the simplest orbit in which one can describe this model. Notice that we have not used the mCYBE for $r$ above, and it will likely play a role in that computation.

\subsection{$H,Q$} \label{s:QHorbit}

In the $(H,Q)$-orbit the Bianchi identities imply that the $Q$-flux represents the structure constants of a Lie algebra $\tilde{\mathfrak{g}}$ with generators $\tilde{T}^i$ and that the $H$-flux satisfies $H_{k[ij} Q_{l]}{}^{mk} = 0$. The latter in turn implies that $H^{ijk} = \tilde{\kappa}^{il}  \tilde{\kappa}^{jm}  \tilde{\kappa}^{kn} H_{lmn}$, with $\tilde{\kappa}^{ij} = \langle \tilde{T}^i , \tilde{T}^j \rangle$ the Killing form, must be a 3-cocycle of $\tilde{\mathfrak{g}}$. Moreover $Q_i{}^{jk} \F_k = 0$ and $Q_k{}^{ij} \F^k =  0$ imply that $\F_k = 0$ if $\tilde{T}^k \not\in Z(\tilde{\mathfrak{g}})$ and $\F^k = 0$ if $\tilde{T}^k \in [ \tilde{\mathfrak{g}}, \tilde{\mathfrak{g}}]$. We also have the orthogonality condition $\F_i \F^i = 0$ as well as $H_{ijk} \F^k = 0$.

Since in this orbit the $H$-flux is non-vanishing we turn to a particular parametrisation of the twist $U$ explained in section \ref{s:AnsatzOrbitsH}. As  all of the components of the adjoint action $M_I{}^J = \llangle m T_I m^{-1} , T^J \rrangle$ are turned on,  we prefer to continue with   $U^{(2)}_I{}^M$ given in \eqref{eq:parU2}, where we know in general the solution for $\omega_{ij}$.
  In terms of the functions $(\rho, b , \beta)$ of \eqref{eq:parU} we  have in general 
\begin{equation} \label{eq:twistsQH}
\begin{aligned}
\rho_i{}^\mu &= (M_{ij} \xi^{jk} + M_i{}^j \delta_j^k) V_k{}^\mu , \\
\beta^{ij} \rho_j{}^\mu &= (M^i{}_j \xi^{jk} + M^{ij} \delta_j^k ) V_k{}^\mu ,\\
b_{ij} \rho_\mu{}^j &= (M_{ij} (\delta + \xi \omega)^j{}_k + M_i{}^j \omega_{jk} ) V_\mu{}^k .
\end{aligned}
\end{equation}
 Recall from section \ref{s:AnsatzOrbitsH} that the $H$-flux equation is solved, up to closed two-forms,  by $\omega = \bar\omega - \Omega^{(2)}$.
Instead of trying to solve the other flux equations given in eq.~\eqref{eq:FF3} directly, it will be more convenient to use the expressions given in \eqref{eq:Xi1}--\eqref{eq:Xi4}.  In this orbit they read
 \begin{align}
  \xi^{kl} H_{ijl}  &= 0 , \label{eq:XiQH1} \\
  \partial_\mu \xi^{ij} + 2  V_{\mu l} \xi^{m[i} Q_m{}^{j]l}  &= 0 , \label{eq:XiQH2} \\
 Q_l{}^{[ij} \xi^{k]l} &=0 \label{eq:XiQH3} .
\end{align}
Notice that $\xi = 0$ is a trivial  possible solution. 
To solve these equations in general, it will be important to consider the  series expansion of the one-form $dm m^{-1}$. We find
\begin{equation}
\begin{aligned}
\partial_\mu m m^{-1} &= V_\mu{}^i T_i + V_{\mu i} \tilde{T}^i =  \sum_{N=0}^\infty \frac{1}{(N+1)!}  \text{ad}_{m^i T_i}^N  \partial_\mu m^j T_j , 
\end{aligned}
\end{equation}
with 
\begin{equation} \label{eq:QHVExpansion}
\begin{aligned}
V_\mu{}^i &= \sum_{N=0}^\infty \frac{1}{(2N+1)!} \partial_\mu m^j ((mH \cdot mQ)^N)_j{}^i , \\
V_{\mu i} &= \sum_{N=0}^\infty \frac{1}{(2N+2)!} \partial_\mu m^j (mH)_{jl} ((mQ \cdot mH)^N)^l{}_i ,
\end{aligned}
\end{equation}
 where we have defined $(mH)_{ij} \equiv m^l H_{lij}$ and $(mQ)^{ij} \equiv m^l Q_l{}^{ij}$, and which can be proven by induction.  Notice that we can always write $ V_{\mu i} = A_\mu^{jk} H_{jki}$
 for some tensor $A$ which carries all the field dependence.  With this observation, it is now easy to show that using eq.~\eqref{eq:XiQH1} and the Bianchi identity $H_{k[ij} Q_{l]}{}^{mk} = 0$ that
 \begin{equation}
 V_{\mu l} \xi^{m[i} Q_m{}^{j]l}  = 0 .
 \end{equation}
 Hence eq.~\eqref{eq:XiQH2} simply requires that $\xi^{ij}$ must be a constant matrix. Then the remaining conditions are algebraic: eq.~\eqref{eq:XiQH3} implies that $\xi$ must be a 2-cocycle for $\tilde{\mathfrak{g}}$ while eq.~\eqref{eq:XiQH1}  can be seen as a compatibility condition requiring that $\xi$ and $H$ must be orthogonal. Therefore, turning on such a particular $\xi$ from $\xi = 0$ can be interpreted as a deformation which is a (novel) solution-generating technique. 

To have the expression of the $(\rho, b , \beta)$-twists it will again be convenient  to consider a series expansion, now of $M_{ij}$. One finds in particular $M_{ij} = \sum_{N=0}^\infty \frac{1}{(2N+1)!} (mH)_{il} ((mQ \cdot m H)^N)^l{}_j$ so that upon \eqref{eq:XiQH1} we have $M_{ij} \xi^{jk} = 0$. Hence
 \begin{equation} \label{eq:twistsQH2}
 \rho_i{}^\mu = M_i{}^j V_j{}^\mu , \qquad \beta^{ij} =  (M^{-1})_l{}^i \xi^{lk}  (M^{-1})_k{}^j + M^{ik}  (M^{-1})_k{}^j  , \qquad b_{ij} = M_i{}^l \omega_{lm} M_j{}^m ,
 \end{equation}
 where  we have used $M_i{}^l M^j{}_l + M_{il} M^{jl} = \delta_i{}^j$ and $M_i{}^l M_{jl} + M_{il} M_j{}^l = 0$, from $M_I{}^L \in O(d,d)$, as well as $M_{ij} \xi^{jk} = 0$. We can now easily calculate the expression for $\dr_{ij}{}^k$ resulting in 
 \begin{equation}
 \dr_{ij}{}^k =  H_{ijl} M^{ln} (M^{-1})_n{}^k .
 \end{equation}
 
 Let us now consider  the conditions in which case a rigid $O(d,d)$ transformation by \eqref{eq:rigidOdd} describes a simpler orbit in disguise (i.e.~$H$ or $Q$ can be turned off at no other cost). We find that we can not turn off the $H$-flux with a constant $b$ and $\beta$ transformation since the condition found from $H'_{ijk} = 0$ together with the consistency condition from $F'_{ij}{}^k = 0$ requires $H_{ijk}$ to be vanishing, which is of course not possible.
The same holds for the $Q$-flux where the condition required from $Q'_i{}^{jk} = 0$ does not match with the conditions found from $R'^{ijk} = 0$ as well as $F'_i{}^{jk} = 0$. 

To complete the discussion of this orbit in general we have for the generalised fluxes $\F_I$ 
\begin{equation}
\F_i = \dr_{ij}{}^j + 2 \partial_i \bar{\lambda} , \qquad \F^i = Q_j{}^{ji} + 2 \beta^{ij} \partial_j \bar{\lambda} ,
\end{equation}
with $\dr_{ij}{}^k \partial_i \bar{\lambda} = 0$ such that $\F_i$ is constant.
From the requirement that $\F_i$ is constant, $\partial_i \F^j = 0$, we therefore also must have that  $Q_i{}^{jk} \partial_k \bar{\lambda} = 0$. {From the Bianchi  $Q_i{}^{jk}\F_k =0$ we then find the condition $Q_{i}{}^{jk} \dr_{kl}{}^l  = Q_i{}^{jk} H_{klm} \beta^{ml} = 0$.} Furthermore $H_{ijk} \F^k = 0$ implies $H_{ijk} Q_l{}^{lk} = 0$, and $Q_k{}^{ij} \F^k = 0$ implies $Q_k{}^{ij} Q_l{}^{lk} + 2 Q_k{}^{ij} \beta^{kl} \alpha_l = 0$.
Notice that the orthogonality condition $\F_i \F^i = 0$ will be  automatically satisfied  because of the previous conditions.

 \textbf{(Asymmetrical) $\lambda$-deformations on group manifolds} ---
It is known from  \cite{Demulder:2018lmj}, when ignoring spectators,  that the WZW model and its $\lambda$-deformation \cite{Sfetsos:2013wia} can be described by the particular twist $U^{(2)}$ with $\xi = 0$ and by a particular choice of fluxes (or, equivalently, a particular pre-Roytenberg algebra).  In the following we will show that, generalising \cite{Demulder:2018lmj}, also the asymmetrical $\lambda$-deformation on group manifolds of \cite{Driezen:2019ykp,Driezen:2020job} fits within the same pre-Roytenberg algebra. We take $\mathfrak{r} = \mathfrak{k} \oplus \mathfrak{k}$ where $\mathfrak{k}$ is a Lie algebra with generators $t_i$ and structure constants $f_{ij}{}^k$.
The bilinear form  $\eta_{IJ}$ is taken to be \cite{Klimcik:2015gba,Demulder:2018lmj}
\begin{equation}
\llangle \{ x_1 , y_1 \} ,\{ x_2 , y_2\} \rrangle = \langle x_1 , x_2 \rangle - \langle y_1 , y_2 \rangle ,
\end{equation}
where $x_i, y_i \in \mathfrak{k}$ and $\kappa_{ij} = \langle t_i , t_j \rangle$ is the Killing form on $\mathfrak{k}$.
 The generators $\tilde{T}^i$ with structure constants $Q_i{}^{jk}$ generate  the diagonal embedding $\tilde{\mathfrak{g}} = \mathfrak{k}_{\text{\tiny diag}}$ in $\mathfrak{r}$ by the map $x \mapsto \{ x, x \}/\sqrt{2}$.
 This subgroup is maximally isotropic. The complementary isotropic subspace, spanned by $T_i$, is the anti-diagonal embedding by the map $x \mapsto \{ x, -x \}/\sqrt{2}$.  Hence, in this case we have
 \begin{equation} \label{eq:LambdaFluxes}
 H_{ijk} = \frac{1}{\sqrt{2}} f_{ij}{}^l \kappa_{lk} , \qquad Q_i{}^{jk} = \frac{1}{\sqrt{2}} \kappa_{il} \kappa^{jm} \kappa^{kn} f_{mn}{}^l  , \qquad F_{ij}{}^k = 0 , \qquad R^{ijk} = 0 .
 \end{equation}
 Notice that the Bianchi identities for $\F_{IJK}$ are automatically satisfied upon the Jacobi identity for $f_{ij}{}^k$ and the ad-invariance of $\kappa_{ij}$. 
 To connect to the $\lambda$-deformed background in terms of the metric and B-field there is a subtlety in choosing a good parametrisation of the group element $m = e^{m^i (y) T_i}$. Following \cite{Demulder:2018lmj} we  take $m = \{ \bar{g} , \bar{g}^{-1} \}$ with $\bar{g} \in \exp (\mathfrak{k} )$ and define a group element $\tilde{g}  \equiv  \bar{g}^2   \in \exp (\mathfrak{k} )$. 
Using this identification in the twist $U^{(2)}_I{}^M$ we find in terms of the general parametrisation~\eqref{eq:parU} that\footnote{For details of this calculation see \cite{Demulder:2018lmj}. Following the logic of \cite{Demulder:2018lmj} we find that to find the same deformed geometry, we should replace  $e = g^{-1}dg$ with $v = dg g^{-1}$  in their frame fields of eq.~(5.63), an overall minus sign is missing in entry $(1,2)$ and $(2,1)$, and we should take $\lambda \rightarrow -\lambda$ in eq.~(5.66). This likely corresponds to a known symmetry of these backgrounds \cite{Itsios:2014lca} and indeed we agree on the final result.}
\begin{equation}
 \rho_i{}^\mu = \frac{1}{\sqrt{2}}  (1+ \AD_{\tilde{g}})_i{}^j \tilde{v}_j{}^\mu  , \qquad \beta = - \kappa^{-1} \frac{1-\AD_{\tilde{g}}}{1+\AD_{\tilde{g}}} , \qquad b = b_0 + \rho^{-1} \omega \rho^{-t} , 
\end{equation}
with
\begin{equation}
b_0 = \frac{1}{4} ( \AD_{\tilde{g}} - \AD^{-1}_{\tilde{g}} ) \kappa ,
\end{equation}
and $\tilde{v} = d\tilde{g} \tilde{g}^{-1}$. 
Deviating  from \cite{Demulder:2018lmj} we now  consider  a different parametrisation denoted by $m = \{ g , g^{-1} \}$ with $g\in \exp \mathfrak{k}$ and which is related to the previous parametrisation as $\AD_{\tilde{g}} =  \AD_g \circ W$ and $\tilde{v} = v = dg g^{-1}$ in which $W$ is a constant \textit{outer} automorphism of $\mathfrak{k}$ which preserves the metric $\kappa$. In particular this means that the group elements $\tilde{g}$ and $g$ are not related by a trivial field redefinition.\footnote{A useful trick to get this generalisation $\AD_{\tilde{g}} =  \AD_g \circ W$ and $\tilde{v} = v = dg g^{-1}$ is to consider a constant $w\in \exp\mathfrak{g}$ such that $\tilde{g} = g w$ and define $W(t_i) = w T_i w^{-1}$. This  $W$ would be an \emph{inner} automorphism and it would be removed by a trivial field redefinition, but the dressing by $W$ in the formulas above is the same also in the case of \emph{outer} automorphisms.}
 The twist functions simply become
\begin{gather}
\rho^{-1} = \frac{1}{\sqrt{2}} (1+ \AD_g W) v^{-1}, \qquad \beta = - \kappa^{-1} \frac{1- \AD_g W}{1+ \AD_g W} , \\ b =  \frac{1}{4} ( \AD_{{g}} W - W^{-1} \AD^{-1}_{{g}} ) \kappa + \rho^{-1} \omega \rho^{-t} .
\end{gather}
{A final subtlety in calculating the background from the twist functions is the choice of the matrix $\dot{E}_A{}^I$, which is constant when we turn off the spectator fields. To obtain the (asymmetrical) $\lambda$-deformed background we must take $\dot{E}_A{}^I$ such that \cite{Klimcik:2015gba,Demulder:2018lmj}
\begin{equation}
 \dot{{\cal H}}^{IJ} = \dot{E}_A{}^I {\cal H}^{AB} \dot{E}_B{}^J = 
 \begin{pmatrix}
\frac{1+\lambda}{1-\lambda} \kappa & 0 \\ 
0 &  \frac{1-\lambda}{1+\lambda} \kappa^{-1}
\end{pmatrix} ,
\end{equation}
in which $\lambda$ is the eponymous deformation parameter.}
The  metric and the B-field can then be extracted from the generalised metric ${\cal H}^{MN} = {U}^{(2)}_I{}^M \dot{{\cal H}}^{IJ} {U}^{(2)}_J{}^N$ resulting, in the coordinate frame, in
\begin{equation}
G = \frac{1-\lambda^2}{2} v O_g \kappa O_g^t v^t , \qquad B = \frac{1}{2} v (O_g \kappa - \kappa O_g^t ) v^t - {\omega} ,
\end{equation}
with $O_g = (1-\lambda \AD_g W)^{-1}$ and where {$d\omega = -\frac{1}{12} f_{ijk} v^i \wedge v^j \wedge v^k$ gives rise to the well-known WZ term. } This coincides  precisely with the metric and B-field of the asymmetrical $\lambda$-model on the group manifold $G = \exp\mathfrak{k}$ \cite{Driezen:2019ykp} and with the original $\lambda$-model when $W=1$ \cite{Sfetsos:2013wia}. Let us point out that here $W$ should not be mistaken  with an automorphism of the pre-Roytenberg algebra as given in \eqref{eq:UtransAuto}.

{ Curiously  this implies that the WZW model background, which can be found by taking the limit $\lambda \rightarrow 0$, is described in  the $(H,Q)$-orbit as well as the $(F,H)$-orbit. This then begs the question if there is  a rigid $O(d,d)$ transformation~\eqref{eq:rigidOdd} that relates
the $(H,Q)$-orbit (with the $H$ and $Q$ fluxes proportional to each other via the Killing form as in \eqref{eq:LambdaFluxes})  to the $(F,H)$-orbit (with $F$ and $H$  again proportional). However, in  general there is  no such rigid $O(d,d)$ transformation, as can be checked for example in the $\alg{su}(2)$ case.}

On the other hand, a rigid $O(d,d)$ transformation does allow to describe the (asymmetrical) $\lambda$-models in the self-dual $(F,Q)$-orbit of Poisson-Lie symmetric backgrounds. By taking $\beta^{ij} = 0$ and $b_{ij} = \langle \tilde{r} t_i , t_j \rangle$, with $\tilde{r}$ a constant antisymmetric operator that satisfies the mCYBE \eqref{eq:mCYBE} for $c^2 = 1$ on $\mathfrak{k}$, we find
  \begin{equation}
  H'_{ijk} = 0, \qquad F'_{ij}{}^k = \sqrt{2}  b_{l[i} f_{j]m}{}^k \kappa^{lm}  , \qquad Q'_i{}^{jk} =\frac{1}{\sqrt{2}} \kappa_{il} \kappa^{jm} \kappa^{kn} f_{mn}{}^l  , \qquad R'^{ijk} = 0, 
  \end{equation}
  where we have used the ad-invariance of $\kappa$. This is now up to analytic continuations (essentially sending $c^2 = 1$ to $c^2=-1$ or vice versa) the Poisson-Lie dual of the non-split $\eta$-deformation \cite{Hoare:2015gda,Sfetsos:2015nya,Klimcik:2015gba,Demulder:2018lmj} of which the $(F,Q)$-fluxes were given in eq.~\eqref{eq:FQfluxesEta}.

Let us emphasise that the generalised fluxes {$\F_{IJK}$} remain invariant  when   turning on a non-trivial automorphism $W$ from the (original) $\lambda$-model representative with $W=1$. Hence this map can be understood as a solution-generating technique  --- recall that in the case of supergravities we have to require that also $\F_I$ stays invariant.\footnote{Notice that in this parametrisation $\lambda$ enters $\dot E$, and for this reason turning on $\lambda$ is not necessarily a solution-generating technique. It is in fact known that the $\lambda$-deformation is not a marginal deformation of the WZW model~\cite{Sfetsos:2013wia}, and in general one has to add RR fluxes to get a supergravity solution~\cite{Hollowood:2014qma,Borsato:2016ose,Sfetsos:2014cea}.} 
Furthermore starting e.g.~with  the $\lambda$-model representative ($\xi = 0)$, and turning on a constant $\xi$   which is orthogonal to $H$ as well as a 2-cocycle for $Q$, is a novel solution-generating technique. In fact, if we transform the twist by a constant $\beta$-shift from the left, the fluxes transform as in~\eqref{eq:rigidOdd}, with $\beta=\xi$ and $b=0$ in this case; then demanding that the fluxes remain invariant produces the two conditions~\eqref{eq:XiQH1} and~\eqref{eq:XiQH3}. Therefore turning on a constant $\xi$ can be understood as the implementation of an automorphism  of the pre-Roytenberg algebra as in~\eqref{eq:UtransAuto}.  Interestingly this deformation may  be viewed as a generalisation of the DTD models (which on its own generalise ordinary NATD) in the $(Q)$-orbit, but now with the addition of $H$-flux.

 \textbf{Comments on finding other representatives} --- 
Let us illustrate that outside of the ansatz considered in this section, the $(H,Q)$-orbit may still capture different non-trivial representatives. Since the generators $\tilde{T}^i$ span a subalgebra $\mathfrak{\tilde{g}}$, another interesting ansatz allowing for systematic progress is the twist considered in  eq.~\eqref{eq:parU3} or eq.~\eqref{eq:parU4}. Taking the latter, and assuming that $\delta_{ij}$ is an ad-invariant bilinear form of $\mathfrak{\tilde{g}}$,  we find for    the equivalent of the $F$-, $Q$- and $R$-flux equations of \eqref{eq:fluxTS} and \eqref{eq:ExprTS}  that
\begin{equation}
\xi^{kl} H_{ijl} = \delta_{il} Q_j{}^{lk} , \qquad Q_i{}^{jk} =  \hat{\partial}_i \xi^{jk} - 3 \delta_{il} \xi^{m[j} Q_m{}^{k]l}  , \qquad 3 \xi^{l[i} Q_l{}^{jk]}  - \xi^{l[i} \xi^{j|m} Q_m{}^{|k]n} \delta_{ln}= 0 ,
\end{equation}
with $\hat{\partial}_i \equiv \delta_{ij} \tilde{V}^{j\mu} \partial_\mu$ and $\hat{\dr}_{ij}{}^k = \delta_{il} Q_j{}^{lk}$. 
In this case notice that the expansion of $\tilde{V}_{\mu i}$ is in terms of $Q$ instead of $H$, namely $\tilde{V}_{\mu i} = \sum_{N=0} \frac{1}{N+1)} \partial_\mu \tilde{m}^j (\tilde{m}Q)^j{}_i$. It is now indeed clear that these equations  may  hold a genuinely different solution. We  leave this problem, and the possibility of other ansatze,  open.

\subsection{$F,H,R$}

In the $(F,H,R)$-orbit the Bianchi identities imply that the $F$-flux represents the structure constants of a Lie algebra $\mathfrak{f}$ whose generators we denote by $t_i$ and which --- as in section \ref{s:FHorbit} --- do not coincide with $T_i$.
Then $H_{ijk}$ and $R_{ijk} = \kappa_{il} \kappa_{jm} \kappa_{kn} R^{lmn}$,  with $\kappa_{ij} = \langle t_i , t_j \rangle$, are 3-cocycles of $\mathfrak{f}$. In fact the $R$-flux satisfies the stronger condition $F_{ij}{}^{[k} R^{lm]j} = 0$  Additionally we have $H_{ijk} R^{klm} = 0$ as well as the Bianchi identities \eqref{eq:BianchiF1} and \eqref{eq:BianchiContr}  for $\F_I$.

Since the $H$-flux is non-vanishing we turn again to methodology of section \ref{s:AnsatzOrbitsH}. Compared to the $(F,H)$-orbit of section \ref{s:FHorbit}, notice that the presence of $R$-flux does not affect the adjoint action $M_I{}^J$ given in \eqref{eq:FHadjoint}. Therefore taking the ansatz $U^{(1)}_I{}^M$ of eq.~\eqref{eq:parU1} we have  in terms of the $(\rho, b, \beta)$-twists again that
\begin{equation}
\rho_i{}^\mu = M_i{}^j V_j{}^\mu , \qquad \beta^{ij} = \xi^{ij} , \qquad b_{ij} = \omega_{ij} + M_{ik} M_j{}^k \ .
\end{equation}
Now, however, the equation for $\dr_{ij}{}^k = - 2 \rho_i{}^\mu \rho_j{}^\nu \partial_{[\mu} \rho_{\nu]}{}^k$ will in principle receive contributions from $V_{\mu i } R^{ijk}$ by using the MC identity \eqref{eq:MCidforV} and the expressions for the derivatives of $M_I{}^J$ \eqref{eq:DerAdjoint}. Before calculating $\dr_{ij}{}^k$ let us first derive several useful properties.
 First notice from the automorphism identities that we have  from the vanishing $Q$-flux that
 \begin{equation}
 M_{il} R^{lmn} = 0 ,
 \end{equation}
 where we have used the fact that $M^i{}_l$ has an inverse. Hence the  automorphism property for $F_{ij}{}^k$  becomes simply $ F_{ij}{}^k = M_i{}^l M_j{}^m (M^{-1})_n{}^k F_{lm}{}^n$. Additionally it will be useful to calculate the expansion of the one-form $V = dm m^{-1}$ in terms of $m = \exp( m^i T_i)$. In general we have
 \begin{equation}
 \partial_\mu m m^{-1} = V_{\mu}{}^i T_i + V_{\mu i} T^i =   \sum_{N=0}^\infty V_\mu^{(M)} , \qquad  V_\mu^{(N)} \equiv \frac{1}{(N+1)!}  \text{ad}_{m^i T_i}^N  \partial_\mu m^j T_j .
 \end{equation}
 Using the commutation relations in $\mathfrak{r}$ we  have the following expressions
 \begin{equation} \label{eq:FHRExpandV}
 \begin{aligned}
V_{\mu}{}^i  &= \sum_{N=0}^\infty \frac{1}{(N+1)!} \partial_\mu m^j ( m  F^N)_j{}^i   \\
V_{\mu i} &=
 \sum_{N=1}^\infty \frac{1}{(N+1)!}  \partial_\mu m^j  \sum_{K=0}^{N-1} (-)^K (m  F^{N-1-K} \cdot m H \cdot (m  F^t)^K)_{ji}
 \end{aligned}
 \end{equation}
 where we have defined $(m F)_i{}^j \equiv m^l F_{li}{}^j$,$(m F^t)^i{}_j \equiv m^l F_{lj}{}^i$ and $(m H)_{ij} \equiv m^l H_{lij}$, and which can be proved by induction  using the relation $V^{(N+1)} = \frac{1}{N+2} \text{ad}_m V^{(N)}$. An important consequence is that using the  Bianchi identities $F_{ij}{}^{[k} R^{lm]j} = 0$ and $H_{ijk}R^{klm} = 0$ we have
 \begin{equation}
 V_{\mu i} R^{ijk} = 0 .
 \end{equation}
 In particular  with $H_{ijk}R^{klm} = 0$ this is easily seen for the case $K = 0$ in \eqref{eq:FHRExpandV}. When $K \neq 0$ it is sufficient to verify that, upon using $F_{ij}{}^{[k} R^{lm]j} = 0$ enough times, eventually an index of $R^{ijk}$ will be contracted with an index of $H_{ijk}$.
Using all of the above in the calculation for $\dr_{ij}{}^k$ we find that the contributions of the $R$-flux will vanish
\begin{equation}
\begin{aligned}
\dr_{ij}{}^k &= F_{ij}{}^k - \rho_i{}^\mu \rho_j{}^\nu  V_{\mu l}  V_{\nu m}  R^{lm n} M^k{}_n , \\
&=  F_{ij}{}^k \ .
\end{aligned}
\end{equation}
The same is true when trying to solve  the $H$-flux equation in terms of the $b$-twist: in the calculation of $d\tilde{b}$ with $\tilde{b}_{\mu\nu} = \rho_\mu{}^i M_{ik} M_j{}^k \rho_\nu{}^j$ the contributions received from $R^{ijk}$ are all of the form $V_{\mu i } R^{ijk}$. Hence, also in this orbit the solution for the $b$-twist is given in eq.~\eqref{eq:bTwistFH}.

From solving the $F$-flux equation of \eqref{eq:FF3} we now have the condition $\beta^{kl} H_{ijl} = 0$,
again like in  the $(F,H)$-orbit. Instead of the parametrisation $\rho_i{}^\mu = M_i{}^j V_j{}^\mu$ we might now as well take $\rho = g^{-1} d g$ with $g\in \exp \mathfrak{f}$ some group element of the Lie group associated to $\mathfrak{f}$.\footnote{We ignore again here the possibility of (outer) automorphisms of which the general discussion is given around eq.~\eqref{eq:UtransAuto}.} Then solving for $Q_i{}^{jk} = 0$ gives  $\beta^{ij} =  \langle r_g t^i , t^j \rangle $, with $r_g = \AD_g^{-1} \circ r \circ \AD_g$, $t^i = \kappa^{ij} t_j$ and $r$ a constant antisymmetric operator. Notice only the difference in generators compared to eq.~\eqref{eq:betaYBdef}. Furthermore $r$ must additionally satisfy
\begin{equation} \label{eq:FHRcondrH}
r^{kl} H_{ijl} = 0 .
\end{equation}
The $R$-flux equation, on the other hand, gives 
\begin{equation} \label{eq:FHRRflux}
( \AD_{g})_l{}^i ( \AD_{g})_m{}^j ( \AD_{g})_n{}^k R^{lmn} = 3  {r}^{l[i} {r}^{j|m} F_{lm}{}^{|k]} ,
\end{equation}
where we have used that $\AD_g$ is an automorphism of $\mathfrak{f}$. Notice  that the matrices $(\AD_g)_i{}^j$ and $M_i{}^j$ coincide if we parametrise $g\in \exp \mathfrak{g}$ as $g = e^{m^i t_i}$, which is most easily seen when writing both in a series expansion. Indeed in that case
\begin{equation}
M_i{}^j = (\AD_g)_i{}^j = \sum_{N=0}^\infty \frac{1}{N!} (mF^N)_i{}^j .
\end{equation}
Additionally, since  $M^{ij}=0$ we   also  have  from \eqref{eq:MAuto} that $M^i{}_l M^j{}_m M^k{}_n R^{lmn} = R^{ijk}$ where $M^i{}_j = (M^{-1})_j{}^i$. These observations combined ensure that eq.~\eqref{eq:FHRRflux} becomes simply
\begin{equation} \label{eq:FHRRflux2}
R^{ijk} = 3  {r}^{l[i} {r}^{j|m} F_{lm}{}^{|k]} .
\end{equation}
When we consider  $R^{ijk} = \alpha \kappa^{il} \kappa^{jm} F_{lm}{}^k$, with $\alpha \in \mathbb{R}$, eq.~\eqref{eq:FHRRflux2}
becomes $\alpha \kappa^{il} \kappa^{jm} F_{lm}{}^k = 3  {r}^{l[i} {r}^{j|m} F_{lm}{}^{|k]}$ which is
  precisely the mCYBE \eqref{eq:mCYBE}  on $\mathfrak{f}$ for $c^2 = \alpha$ as in the $(F,R)$ orbit, cf.~section \ref{s:FRorbit}. 
 
 Interestingly, using a rigid $O(d,d)$ transformation  the $R$-flux can be traded for $Q$-flux when we take $b = 0$ and $\beta = r$ in  \eqref{eq:rigidOdd}, so that this representative can be equally described in the $(F,H,Q)$-orbit. In particular, in that case we will have
 \begin{equation}
H'_{ijk} = H_{ijk}, \qquad F'_{ij}{}^k = F_{ij}{}^k, \qquad Q'_i{}^{jk} = - 2 F_{il}{}^{[j} r^{k]l} , \qquad R'^{ijk} = 0 \ ,
 \end{equation}
 so that, as in the $(F,R)$-orbit of section \ref{s:FRorbit}, the $Q$-flux are the structure constants of $\mathfrak{g}_r$ whose Lie bracket was defined in \eqref{eq:gRLiebracket}. 
 
On a different note, the $(F,H,R)$-orbit that we are describing is a simpler orbit  in disguise if we can turn off at least one of the three types of fluxes. The possibilities that we have are: \textit{(i)} we can turn off the $H$-flux if it is of the form $H_{ijk} = - 3 F_{[ij}{}^l b_{k]l} - b_{il}b_{jm}b_{kn}R^{lmn}$ for some constant antisymmetric matrix $b$ satisfying $  2 F_{il}{}^{[j} \beta^{k]l} -  b_{il} R^{ljk} + 2 b_{il} b_{mn} R^{ln[j} \beta^{k]m} = 0$ for some constant antisymmetric matrix $\beta$. The $F$- and $R$-flux will obtain in that case a shift, $F'_{ij}{}^k = F_{ij}{}^k + b_{il} b_{jm} R^{lmk}$ and $R'^{ijk} = R^{ijk} +\frac{3}{2} \beta^{l[i} R^{jk]m} b_{ml}$, which vanishes when we take $\beta = 0$ (since then we must have $b_{il} R^{ljk} = 0$);
\textit{(ii)} to turn off the $R$-flux we find from  $Q'_i{}^{jk} = 0$ and taking $b = 0$ for simplicity in \eqref{eq:rigidOdd} the condition $H_{imn} \beta^{mj} \beta^{nk}  - 2 F_{il}{}^{[j} \beta^{k]l} = 0$, so that the $R$-flux should be of the form $R^{ijk} = \beta^{l[i} \beta^{j|m} F_{lm}{}^{|k]}$. In that case $H$ stays invariant while $F$ receives a shift $F'_{ij}{}^k = F_{ij}{}^k - H_{ijl} \beta^{lk}$. {Notice that while in the previous discussion we did take the $R$-flux of this form with $\beta = \sqrt{3} r$, the condition \eqref{eq:FHRcondrH} also implies $F_{il}{}^{[j} r^{k]l} = 0$ which is inconsistent with the mCYBE unless $\alpha=0$ from the very beginning, which is actually not the case that we want to consider.} Finally \textit{(iii)} we do not need to discuss the possibility of turning off the $F$-flux since in that case we would end up in the $(H,R)$-orbit which, as discussed before, can not be realised on the strong constraint with the methods of section~\ref{s:AnsatzOrbitsH}, while here we do find possible representatives.

Finally let us close the discussion of this orbit by briefly commenting on the generalised fluxes $\F_I$. We have
\begin{equation}
\F_i = F_{ij}{}^j + 2 \partial_i \bar{\lambda} , \qquad \F^i = - \beta^{jl} F_{jl}{}^i  + 2 \beta^{il} \partial_l \bar{\lambda} \ .
\end{equation}
Constancy of $\F_i$ implies $F_{ij}{}^k \partial_k \bar{\lambda} = 0$ while one can verify that constancy of $\F^i$ is guaranteed by the Bianchi $F_{ik}{}^j \F^k = 0$. In addition, this condition as well as $H_{ijk} R^{klm} = 0$ implies also the  Bianchi $\F_i \F^i =  0$. At last we point out that  when writing down  all  the Bianchi identities for $\F_I$ explicitly one finds immediately that they are satisfied when $\mathfrak{f}$ is semisimple, the dilaton is isometric, and $\alg r$ as well as $\mathfrak{f}$ are unimodular.

\subsection{$H,Q,R$}
In the $(H,Q,R)$-orbit, the Bianchi identities imply that $Q_i{}^{jk}$ are structure constants for $\tilde{\mathfrak{g}} = \text{span} (\tilde{T}^i)$ and that the $R$-flux is a 3 cocycle of $\tilde{\mathfrak{g}}$. Furthermore we have $H_{k[ij} Q_{l]}{}^{mk} = 0$ and $H_{ijk} R^{klm} = 0$. Within the ansatz of  \ref{s:AnsatzOrbitsH} this again implies several simplifications. First notice that the $R$-flux will not alter the series expansion of the one-form $dm m^{-1}$ given in \eqref{eq:QHVExpansion} as well as the expansion of $M_{ij}$. Hence, upon the Bianchi $H_{ijk} R^{klm} = 0$ we have
\begin{equation} \label{eq:QHRRcondition}
V_{\mu l } R^{ljk} = 0 , \qquad M_{il} R^{ljk} = 0 .
\end{equation}
Let us now use  the parametrisation $U^{(2)}$ of \eqref{eq:parU2} where, recall,  we have the general solution of the $H$-flux equation in terms of the two-form $\omega$. For the expressions of the other flux equations in terms of the yet unknown $\xi^{ij}$ we refer to the most general orbit discussed in section \ref{s:AnsatzOrbitsH} by setting $F=0$ in the equations \eqref{eq:Xi1}--\eqref{eq:Xi4}. Let us point out here that due to the presence of $R$-flux,  $\xi = 0$ will not be a solution. Using \eqref{eq:QHRRcondition} the expressions simplify to
\begin{align}
&  \xi^{kl} H_{ijl} = 0 , \label{eq:QHRXi1} \\
& \partial_\mu \xi^{ij} + 2  V_{\mu l} \xi^{m[i} Q_m{}^{j]l} =0, \label{eq:QHRXi2} \\
&3 Q_l{}^{[ij} \xi^{k]l}  = R^{ijk} \label{eq:QHRXi3}
\end{align}
Similarly as in section \ref{s:QHorbit} the Bianchi identity $H_{k[ij} Q_{l]}{}^{mk} = 0$ implies that $V_{\mu l} \xi^{m[i} Q_m{}^{j]l} = 0$ so that $\xi^{ij}$ must simply be a constant matrix. 
However, given a rigid $O(d,d)$ transformation it is possible to turn off the $R$-flux for this particular constant $\xi^{ij}$. In order to do so one should take in \eqref{eq:rigidOdd} the constant matrices $b_{ij} = 0$ and $\beta^{ij} = - \xi^{ij}$ such that $H_{ijk}\beta^{kl} = 0$. This rigid $O(d,d)$  will leave the other fluxes invariant and thus we are in fact describing the $(H,Q)$-orbit  (which, recall, is a genuine orbit modulo rigid $O(d,d)$). Concluding, within the ansatz of section \ref{s:AnsatzOrbitsH} we cannot describe a particular non-trivial representative in the $(H,Q,R)$-orbit and, therefore, we cannot describe a rigid $T$-transformation from the $(F,H,R)$-orbit as a solution-generating technique. It would be interesting to explore other ansatze for this purpose.

\subsection{$F,H,Q$}
The $(F,H,Q)$-orbit describes what is known as a quasi-Manin triple.  The generators $\tilde{T}^i$ span a subalgebra $\tilde{\mathfrak{g}}$ of $\mathfrak{r}$ with structure constants $Q_i{}^{jk}$, while the generators $T_i$ do not. Furthermore we have the Bianchi $F_{[ij}{}^k H_{lm]k} = 0$ as well as \eqref{eq:Jacobi2} and \eqref{eq:Jacobi3}. To discuss this orbit we use the ansatz of section \ref{s:AnsatzOrbitsH} and  the parametrisation $U^{(2)}$ of \eqref{eq:parU2} for the twist.
The general solution of the $H$-flux equation was given  in terms of $\omega_{ij}$ in \eqref{eq:EqForOmegaparU2}. 
The equations to be solved for the unknown $\xi^{ij}$ can  be found in section \ref{s:AnsatzOrbitsH} by setting $R = 0$ in \eqref{eq:Xi1}--\eqref{eq:Xi4}. They are
\begin{align}
& 2 \xi^{kl} H_{ijl} + \xi^{km} \Lambda_{ml} F_{ij}{}^l  = 0 , \label{eq:FQHXi1} \\
& \hat{\partial}_i \xi^{jk} + 2  \Lambda_{il} \xi^{m[j} Q_m{}^{k]l}  - \xi^{m[j} \xi^{k]n} H_{imn} = 0 , \label{eq:FQHXi2} \\
&3 Q_l{}^{[ij} \xi^{k]l} + 3 \xi^{l[i} \xi^{j|m} F_{lm}{}^{|k]}  - \xi^{l[i} \xi^{j|m} \xi^{|k]n} \Lambda_{np} F_{lm}{}^p = 0 \label{eq:FQHXi3}
\end{align}
 In the first place,  we always have  the trivial solution $\xi = 0$. Recall that no genuine representatives exist in the $(F,Q,R)$-orbit, so that we cannot employ a rigid $T$-transformation as a solution-generating technique in this case. {On the other hand, it is possible to consider  a rigid $\beta$-transformation relating this representative and the one of the $(F,H,R)$-orbit.} It would be interesting to find also non-trivial solutions for $\xi^{ij}$ within this orbit in the hope of having solution-generating techniques mapping cases with different $\xi$'s. We leave this problem open.

\section{Type II superstring and Ramond-Ramond fields}\label{sec:RR}
Let us now include also the RR fields in the discussion of the solution-generating techniques,
which in general may be relevant for type II backgrounds.
RR fields in the doubled formulation were discussed in various works, see for example~\cite{Hohm:2011dv,Coimbra:2011nw,Jeon:2012kd}. Here we will employ the spinorial formulation of~\cite{Hohm:2011dv} following the rewriting of~\cite{Geissbuhler:2013uka}. In this section we will review only the essential ingredients of the construction, and we refer to appendix~\ref{app:RR} for more details. 
One uses  a democratic formulation~\cite{Fukuma:1999jt,Bergshoeff:2001pv} where all even (odd) forms from 1 to $D=10$ are used for the RR fields strengths of type IIA (IIB). The RR potentials are encoded in a spinor $\ket{\Csp}$ and the RR field strengths in a spinor\footnote{Our $\ket{\Fsp}$ is $\mathcal G$ of~\cite{Geissbuhler:2013uka}. Our $\ket{\bm F}$ and  $\ket{\hat{\bm F}}$ appearing later are respectively $\ket{F}$ and $\ket{\hat F}$ of~\cite{Hohm:2011dv}.} $\ket{\Fsp}$. 
Given Gamma matrices $\bm{\Gamma^A}$ in $2D$-dimensions satisfying the Clifford algebra relations $\{\bm{\Gamma^A},\bm{\Gamma^B}\}=2\eta^{\bm {AB}}$ (notice the flat indices) we write $\bm{\Gamma^A}=\sqrt 2\bm{\psi^A}$ so that $\{\bm{\psi_a},\bm{\psi^b}\}=\delta_{\bm a}^{\bm b}$, $\{\bm{\psi_a},\bm{\psi_b}\}=0$, $\{\bm{\psi^a},\bm{\psi^b}\}=0$ are anticommutation relations for the fermionic oscillators $\bm{\psi_a},\bm{\psi^a}$. Starting from the Clifford vacuum $\ket{0}$ such that $\bm{\psi_a}\ket{0}=0$ for all $\bm a$, we  rewrite the spinor $\ket{\Fsp}$ as 
\be\label{eq:FF}
\ket{\Fsp}=\sum_{p=0}^D\frac{e^{\bm \phi}}{p!}\hat{\bm F}_{\bm m_1\cdots \bm m_p}\bm e_{\bm a_1}{}^{\bm m_1}\cdots \bm e_{\bm a_p}{}^{\bm m_p}\bm{\psi^{a_1}}\cdots\bm{\psi^{a_p}}\ket{0}.
\ee
This rewriting relies on the one-to-one map between spinors $\ket{\Fsp}$ and polyforms $\hat{\bm F}$, where $\hat{\bm F}=\sum_{p=0}^D\frac{1}{p!}\hat{\bm F}_{\bm m_1\cdots \bm m_p}d\bm{x^{m_1}}\cdots d\bm{x^{m_p}}$ on $\mathbb R^{1,D-1}$. Here we are using  $\bm e_{\bm a}{}^{\bm m}$ which is the (inverse) vielbein for the metric $\bm{G_{mn}}$.
Importantly, $\hat{\bm F}_{\bm m_1\cdots \bm m_p}$ are the RR field strengths that are commonly used in type II supergravity (they are the $\hat F$ of~\cite{Hohm:2011dv}), and they are the ones that appear in the quadratic couplings of the fermions in the Green-Schwarz formulation of the superstring (they are the $F$ of~\cite{Wulff:2013kga}).
The two spinors $\ket{\Fsp}$ and $\ket{\Csp}$ are related by
\be\label{eq:Fdc}
\ket{\Fsp}=(\bm{\psi^A}\partial_{\bm A}-\tfrac16 \FF_{\bm{ABC}}\bm{\psi^{ABC}}-\tfrac12 \FF_{\bm A}\bm{\psi^A})\ket{\Csp}=\slashed\nabla\ket{\Csp},\qquad
\slashed\nabla\equiv\slashed\partial-\slashed{\FF}^{(3)}
-\tfrac12 \slashed{\FF}^{(1)},
\ee
where $\slashed{\FF}^{(n)}$ includes the $1/n!$ factor.
Gauge transformations of RR potentials read as $\delta_{\bm\lambda}\ket{\Csp}=\slashed\nabla\ket{\bm \lambda}$ and Bianchi identities as $\slashed\nabla\ket{\Fsp}=0$. Notice that these Bianchi identities are a consequence of $\slashed\nabla^2=0$ which holds on the strong constraint.
After defining $\bm \Psi_-=(\bm \psi_0+\bm \psi^0)(\bm \psi_1-\bm \psi^1)\cdots (\bm \psi_9-\bm \psi^9)$ we also impose the self-duality condition $\bm \Psi_-\ket{\Fsp}=\ket{\Fsp}$. Notice that this differs from~\cite{Geissbuhler:2013uka}, see appendix~\ref{app:RR}.
The self-duality condition translates into the duality conditions for the $p$-forms as\footnote{The Hodge dual is defined as
$(*\bm A)_{\bm m_1\cdots \bm m_p}=\frac{1}{(10-p)!}\bm G_{\bm m_1\bm n_1}\cdots \bm G_{\bm m_p\bm n_p}\bm \varepsilon^{\bm k_{p+1}\cdots \bm k_{10}\bm n_1\cdots \bm n_p}\bm A_{\bm k_{p+1}\cdots \bm k_{10}}$
where $\bm \epsilon^{1\cdots 10}=1$, $\bm \epsilon_{1\cdots 10}=-1$, $\bm \varepsilon^{\bm m_1\cdots \bm m_{10}}=\frac{1}{\sqrt{-\bm G}}\bm \epsilon^{\bm m_1\cdots \bm m_{10}}$, $\bm \varepsilon_{\bm m_1\cdots \bm m_{10}}=\sqrt{-\bm G}\bm \epsilon_{\bm m_1\cdots \bm m_{10}}$.
We also have $**\omega^{(p)}=-(-1)^{p(10-p)}s \omega^{(p)}$ where the additional minus sign is due to the Lorentz signature.}
$\hat {\bm F}^{(p)}=-(-1)^{\frac12 p(p+1)}*\hat {\bm F}^{(10-p)}$.

The transformation rules of RR fields $\hat{\bm F}_{\bm m_1\cdots \bm m_p}$ under the $O(D,D)$ solution-generating techniques are found by the observation that $\ket{\Fsp}$ is in fact invariant under these transformations. Notice that knowing that the generalised fluxes $\FF$ are invariant under the $O(D,D)$ solution-generating techniques, it is obvious that keeping also $\ket{\Fsp}$ invariant ensures that even when including the RR sector we still have a solution of the supergravity equations, and  that we correctly satisfy the Bianchi identities and the constraints. The fact that we can identify a dynamical field (in this case $\ket{\Fsp}$) that remains invariant under the transformation is again a confirmation of the usefulness of the DFT formulation that we are employing.

The fact that $\ket{\Fsp}$ is invariant does not mean that the RR fields $\hat{\bm F}_{\bm m_1\cdots \bm m_p}$  remain invariant. This is similar to what we saw in the NSNS sector: while the generalised fluxes are invariant, the background metric $\bm{ G_{mn}}$, the field $\bm{B_{mn}}$ and the dilaton $\bm \phi$ do transform.
To present the transformation rules it is useful to define $\ket{\bm{F}}\equiv e^{-\dd}S_{[\EE]}^{-1}\ket{\Fsp}$ where $S_{[\EE]}$ is a $Spin(D,D)$ element corresponding to the $O(D,D)$ element $\EE$. See the appendix for more details.  Another useful rewriting of the spinor for the RR field strenghts is $\ket{\hat{\bm F}}\equiv S_{[\bm B]}\ket{\bm F}$, where we use the $Spin(D,D)$ element $S_{[\bm B]}\equiv \exp(-\tfrac12 \bm{B_{mn}\psi^m\psi^n})$ with $\bm{B_{mn}}$ the B-field of the supergravity background.
We will then write $\ket{{\bm F}}=\sum_{p=1}^D\frac{1}{p!}\bm { \bm F}_{\bm m_1\cdots \bm m_p}\bm{\psi^{m_1}}\cdots\bm{\psi^{m_p}}\ket{0}$ and $\ket{\hat{\bm F}}=\sum_{p=1}^D\frac{1}{p!}\bm \hat{ \bm F}_{\bm m_1\cdots \bm m_p}\bm{\psi^{m_1}}\cdots\bm{\psi^{m_p}}\ket{0}$,
which is compatible with~\eqref{eq:FF}.
From the invariance of $\ket{\Fsp}$ and using that $\EE=\dot\EE\bm U$ and $\dd=\dot \dd+\lambda$, it follows that 
\be
\ket{\dot{ \bm F}}\equiv e^\lambda S_{[\bm U]}\ket{\bm F},
\ee
is also invariant under the $O(D,D)$ transformations. If we use primes to denote the new background related by the $O(D,D)$ transformation, this fact can be exploited to compute the RR fields of the new solution just by identifying $\ket{\dot{ \bm F}}= e^{\lambda'} S_{[\bm U']}\ket{\bm F'}$.

From an operational point of view, starting from the polyform $\hat{\bm F}=\sum_{p=1}^D\frac{1}{p!}\hat{\bm F}_{\bm m_1\cdots \bm m_p}d\bm{x^{m_1}}\cdots d\bm{x^{x_p}}$ of a supergravity solution, we can get ${\bm F}=\exp(\bm B^{(2)})\wedge\hat{\bm F}$ with $\bm B^{(2)}\equiv \tfrac12 \bm{B_{mn}}d\bm{x^m}\wedge d\bm{x^n}$, which is a consequence of the relation between the corresponding spinors.
To obtain ${\dot{\bm F}}$ we write the $Spin(D,D)$ element $S_{[\bm U]}=S_{[\bm \rho]} S_{[\bar{\bm \beta}]}S_{[\bar{\bm  b}]}^{-1}$ which follows from the rewriting of the twist $U$ as 
\be
U=
\left(
\begin{array}{cc}
1 & \beta\\
0 & 1
\end{array}
\right)
\left(
\begin{array}{cc}
1 & 0\\
b & 1
\end{array}
\right)
\left(
\begin{array}{cc}
\rho^t & 0\\
0 & \rho^{-1}
\end{array}
\right)
=
\left(
\begin{array}{cc}
\rho^t & 0\\
0 & \rho^{-1}
\end{array}
\right)
\left(
\begin{array}{cc}
1 & \bar\beta\\
0 & 1
\end{array}
\right)
\left(
\begin{array}{cc}
1 & 0\\
\bar b & 1
\end{array}
\right)
\ee 
where $\bar\beta=\rho^{-t}\beta\rho^{-1}$ and $\bar b=\rho b\rho^t$ have curved indices.
We prefer this rewriting because it simplifies the translation of the action on polyforms. In fact
\be
\ket{\bm F^{(\bar b)}}=S_{[\bar b]}^{-1}\ket{\bm F}=e^{\frac12 \bar b_{\mu\nu}\bm \psi^\mu\bm \psi^\nu}\ket{\bm F}
\quad\implies\quad
\bm F^{(\bar b)}=e^{\bar b\wedge}\bm F=\bm F+\bar b\wedge \bm F+\frac{1}{2!}\bar b\wedge \bar b\wedge \bm F+\ldots,
\ee
where $\bar b\equiv \frac12 \bar b_{\mu\nu}dy^\mu\wedge dy^\nu$ and
\be
\ket{\bm F^{(\bar\beta)}}=S_{[\bar\beta]}\ket{\bm F}=e^{\frac12 \bar\beta^{\mu\nu}\psi_\mu\psi_\nu}\ket{\bm F}
\quad\implies\quad
\bm F^{(\bar\beta)}=e^{\bar\beta\vee}\bm F=\bm F+\bar \beta\vee \bm F+\frac{1}{2!}\bar \beta\vee \bar\beta\vee \bm F+\ldots,
\ee
where $\bar \beta\vee \bm F= \frac12 \bar\beta^{\mu\nu}\iota_\mu\iota_\nu \bm F$ and $\iota_{\bm m}dx^{\bm n}=\delta_{\bm m}^{\bm n}$, $\iota_{\bm m}(\omega^{(p)}\wedge\chi^{(q)})=\iota_{\bm m}\omega^{(p)}\wedge\chi^{(q)}+(-1)^p\omega^{(p)}\wedge\iota_{\bm m}\chi^{(q)}$.
Finally one has
\be
\begin{aligned}
S_{[\bm \rho]}\ket{\bm F}&=(\det\rho)^{1/2}\exp(-\psi^{\bm m}\bm R_{\bm m}{}^{\bm n}\psi_{\bm n})
\sum_p\frac{1}{p!}\bm F_{\bm m_1\cdots \bm m_p}\bm \psi^{\bm m_1}\cdots\bm \psi^{\bm m_p}\ket{0}\\
&=(\det\rho)^{1/2}
\sum_p\frac{1}{p!}\bm F_{\bm m_1\cdots \bm m_p}\bm \rho_{\bm i_1}{}^{\bm m_1}\cdots \bm \rho_{\bm i_p}{}^{\bm m_p}\bm \psi^{\bm i_1}\cdots\bm \psi^{\bm i_p}\ket{0}
\end{aligned}
\ee
where $\bm \rho=\exp \bm R$ acts as the identity on spectator coordinates $\dot x^{\dot \mu}$ and coincides with $\rho$ on the $y$-block.
In other words the only action of $S_{[\bm \rho]}$ is to translate curved indices $\mu,\nu$ into algebra-like indices $i,j$, and multiply by $(\det\rho)^{1/2}$. Notice that this factor cancels in $\ket{\dot{\bm F}}$ once we rewrite $\lambda=\bar\lambda-\frac12\log\det\rho$.
With the above formulas and starting from $\check{\bm F}, U,\lambda$ one can obtain $\dot{\bm F}$, and given another representative with $U',\lambda'$ one can similarly compute ${\bm F}'$ and $\hat{\bm F}'$ by the same formulas.

As we argued in the previous sections, the solution-generating techniques that we are studying cover not only those that leave the generalised fluxes invariant. They also cover those that relate different sets of generalised fluxes by constant $O(D,D)$ transformations, the prominent example being the rigid $T$-transformation implemented by the matrix in~\eqref{eq:T}. Let us show how the RR fields transform when this $T$-transformation is involved. When the fluxes are related as in $\F'_{A'B'C'}=T_{A'}{}^AT_{B'}{}^BT_{C'}{}^C\F_{ABC}$ and $\F'_{A'}=T_{A'}{}^A\F_{A}$, from~\eqref{eq:Fdc} and using $S_{[T]}\bm\psi^{\bm A}S_{[T]}^{-1}=\bm \psi^{\bm B}\bm{T_B{}^A}$ one sees that we have the relations $\ket\Fsp'=S_{[T]}\ket{\Fsp}$ and $\ket\Csp'=S_{[T]}\ket{\Csp}$, provided that $\ket\Csp$ is taken to be independent of the coordinates that are being dualised.\footnote{Notice that the ansatz for the spinor taken in~\cite{Catal-Ozer:2017cqr} is equivalent to this condition, and a similar condition is taken also in~\cite{Demulder:2018lmj,Sakatani:2019jgu}.}
Notice that the chirality of the spinors remains the same if we dualise in even $d$ dimensions, and changes in odd $d$. In addition notice that this transformation\footnote{In the presence of RR fields we assume that  we do not dualise the time direction.} gives $\Psi'_-=S_T\Psi_-S_T^{-1}=(-1)^d\Psi_-$, therefore in $d$-even dimensions it preserves the self-duality condition, while in $d$ odd it changes the sign, see also~\cite{Hohm:2011dv}.

To obtain the transformation rules of RR fields one uses the fact that the relation to the dualised model is via a twist
\be
\tilde U=
\left(
\begin{array}{cc}
0 & 1\\
1 & 0
\end{array}
\right)
\left(
\begin{array}{cc}
1 & \beta\\
0 & 1
\end{array}
\right)
\left(
\begin{array}{cc}
1 & 0\\
b & 1
\end{array}
\right)
\left(
\begin{array}{cc}
\rho^t & 0\\
0 & \rho^{-1}
\end{array}
\right)
=
\left(
\begin{array}{cc}
\rho^{-1} & 0\\
0 & \rho^{t}
\end{array}
\right)
\left(
\begin{array}{cc}
0 & 1\\
1 & 0
\end{array}
\right)
\left(
\begin{array}{cc}
1 & \bar\beta\\
0 & 1
\end{array}
\right)
\left(
\begin{array}{cc}
1 & 0\\
\bar b & 1
\end{array}
\right)
\ee 
that now has an additional matrix implementing the $T$-transformation. Also in this case it is preferable to pull the $GL(d)$ block with $\rho$ to the left, because this rewriting permits to implement the action of the T-duality matrix (the second block in the last equation) on polyforms in a simple way.
Notice that because of the above rewriting we can think of the operation as a dualisation of $d$ \emph{coordinates} not just in the case of abelian T-dualities, but also in their generalisations. When dualising more than one coordinate we can think of it as a factorised product of single T-dualities along the $m$ direction implemented by $S_{[\mu]}=(\psi^\mu-\psi_\mu)(-1)^{N_F}$. Under such transformation one has in polyform notation
\be
\ket{\bm F^{(\mu)}}=S_{[\mu]}\ket{\bm F}
\quad\implies\quad
\bm F^{(\mu)}=\bm F\wedge dy^\mu+\bm F\vee dy^\mu,
\ee
where if $A$ is a $p$-form $A\vee dy^\mu=(-1)^{p-1}dy^\mu\vee A=(-1)^{p-1}\iota_\mu A$.
A similar point of view to obtain the transformation rules of RR fields under PL duality was used in~\cite{Sakatani:2019jgu}.

\section{Conclusions and outlook}\label{sec:concl}

We have discussed an ansatz for the generalised vielbein of DFT by demanding that it takes a ``twisted'' form, as in generalised Scherk-Schwarz reductions of $D$-dimensional backgrounds on $d$-dimensional spaces, and that the twist $U$ gives rise to constant generalised fluxes in the $d$-dimensional space when imposing the strong constraint of DFT.
The results are organised as in Figure~\ref{fig:orbits} into orbits depending on  which of the different fluxes $F,H,Q$ and $R$ are turned on,  and in general they can be related by rigid $T$-transformations by $F\leftrightarrow Q$ and $H\leftrightarrow R$, or more generic $O(d,d)$ transformations.
Our classification of the representatives is complete when the $H$-flux is vanishing, while we have employed particular  methods  for the cases with non-trivial $H$-flux which may  not cover all possible representatives.

When an orbit contains more than one representative, or when by means of a rigid $O(d,d)$ transformation (including a rigid $T$-transformation) we can relate it to another orbit that admits representatives, we can view the maps relating two of them as a solution-generating technique in (super)gravity, as well as a canonical transformation at the level of the $\sigma$-models. Therefore when one of the two $\sigma$-models is classically integrable it follows that the other enjoys the same property.
Our results add new possibilities to the known zoo of generalised T-dualities and Yang-Baxter deformations.
Generic $O(d,d)$ transformations relating possibly different orbits may be seen as generalisations of the so called PL-\emph{plurality}, which is traditionally defined as the possibility of decomposing in different ways the same Drinfel'd double.
Let us remark that in this paper we have only imposed the $O(d,d)$ symmetry of the background (i.e.~the fact that it takes the form of a generalised Scherk-Schwarz ansatz and that it gives rise to constant $\F_{IJK},\F_I$). At no point we have imposed the DFT (or supergravity) equations of motion, and this is actually not relevant if one is only interested in canonical transformations of $\sigma$-models. In general, one may have to choose appropriate RR fields and/or spectator background fields in order to solve the DFT/supergravity
equations.
Similar comments apply to the integrability of the $\sigma$-models giving rise to the backgrounds that we describe. At no point we imposed the existence of a Lax connection, and in general this may introduce additional conditions.

Let us now summarise our classification.
In the $(F)$-orbit one can independently switch on a twist $\beta$ satisfying the CYBE, which corresponds to possible (homogeneous) YB deformations~\cite{Klimcik:2002zj,Klimcik:2008eq,Kawaguchi:2014qwa} of isometric backgrounds.
This orbit is related by rigid $T$-transformations to the $(Q)$-orbit, which contains backgrounds that arise from applying NATD~\cite{delaOssa:1992vci} to the ones in the $(F)$-orbit, and more generally include the DTD models of~\cite{Borsato:2016pas,Borsato:2018idb}.
The $(F,Q)$ orbit contains representatives that are PL-symmetric~\cite{Klimcik:1995ux,Klimcik:1995dy}, and it allows also for novel deformations that can be understood as the natural generalisation of the homogeneous YB-deformations, now without the need of having isometries in the initial background.  Because the fluxes remain invariant, these deformations preserve the integrability of the $\sigma$-model. When we perform  this generalised deformation combined with a rigid $T$-transformation, the resulting map may be understood as a generalised notion of T-duality. 
The $(R)$-orbit is empty, it contains no non-trivial representative when demanding the strong constraint. The $(F,R)$ orbit contains the so-called inhomogeneous YB-model (or $\eta$-model)~\cite{Klimcik:2002zj,Klimcik:2008eq}, slightly generalised in our treatment by the presence of spectators. The $(Q,R)$ and $(F,Q,R)$ orbits are trivial in the sense that one can turn off the $R$-flux by a rigid $O(d,d)$ transformation.

When the $H$-flux is non-zero our classification is not exhaustive, because the methods of section~\ref{s:AnsatzOrbitsH} that we employ in these cases do not guarantee that we are covering all possible representatives. Nevertheless we identify interesting possibilities for the $(H), (F,H), (H,Q)$ and $(F,H,R)$ orbits. The $(H,R)$ and $(H,Q,R)$ orbits seem trivial within these methods since the $R$-flux can again be turned off by a rigid $O(d,d)$. The $(F,H)$-orbit contains the PCM+WZ model as well as a novel generalisation of the homogeneous YB-deformation in which the YB-operator is constrained to be compatible with $H$. When applied to the integrable PCM+WZ model, this deformation preserves integrability. The $(H,Q)$ orbit contains the integrable (asymmetrical) $\lambda$-model~\cite{Sfetsos:2013wia,Driezen:2019ykp} as well as  additional novel deformations by 2-cocycles  compatible with the $H$-flux, which are reminiscent  of the deformations of NATD in the $(Q)$-orbit that were called DTD. {The map between an $(F,R)$ representative and the generalised DTD representative of $(H,Q)$ by means of a rigid $T$-transformation can be understood as another generalised T-duality.} In the $(F,H,R)$-orbit we can describe a generalisation of the inhomogeneous YB-model\footnote{We stress that we cannot view it as a solution-generating technique from the PCM+WZ model since we are going from the $(F,H)$ to the $(F,H,R)$ orbit by turning on $R$.} as well as the fact that a rigid $O(d,d)$ relates it to the $(F,H,Q)$-orbit. Within the methods that we use, the $(F,H,Q)$-orbit admits at least the solutions of \cite{Demulder:2018lmj}.
The question of having  more general representatives in this orbit, as well as representatives in the most general $(F,H,Q,R)$-orbit, remains open. Finally in the $(H)$-orbit we identified possible deformations of the torus with $H$-flux without relying on the methods of section~\ref{s:AnsatzOrbitsH}. This confirms that it should be possible to  go beyond these methods, and it would  certainly be interesting to do so in a systematic way   in the other orbits with non-vanishing $H$-flux.
It would also be interesting to look at this classification  from the point of view of para-Hermitian geometry~\cite{Hassler:2019wvn}. Let us stress that all the maps connecting different representatives of the same orbit (up to rigid $O(d,d)$) can be used to generate integrable $\sigma$-models when starting from another known one. 

A general observation that we made, which generalises  previously known results, is that  a constant automorphism $W$ of the pre-Roytenberg algebra $\mathfrak{r}$ (modulo gauge transformations) may generate a new representative  from a known representative for the twist. They must not be mistaken with the  rigid $O(d,d)$ transformations of~\eqref{eq:twistRedundancy}, since these automorphisms do not involve a compensating transformation of the spectator background.  For instance, in the $(F)$-orbit the automorphism reduced to the subalgebra generated by $F_{ij}{}^k$ must be an outer automorphism. An interesting possibility is that for non-semisimple algebras outer automorphisms may involve continuous parameters and can therefore be seen as deforming the background.

While we have focused on the NSNS sector in most of the paper, in section~\ref{sec:RR} we have explained how to obtain the transformation rules of RR fluxes by demanding that the maps under consideration are in fact mapping type II solutions to type II solutions.

Let us now comment on several interesting open questions.

While  we have discussed in our classification of representatives  various known integrable  2-dimensional $\sigma$-models, it would be interesting to rewrite in this language also others such as the bi-YB-deformation of~\cite{Delduc:2017fib,Klimcik:2019kkf,Klimcik:2020fhs} and the deformations of~\cite{Sfetsos:2017sep}.

In this paper we have not analysed the special case of solution-generating techniques involving (super)cosets, and this would be a very interesting future direction. In the (super)coset case the spectator background fields are expected to project on the coset part of the algebra, and can therefore give rise to solution-generating techniques even when demanding  conditions that are weaker than the invariance of the $O(d,d)$ generalised fluxes $\F_{IJK},\F_I$, see~\cite{Araujo:2017enj,Hronek:2020skb} for an observation along these lines. In this paper we cannot view the inhomogeneous YB-deformations (the $\eta$-deformation) as a solution-generating technique from $\eta=0$ to $\eta\neq 0$, because in general it entails going from the $(F)$-orbit to the $(F,R)$-orbit, but it is possible that the (unimodular) $\eta$-deformation of the superstring~\cite{Hoare:2018ngg} turns out to be a solution-generating technique because of these additional features of the supercoset. The transformation rules of the NSNS and RR fields~\cite{Borsato:2016ose} are in fact strongly suggesting the underlying $O(D,D)$ structure also in this case.
Similar comments apply to the $\lambda$-deformation of the superstring~\cite{Hollowood:2014qma}, as well as to the construction of~\cite{Sfetsos:2014cea}.

Another possible future direction, which is also necessary for the point above, is to discuss solution-generating techniques that involve \emph{super}algebras. It is likely that for this purpose the formulation of~\cite{Jeon:2012kd} for type II superstrings in DFT language will be more useful than the formulation of~\cite{Hohm:2011dv}.


It would  be interesting to relax some assumptions that we have made. In particular, the solution-generating techniques discussed in this paper arise by demanding that the generalised fluxes (and their flat derivatives) remain invariant under the map.
To find more general solution-generating techniques in supergravity, one may try to look for more complicated symmetries of the DFT equations of motion that do not
necessarily leave the generalised fluxes invariant, see~\cite{Hronek:2020skb} for a step in this direction. 
Additionally we were interested in backgrounds that satisfy the strong constraint of DFT, namely those whose fields depend only on the physical coordinates $y$ and not on the dual $\tilde y$. In the context of gauged DFT it is possible to relax the strong constraint as done in~\cite{Dibitetto:2012rk}, and it would be interesting to look  at this generalisation of the classification as well.
One may also relax our analysis by looking for solution-generating techniques in the context of the generalised supergravity equations of~\cite{Arutyunov:2015mqj,Wulff:2016tju}.

\section*{Acknowledgements}
We thank S.~Demulder, B.~Hoare, S.~Lacroix, D.~Osten and L.~Wulff for useful discussions, and we are very grateful especially to D.~Marqu\'es for discussions and for inspiring this project. We thank D.~Marqu\'es, D.~Thompson and L.~Wulff for comments on the manuscript.
This work was supported by the fellowship of ``la Caixa Foundation'' (ID 100010434) with code LCF/BQ/PI19/11690019,
by AEI-Spain (FPA2017-84436-P and Unidad de Excelencia Mar\'\i a de Maetzu MDM-2016-0692), by Xunta de Galicia-Conseller\'\i a de Educaci\'on (Centro singular de investigaci\'on  de  Galicia  accreditation  2019-2022), and  by the European Union FEDER.

\appendix

\section{Notation}\label{app:notation}
We follow the convention of using a boldface notation for the objects (and their indices) of the full $D$-dimensional space, both before and after the doubling (e.g.~we have the metric $\bm{G_{mn}}$ and the generalised fluxes $\bm{\F_{ABC}}$), and the same notation but without boldface for the corresponding quantities in $d$ dimensions, (e.g.~$G_{mn}$ and $\F_{ABC}$). A recap on our indices conventions is as follows:

\begin{tabular}{rl}
$\bm{m,n},\ldots$ & Curved $D$-dimensional indices of coordinates $\bm{x^m}$\\
$\mu,\nu,\ldots, \alpha , \beta , \ldots$ & Curved $d$-dimensional indices of $y^\mu$\\
$\dot\mu,\dot\nu,\ldots$ &  Curved $(D-d)$-dimensional indices of spectator coordinates $\dot x^{\dot \mu}$\\
 & \\
$\bm{M,N},\ldots$ &  Curved $O(D,D)$ indices\\
${M,N},\ldots$  & Curved $O(d,d)$ indices\\
& \\
$\bm{A,B},\ldots$ &  Flat $O(1,D-1)\times O(D-1,1)$-dimensional indices\\
& \\
${I,J},\ldots$  & Algebra indices in $2d$ dimensions 
\\
${i,j},\ldots$  & Algebra indices in $d$ dimensions 
\end{tabular}

\section{Brief recap on DFT and gDFT}\label{app:DFT}
The generalised vielbein may be parametrised as
\begin{equation}
\bm{E_A{}^M}=
\frac{1}{\sqrt2}
\left(
\begin{array}{cc}
\bm{e^{(+)an}(G-B)_{nm}} & \bm{e^{(+)am}}\\
-\bm{e^{(-)}_a{}^n(G+B)_{nm}} & \bm{e^{(-)}_a{}^m}
\end{array}
\right)\,.
\label{eq:E}
\end{equation}
Here $\bm{e^{(\pm)}}$ are two vielbeins for the metric $
\bm{G_{mn}}$ and $\bm{B_{mn}}$ is the Kalb-Ramond field.
The generalised vielbein satisfies the following relations with the $O(D,D)$ metric~\eqref{eq:eta} and the generalised metric~\eqref{eq:H}
\begin{equation}
\bm{\eta^{MN}}=\bm{E_A{}^M\eta^{AB}E_B{}^N}\,,
\qquad\qquad
\bm{\mathcal H^{MN}}=\bm{E_A{}^M\mathcal H^{AB}E_B{}^N}\,,
\end{equation}
where
\begin{equation}
\bm{\eta_{AB}}=
\left(
\begin{array}{cc}
	\bm{\bar\eta^{ab}} & \mathbf 0\\
	\mathbf 0 & -\bm{\bar\eta_{ab}}
\end{array}
\right)\,,\qquad\qquad
\bm{\mathcal H^{AB}}=
\left(
\begin{array}{cc}
	\bm{\bar\eta_{ab}} & \mathbf 0\\
	\mathbf 0 & \bm{\bar\eta^{ab}}
\end{array}
\right)\,,
\end{equation}
Curved indices $\bm{M,N,\ldots}$ are raised and lowered with $\bm{\eta^{MN}}$ and $\bm{\eta_{MN}}$, while flat indices $\bm{A,B,\ldots}$ with $\bm{\eta^{AB}}$ and $\bm{\eta_{AB}}$.
The generalised dilaton is $\bm d=\bm \phi-\tfrac14\log(-\det\bm G)$ with $\bm \phi$ the usual dilaton.
The generalised fluxes are defined as
\be
\FF_{\bm{ABC}}=3 \OO_{\bm {[ABC]}},  \qquad\qquad
\FF_{\bm A}=\OO^{\bm B}{}_{\bm{BA}}+2 \EE_{\bm A}{}^{\bm M}\partial_{\bm M}\dd
\ee
in terms of the generalized Weitzenb\"ock connection
\be
\OO_{\bm {ABC}}= \EE_{\bm A}{}^{\bm M}\partial_{\bm M} \EE_{\bm B}{}^{\bm N} \EE_{\bm{ CN}}, 
\ee
and they satisfy the following Bianchi identities 
\be
\begin{aligned}
\partial_{[{\bm A}}\FF_{\bm{ BCD}]}-\frac34\FF_{[\bm{ AB}}{}^{\bm E}\FF_{\bm{ CD}]E}&=\bm{\mathcal Z}_{\bm{ ABCD}},\\
2\partial_{[\bm A}\FF_{\bm{ B}]}+(\partial^{\bm C}-\FF^{\bm C})\FF_{\bm{ ABC}}&=\bm{\mathcal Z}_{\bm{ AB}}
,\\
\partial^{\bm A}\FF_{\bm A}-\tfrac12 \FF^{\bm A}\FF_{\bm A}+\tfrac{1}{12}\FF^{\bm {ABC}}\FF_{\bm{ ABC}}&=\bm{\mathcal Z}
\end{aligned}
\ee
where
\be
\begin{aligned}
\bm{\mathcal Z}_{\bm{ ABCD}}&\equiv -\frac34\OO_{\bm{ E[AB}}\OO^{\bm E}{}_{\bm{ CD}]},\\
\bm{\mathcal Z_{\bm{ AB}}}&\equiv (\partial^{\bm M}\partial_{\bm M}\EE_{[\bm A}{}^{\bm N})\EE_{\bm B]\bm N}
-2\OO^{\bm C}{}_{\bm{ AB}}\partial_{\bm C}\dd,\\
\bm{\mathcal Z}&\equiv -2\partial^{\bm A}\dd\partial_{\bm A}\dd+2\partial^{\bm A}\partial_{\bm A}\dd+\tfrac14 \OO^{\bm{ ABC}}\OO_{\bm{ ABC}}.
\end{aligned}
\ee
On the strong constraint one has $\bm{\mathcal Z}_{\bm{ ABCD}}=0$, $\bm{\mathcal Z}_{\bm{ AB}}=0$ and $\bm{\mathcal Z}=0$.

A generalised diffeomorphism is implemented on tensors $\bm{V_{ M}{}^{ N}}$ and on the generalised dilaton (transforming as a density) as
\be\label{eq:genLie}
\begin{aligned}
&\delta_{\bm\xi}\bm{V_{ M}{}^{ N}}=
\hat{\mathcal L}_{\bm\xi}\bm{V_{ M}{}^{ N}}=
\bm{\xi^{ P}}\partial_{\bm P}\bm{V_{ M}{}^{ N}}+(\partial^{\bm N}\bm{\xi_{ P}}-\partial_{\bm P}\bm{\xi^{ N}})\bm{V_{ M}{}^{ P}}
+(\partial_{\bm M}\bm{\xi^{ P}}-\partial^{\bm P}\bm{\xi_{ M}})\bm{V_{ P}{}^{ N}},\\
&\delta_{\bm\xi}e^{-2\dd}=\hat{\mathcal L}_{\bm\xi}e^{-2\dd}=\partial_{\bm M}(\bm{\xi^{ M}}e^{-2\dd}),\qquad\iff\qquad
\delta_{\bm\xi}\dd=\hat{\mathcal L}_{\bm\xi}\dd= \bm{\xi^{ M}}\partial_{\bm M}\dd-\frac12 \partial_{\bm M}\bm{\xi^{\bf M}},
\end{aligned}
\ee
where $\hat{\mathcal L}_{\bm\xi}$ is the generalised Lie derivative and $\bm{\xi^M}$ the parameter of the transformation. On the strong constraint the generalised fluxes transform as scalars under generalised diffeomorphisms. In terms of the generalised Lie derivative they may be written also as
\be
\FF_{\bm {ABC}}=\hat{\mathcal L}_{\EE_{\bm A}}\EE_{\bm B}{}^{\bm M}\EE_{\bm{ CM}},\qquad
\FF_{\bm A}=2\hat{\mathcal L}_{\EE_{\bm A}}\dd.
\ee
The strong constraint is again a sufficient condition also for the closure of the algebra of generalised diffeomorphisms, so that $[\hat{\mathcal L}_{\bm\xi_1},\hat{\mathcal L}_{\bm\xi_2}]=\hat{\mathcal L}_{\bm\xi_{12}}$ with $\bm\xi_{12}=[\bm\xi_1,\bm\xi_2]_{(\text C)}^{\bm M}=\frac12(\delta_{\bm\xi_1}\bm\xi_2^{\bm M}-\delta_{\bm\xi_2}\bm\xi_1^{\bm M})=2\bm\xi_{[1}^{\bm P}\partial_{\bm P}\bm\xi_{2]}^{\bm M}+\partial^{\bm M}\bm\xi_{[1{\bm P}}\bm\xi_{2]}^{\bm P}$ given by the C-bracket.

The point of view of this paper is similar to the setup of gauged DFT~\cite{Hohm:2011ex,Grana:2012rr}. Rather than the original interpretation of~\cite{Hohm:2011ex}, where the $O(D,D)$ theory is gauged by shifting the generalised fluxes by some constant gaugings, we are closer to the interpretation of~\cite{Grana:2012rr} where an appropriate generalised Scherk-Schwarz (gSS) reduction from $D$ to $(D-d)$ dimensions essentially gives the same construction, i.e.~generalised fluxes for the $(D-d)$-dimensional theory that are gauged by the fluxes of the $d$-dimensional space.
If tensors are decomposed as in the gSS ansatz $\bm{V_{ M}{}^{ N}(x)}=\dot{\bm V}_{\bm I}{}^{\bm J}(\dot x)\bm{U^{ I}{}_{ M}}(y)\bm{U_{J}{}^{ N}}(y)$ then generalised diffeomorphisms with parameter
$\bm{\xi^{ M}}=\dot {\bm\xi}^{\bm I}\bm{U_{ I}{}^{ M}}$
respect the ansatz in the sense that $\delta_{\bm\xi} \bm{V_{ M}{}^{ N}}=\dot\delta_{\dot{\bm\xi}}\dot{\bm V}_{\bm I}{}^{\bm J}\bm{U^{ I}{}_{ M}}\bm{U_{ J}{}^{ N}}$
where in the case of a vector
$\dot\delta_{\dot{\bm\xi}}\dot {\bm{V^{ I}}}=\delta_{\dot {\bm\xi}}\dot{ \bm{V^{ I}}}+\F^{\bm I}{}_{\bm{ JK}}\dot{\bm \xi}^{\bm J}\dot{\bm V}^{\bm K}$.
Imposing the strong constraint both on the ``external space'' (with coordinates $\dot x$) and ``internal space'' (with coordinates $y$) is a sufficient condition for closure of the algebra of the $\dot\delta_{\dot{\bm\xi}}$ transformations, since $[\dot\delta_{\dot{\bm\xi}_1},\dot\delta_{\dot{\bm\xi}_2}]=\dot\delta_{\dot{\bm\xi}_{12}}$ where $\dot{\bm\xi}_{12}^{\bm I}=[\dot{\bm{ \xi}}_1,\dot{\bm \xi}_2]_{(\F)}^{\bm I}\equiv [\dot{\bm \xi}_1,\dot{\bm \xi}_2]_{(\text C)}^{\bm I}+\F^{\bm I}{}_{\bm{ JK}}\dot{\bm{\xi}}_1^{\bm J}\dot{\bm\xi}_2^{\bm K}$. Notice the relation to the C-bracket $[\bm\xi_1,\bm\xi_2]_{(\text C)}^{\bm M}=[\dot{\bm \xi}_1,\dot {\bm\xi}_2]_{(\F)}^{\bm I}\bm U_{\bm I}{}^{\bm M}$.
While the strong constraint on the internal space is a sufficient condition to define consistently gauged DFT, it may be relaxed, as long as $\mathcal Z_{IJKL}=\mathcal Z_{IJ}=\mathcal Z=0$.

\subsection{Geometric interpretation}\label{app:geom}
The fact that the solution-generating techniques that we discuss do not modify the generalised fluxes when written in flat indices, still allows for the possibility of finding maps between geometric and (globally) non-geometric backgrounds.
(Non)geometry is in fact captured by the fluxes written in \emph{curved} indices --- otherwise their definition in flat indices is dependent on the chosen double-Lorentz gauge, see~\cite{Grana:2008yw} for a discussion. We refer to~\cite{Plauschinn:2018wbo} for a comprehensive review on non-geometric backgrounds in string theory.
After going from flat to curved indices, the independent components of $\FF_{\bm M \bm N \bm P}$ are usually divided into geometric fluxes $\bm F_{\bm m\bm n}{}^{\bm p},\bm H_{\bm m\bm n\bm p}$ and non-geometric fluxes $\bm Q_{\bm m}{}^{\bm n\bm p},\bm R^{\bm m\bm n\bm p}$~\cite{Grana:2008yw}, see also~\cite{Geissbuhler:2013uka,Aldazabal:2013sca}.  
These fluxes are often used to distinguish between 
\begin{itemize} 
\item \emph{geometric backgrounds}, i.e.~backgrounds that are well defined globally when using diffeomorphisms and B-field gauge transformations as transition functions in different patches. They have $\bm Q_{\bm m}{}^{\bm n\bm p}=\bm R^{\bm m\bm n\bm p}=0$.
\item \emph{globally non-geometric but locally geometric backgrounds}, e.g.~T-folds, that require T-duality transformations to glue different patches. They have $\bm Q_{\bm m}{}^{\bm n\bm p}\neq 0$ and $\bm R^{\bm m\bm n\bm p}=0$.
\item \emph{locally non-geometric backgrounds}, i.e.~depending on dual coordinates. They have $\bm Q_{\bm m}{}^{\bm n\bm p}\neq 0$ and $\bm R^{\bm m\bm n\bm p}\neq 0$.
\end{itemize}
We do not consider locally non-geometric backgrounds because we always impose the strong constraint.
If $\n\FF_{\bm A\bm B\bm C}=\FF_{\bm A\bm B\bm C}$ then the generalised fluxes in curved indices are related as
\be
\n \FF_{\bm M \bm N \bm P}=\bm h_{\bm M}{}^{\bm Q}\bm h_{\bm N}{}^{\bm R}\bm h_{\bm P}{}^{\bm S} \FF_{\bm Q \bm R \bm S},
\quad
\text{where}
\qquad
\bm h_{\bm M}{}^{\bm Q}=\n \EE_{\bm M}{}^{\bm A}\EE_{\bm A}{}^{\bm Q}=\n{\bm U}_{\bm M}{}^{\bm I}\bm U_{\bm I}{}^{\bm Q}.
\ee
Notice that because of the assumptions we made on the twists $\bm U,\n{\bm  U}$, the matrix $\bm h_{\bm M}{}^{\bm Q}$ is of block diagonal form as $\bm U$ in~\eqref{eq:split-U}, i.e.~it acts as the identity in the block for the (doubled) $\dot x$ coordinates, while it is 
\be
h_{ M}{}^{ Q}=\n U_{ M}{}^{ I}U_{I}{}^{ Q},
\ee
in the block for the $y$-coordinates.
If $h_{ M}{}^{ Q}$ has off-diagonal components $h_{\mu\rho},h^{\mu\rho}$ then it is possible that the type of fluxes (in curved indices) change after the transformation. To show examples we say how they change in  solution-generating techniques involving the $(F)$ and $(Q)$-orbits.
Notice that 
\be
\FF_{\bm M \bm N \bm P}= \bm U_{\bm M}{}^{\bm I} \bm U_{\bm N}{}^{\bm J} \bm U_{\bm P}{}^{\bm K}
(\dot {\bm \F}_{\bm {IJK}}+\F_{\bm {IJK}}),
\ee
where $\dot {\bm \F}_{\bm {IJK}}=\dot \EE_{\bm I}{}^{\bm A}\dot \EE_{\bm J}{}^{\bm B}\dot \EE_{\bm K}{}^{\bm C}\dot {\bm \F}_{\bm A\bm B\bm C}$ and $\dot \F_{\bm {IJK}}=3\delta_{[\bm{I}}{}^{\dot \mu}\dot E_{\bm{J|}}{}^{\bm{B}}\partial_{\dot \mu}\dot E_{\bm{B|K}]}$.
Because of the derivative at least one leg is dotted in $\dot \F_{\bm {IJK}}$, and the corresponding $U$ and $h$ act as the identity on that leg. The other two legs are not necessarily in the dotted directions, so that $U$ and $h$ can act non-trivially on them. Therefore let us stress that in general it is important to include the contribution of $\dot \F_{\bm {IJK}}$ to understand how the fluxes change under the transformation.\footnote{In the standard example of the T-duality chain on the 3-torus with $H$-flux done for example  in~\cite{Shelton:2005cf} one dualises one leg at a time, and therefore the contribution to the flux that is discussed in each step is indeed coming from $\dot \F_{\bm {IJK}}$, with two legs undotted.} However, in order to have a more concrete discussion, we will  look in the following only at the piece coming from $\F_{\bm{IJK}}$,  and we remind that this is non-zero only for $\bm{IJK}=IJK$. 
We are therefore looking at flux components with all legs along $y$-directions, and these do not mix with the contributions from $\dot \F$ that have at least one leg along dotted coordinates.
Hence we will analyse only these contributions to the fluxes
\be
\F_{ M  N  P}= U_{ M}{}^{ I} U_{ N}{}^{ J} U_{ P}{}^{ K}
\F_{ {IJK}},\qquad
\n \F_{ M  N  P}= \n U_{ M}{}^{ I} \n U_{ N}{}^{ J} \n U_{ P}{}^{ K}
\F_{ {IJK}},
\ee
that are related as
\be
\n \F_{ M  N  P}= h_{ M}{}^{ Q} h_{ N}{}^{ R} h_{ P}{}^{ S}
\F_{ {QRS}}.
\ee
Separating the components we have 
\be\label{eq:new-geom-F}
\begin{aligned}
\n\F_{\mu\nu\rho}&=
h_{\mu}{}^\alpha h_{\nu}{}^\beta h_{\rho}{}^{\gamma}\F_{\alpha\beta\gamma}
+3h_{[\mu}{}^\alpha h_{\nu}{}^\beta h_{\rho]\gamma}\F_{\alpha\beta}{}^\gamma
+3 h_{[\mu}{}^\alpha h_{\nu |\beta}h_{|\rho]\gamma}\F_{\alpha}{}^{\beta\gamma}
+ h_{\mu\alpha} h_{\nu \beta}h_{\rho\gamma}\F^{\alpha\beta\gamma},\\
\n\F_{\mu\nu}{}^\rho&=
h_\mu{}^\alpha h_\nu{}^\beta h^{\rho\gamma}\F_{\alpha\beta\gamma}
+(h_\mu{}^\alpha h_{\nu}{}^\beta h^\rho{}_\gamma+2h_{[\mu}{}^\beta h_{\nu]\gamma}h^{\rho\alpha})\F_{\alpha\beta}{}^\gamma\\
&\qquad+(h_{\mu\beta}h_{\nu\gamma}h^{\rho\alpha}+2h_{[\mu}{}^\alpha h_{\nu]\beta}h^\rho{}_\gamma)\F_\alpha{}^{\beta\gamma}
+h_{\mu\alpha}h_{\nu\beta}h^\rho{}_\gamma \F^{\alpha\beta\gamma},\\
\n \F_\mu{}^{\nu\rho}&=
h_\mu{}^\alpha h^{\nu\beta}h^{\rho\gamma}\F_{\alpha\beta\gamma}
+(h_{\mu\gamma}h^{\nu\alpha}h^{\rho\beta}+2h_\mu{}^\alpha h^{[\nu|\beta}h^{|\rho]}{}_\gamma)\F_{\alpha\beta}{}^\gamma\\
&\qquad+(h_\mu{}^\alpha h^\nu{}_\beta h^\rho{}_\gamma+2h_{\mu\beta}h^{[\nu}{}_\gamma h^{\rho]\alpha})\F_\alpha{}^{\beta\gamma}
+h_{\mu\alpha}h^\nu{}_{\beta}h^\rho{}_\gamma \F^{\alpha\beta\gamma},\\
\n \F^{\mu\nu\rho}&=
h^{\mu\alpha}h^{\nu\beta}h^{\rho\gamma}\F_{\alpha\beta\gamma}
+3h^{[\mu|\alpha}h^{|\nu|\beta} h^{|\rho]}{}_{\gamma}\F_{\alpha\beta}{}^\gamma
+ 3h^{[\mu|\alpha}h^{|\nu}{}_\beta h^{\rho]}{}_\gamma\F_\alpha{}^{\beta\gamma}
+h^\mu{}_{\alpha}h^\nu{}_{\beta}h^\rho{}_\gamma \F^{\alpha\beta\gamma}
\end{aligned}
\ee

\section{On the $O(d,d)$ parametrisation of the twist}\label{app:parU}
Here we explain why we take~\eqref{eq:parU} as a parametrisation of the twist $U$. The $O(d,d)$ group is generated by the matrices~\cite{Aldazabal:2013sca}
\be
R_M{}^N=\left(
\begin{array}{cc}
(\rho^t)^\mu{}_\nu & 0\\
0 & (\rho^{-1})_\mu{}^\nu
\end{array}
\right),\quad
J_M{}^N=\left(
\begin{array}{cc}
\delta^\mu{}_\nu & 0\\
b_{\mu\nu} & \delta_\mu{}^\nu
\end{array}
\right),\quad
T_M^{[\sigma]N}=\left(
\begin{array}{cc}
\delta^\mu{}_\nu-u^\mu_{(\sigma)\nu} & u_{(\sigma)}^{\mu\nu}\\
u^{(\sigma)}_{\mu\nu} & \delta_\mu{}^\nu-u_\mu^{(\sigma)\nu}
\end{array}
\right),
\ee
where $\rho\in GL(d)$, $b^t=-b$ and $u_{(\sigma)}=u^{(\sigma)}=\text{diag}(0,\ldots,0,1,0,\ldots,0)$, where the 1 is at the position $\sigma \in \{1, \ldots , d\}$.
The matrices $R$ and $J$ generate two separate subgroups, which together  generate a larger subgroup of $O(d,d)$ that we call $G_{\text{geom}}$. Notice that 
$
RJ=J'R,
$
with  $b'=\rho^{-1}b\rho^{-t}$.
Hence we can always use this move to write any element of $G_{\text{geom}}$ e.g.~in the order $JR$.
Each $T^{[\sigma]}$ can be understood as a T-duality along the direction $\sigma$. When we involve the matrices $T^{[\sigma]}$ we generate also different types of matrices. A distinguished one is given by the multiplication of all the $T^{[\sigma]}$
\be
T_M{}^N=(T^{[1]}T^{[2]}\ldots T^{[d]})_M{}^N=\left(
\begin{array}{cc}
0 & \delta^{\mu\nu}\\
\delta_{\mu\nu} & 0
\end{array}
\right),
\ee
which we will call a rigid $T$-transformation since it is constant.  Notice that $T^2=1$ and that $T$ is not equivalent to the $O(d,d)$ metric $\eta$, because of the position of the indices in the definitions. 
We also have matrices $S\equiv TR$
\be
S_M{}^N=\left(
\begin{array}{cc}
0 & (\sigma^{-1})^{\mu\nu}\\
(\sigma^t)_{\mu\nu} & 0
\end{array}
\right),\quad
\ee
with $\sigma\in GL(d)$,  which form a subgroup on their own. Notice that $TRT=R'$ where $\rho'=\rho^{-t}$. Finally, we also have matrices $K\equiv TJT$
\be
K_M{}^N=\left(
\begin{array}{cc}
\delta^\mu{}_\nu & \beta^{\mu\nu}\\
0 & \delta_\mu{}^\nu
\end{array}
\right),
\ee
with $\beta^t=-\beta$, which again form a subgroup on their own.
More complicated elements occur if instead of $T$ we multiply by  single $T^{[\sigma]}$ matrices. However, we will not consider this case here since we will assume that we have already reduced from $D$ to $d$ dimensions, so that T-duality is implemented on all the $d$ coordinates.

One way to classify the possibilities in the parametrisation of the most general $O(d,d)$ element is to observe whether they involve an even or an odd number of $T$'s, which we consequently call ``even" and ``odd" elements respectively.
Let us first analyse the most generic parametrisation for an even twist and argue that they are generated by the matrices $R,J,K$. 
As we  have remarked  $TRT=R'$  and we can pull any $R$ to the right of any $J$. Notice that the  same considerations apply for $K$.
Therefore,  any even twist can be written as a generic product of $J$'s and $K$'s times a single $R$. 
Finally  notice that we can always write 
$
JK=K'J'R'
$
where
\be
\rho'=(1+b\beta)^{-1},\qquad
b'=b(1+\beta b),\qquad
\beta'=\beta(1+b\beta)^{-1},
\ee
\emph{if we assume} that $1+b\beta$ is invertible. Therefore, under this assumption we can always re-arrange the order of  the matrices $J$ and $K$ in the product, at the cost of introducing a matrix of type $R$.
We can therefore conclude that any twist $U$ of even type can be written as the product
\be
U_{\text{even}}=KJR,
\ee
for some $K,J,R$.
Notice, however, because of the previous assumption  we may be excluding twists involving products of the type $JK$ with $1+b\beta$ not invertible.
Let us finally discuss the most generic parametrisation for odd twists. As they must be generated by an odd number of $T$'s we can write them as a generic product of $U_{\text{even}}$'s and $T$'s (odd). By noticing that 
$
TKJR=(TKT)(TJT)(TRT)T=J'K'R'T=K''J''R''T
$
where we used $T^2=1$, we conclude that we can always pull $T$ to the left, for example. Then the most generic parametrisation for an odd twist is
\be
U_{\text{odd}}=TKJR.
\ee
The only effect of the rigid $T$ matrix in the odd twist is to map an orbit to its dual.

\section{Details on RR fields and type II}\label{app:RR}
In this appendix we collect some additional details that are useful for the DFT formulation of type II supergravity. We closely follow~\cite{Hohm:2011dv} and~\cite{Geissbuhler:2013uka} and highlight the differences when present.
In this appendix we will always assume that $D=10$. We will also remove the boldface from all objects and indices, since the reduction from $D$ to $d$ dimensions is not relevant here, and we prefer to have a simpler notation.

Given the Clifford algebra $C(D,D)$  generated by the Gamma matrices ${\Gamma_M}$ satisfying $\{{\Gamma_M},{\Gamma_N}\}=2{\eta_{MN}}$ (notice the \emph{curved} indices as in~\cite{Hohm:2011dv}) we can define ${\psi_M}={\Gamma_M}/\sqrt{2}$ so that we can have a representation in terms of fermionic oscillators ${\psi^m},{\psi_m}$ satisfying $\{{\psi_m},{\psi^n}\}=\delta_{{m}}^{{n}}$, $\{{\psi_m},{\psi_n}\}=0$, $\{{\psi^m},{\psi^n}\}=0$. The Clifford vacuum $\ket{0}$ is defined such that ${\psi_m}\ket{0}=0$ for all ${m}$. A generic spinor can be rewritten as $\ket{ \chi}=\sum_{p=0}^D\frac{1}{p!} C_{ m_1\cdots  m_p} \psi^{ m_1}\cdots \psi^{ m_p}\ket{0}$ which is consequence of the one-to-one map between spinors $ \chi$ and polyforms $ C=\sum_{p=0}^D\frac{1}{p!} C_{ m_1\cdots  m_p}d x^{ m_1}\cdots d x^{ m_p}$ on $\mathbb R^{1,D-1}$.
One fixes the normalisation $\braket{0|0}=1$ and defines conjugation such as $( \psi^{ m_1}\cdots \psi^{ m_p}\ket{0})^\dagger=\bra{0} \psi_{ m_p}\cdots \psi_{ m_1}$.

$Pin(D,D)$ is a subgroup of $C(D,D)$, see~\cite{Hohm:2011dv} for its definition. 
The algebras of $O(D,D)$ and $Pin(D,D)$ are isomorphic and if we call ${J_{MN}}$ the generators of the algebra of $O(D,D)$ satisfying  $[{J_{MN}},{J_{PQ}}]={\eta_{MP}J_{QN}}-{\eta_{NP}J_{QM}}-{\eta_{MQ}J_{PN}}+{\eta_{NQ}J_{PM}}$ then in the spinor representation we can identify ${J^{MN}}\equiv\tfrac12 {\Gamma^{MN}}$, where $\Gamma^{ M_1\cdots  M_p}$ is the totally antisymmetric product of $p$ Gamma matrices, with the factor $1/p!$ included, e.g. ${\Gamma^{MN}}=\tfrac12[{\Gamma^M},{\Gamma^N}]$.
We denote by $\rho$ the group homomorphisms $\rho:Pin(D,D)\to O(D,D)$. $Pin(D,D)$ is in fact the double cover of $O(D,D)$ because $\rho(S)=\rho(-S)$ for $S\in Pin(D,D)$. We have 
\be
S\Gamma^MS^{-1}=\Gamma^Nh_N{}^M, \qquad\text{where}\qquad \rho(S)=h_\bullet{}^\bullet.
\ee
This is a change of conventions compared to~\cite{Hohm:2011dv}, where they define $\rho$ such that $\rho(S)=h^\bullet{}_\bullet$. Notice that this amounts to swapping the two diagonal blocks of $h$, and the two off-diagonal blocks.
$Spin (D,D)$ is the subgroup generated by the $S$ made out of only an even number of Gamma matrices and under the group homomorphism $\rho$ it is mapped to $SO(D,D)$.

There are some distinguished elements of $Spin^+(D,D)$ that are important for the rest of the section. Their definition and the corresponding $SO^+(D,D)$ elements are
\begin{center}
\begin{tabular}{ccc}
$S_{[b]}=\exp(-\tfrac12 b_{mn}\psi^m\psi^n)$,&
$S_{[\beta]}=\exp(\tfrac12 \beta^{mn}\psi_m\psi_n)$,&
$S_{[r]}=(\det r)^{-1/2}\exp(\psi^mR_m{}^n\psi_n)$,\\
& & \\
$h_{[b]}=\left(\begin{array}{cc}
1&0\\
-b&1
\end{array}\right)$, &
$h_{[\beta]}=\left(\begin{array}{cc}
1&\beta\\
0&1
\end{array}\right)$, &
$h_{[r]}=\left(\begin{array}{cc}
r^{-T}&0\\
0&r
\end{array}\right)$.
\end{tabular}
\end{center}
Above we have $r_m{}^n=(\exp R)_m{}^n\in GL^+(D)$, where the plus stands for elements with a positive determinant. 
Other useful elements of $Pin(D,D)$ that are not in $Spin(D,D)$ are $S^\pm_m=\psi^m\pm\psi_m$. They satisfy $(S^\pm_m)^2=\pm 1$ and
\be
\rho(S^+_m)=-\left(\begin{array}{cc}
1-u_m&-u_m\\
-u_m&1-u_m
\end{array}\right),\qquad\qquad
\rho(S^-_m)=-\left(\begin{array}{cc}
1-u_m&u_m\\
u_m&1-u_m
\end{array}\right),
\ee
where $u_m=\text{diag}(0,\ldots,0,1,0,\ldots,0)$ has 1 only at position $m$.
It is nevertheless preferable to implement one factorised T-duality along the direction $m$ as in~\cite{Sakatani:2019jgu} by $S_{[m]}=(\psi^m-\psi_m)(-1)^{N_F}$ with $(-1)^{N_F}$ to be defined below. In fact this corresponds to $S_{[m]}\psi^MS_{[m]}^{-1}=\psi^NT^{[m]}_N{}^M$ where $T^{[m]}$ implements T-duality along direction $m$, and acts as the identity on other coordinates, without other unwanted signs.

The charge conjugation matrix is defined as $C=(\psi^0-\psi_0)(\psi^1-\psi_1)\cdots (\psi^9-\psi_9)$ and it satisfies  $C^{-1}=-C$.
One has the relations $C\psi_mC^{-1}=\psi^m$, $C\psi^mC^{-1}=\psi_m$ and $C\Gamma^MC^{-1}=(\Gamma^M)^\dagger$, $C\Gamma_MC^{-1}=(\Gamma_M)^\dagger$.
Defining the T-duality matrix as in~\eqref{eq:T} one sees that $C\Gamma_MC^{-1}=T_M{}^N\Gamma_N$ and therefore $\rho(C)=T$. Under  conjugation $C^\dagger =C^{-1}$, and we have that $S^\dagger=CS^{-1}C^{-1}$ if $S\in Spin^+(D,D)$, while $S^\dagger=-CS^{-1}C^{-1}$ if $S\in Spin^-(D,D)$. Moreover $(S^+_m)^\dagger=CS^+_mC^{-1}$.

One can define chiral spinors from the number operator $N_F=\sum_m\psi^m\psi_m$ that gives $N_F\ket{\chi}_p=p\ket{\chi}_p$ and $(-1)^{N_F}\ket{\chi}=\sum_{p=0}^D(-1)^p\frac{1}{p!}C_{m_1\cdots m_p}\psi^{m_1}\cdots \psi^{m_p}\ket{0}$. Chiral projections of spinors are therefore defined as $\ket{\chi_\pm}=\tfrac12(1\pm(-1)^{N_F})\ket{\chi}$ so that  $(-1)^{N_F}\ket{\chi_\pm}=\pm\ket{\chi_\pm}$. We have that $\chi_+$ is mapped to even forms and $\chi_-$ to odd forms. Notice  that the action of  $Spin(D,D)$ preserves the chirality.

The generalised metric  $\mathcal H$ is an element of $SO^-(D,D)$ because it has determinant 1 and we are in Lorentz signature. One therefore takes an $\mathbb S\in Spin^-(D,D)$ such that $\rho(\mathbb S)=\mathcal H^{\bullet\bullet}$. Notice again the change in conventions,  in~\cite{Hohm:2011dv} one has $\rho(\mathbb S)=\mathcal H_{\bullet\bullet}$.
Here it is $\mathbb S$ rather than $\mathcal H$ that is considered the fundamental field. We have $\mathbb S=\mathbb S^\dagger =-C\mathbb S^{-1} C^{-1}$. 
Let us further define  $S_{[\bar\eta]}=\psi^0\psi_0-\psi_0\psi^0$, then $S_{[\bar\eta]}=S_{[\bar\eta]}^\dagger=S_{[\bar\eta]}^{-1}\in Spin^-(D,D)$.
If $G_{mn}$ is the NSNS metric of the background and $G=e\bar\eta e^T$ in terms of a standard vielbein $e$ and the Minkowski metric $\bar \eta$, then we can define $S_{[G]} \equiv S_{[e]}S_{[\bar \eta]}S_{[e]}^\dagger$ where $S_{[e]}=S_{[r=e]}$, and where we used the definition of $S_{[r]}$ above. We then have  $\rho(S_{[G]}^{-1})=\text{diag}(G,G^{-1})$.
Given the NSNS field $B_{mn}$ of the background we can further define $S_{[B]}\equiv S_{[b=B]}$ using the definition of $S_{[b]}$ above. 
Then if we define $S_{\mathcal H}\equiv S_{[B]}^\dagger S_{[G]}^{-1} S_{[B]}$ we have $\rho(S_{[\mathcal H]})=\mathcal H^{\bullet\bullet}$.
Notice that because of our identification $\rho(S)=h_\bullet{}^\bullet$, if we consider an $O(D,D)$ transformation $\mathcal H'= h^T\mathcal Hh$ on the generalised metric, then it corresponds to $\mathbb S'=S^\dagger \mathbb SS$ which again differs from~\cite{Hohm:2011dv}.
For later convenience it will be useful to define also $\Kappa=C^{-1}\mathbb S$, so that $\rho(\Kappa)=\rho(C^{-1})\rho(\mathbb S)=T\mathcal H^{\bullet\bullet}=\mathcal H_\bullet{}^\bullet$. 

To write  the contribution of RR fields to the DFT action according to~\cite{Hohm:2011dv} one uses a spinor $\ket{\chi}$ that encodes the RR potentials. The contribution of RR fields to the Lagrangian is
\be
\mathcal L_{RR}=\frac14 (\slashed\partial\chi)^\dagger\mathbb S\slashed\partial\chi=\frac18\partial_M\bar\chi\Gamma^M\Kappa\Gamma^N\partial_N\chi=\frac14 
\overline{(\slashed\partial\chi)}\Kappa\slashed\partial\chi,
\ee
where we dropped the bra/ket notation and we defined $\bar\chi=\chi^\dagger C$ and $\slashed{\partial}=\tfrac{1}{\sqrt{2}}\Gamma^M\partial_M=\psi^m\partial_m+\psi_m\tilde\partial^m$. Using the $C(D,D)$ relations and the weak constraint it is easy to see that $\slashed\partial^2=0$. 
Reality of the action follows from $\mathbb S^\dagger=\mathbb S$ and the fact that $\chi $ is Grassmann even. 
Under the constant $Spin(D,D)$ transformation $\chi\to S^{-1}\chi$ and $X^M\to X^Nh_N{}^M$ (where $\rho(S)=h$ so that  $\partial_M\to (h^{-1})_M{}^N\partial_N$) one has $\slashed\partial\chi\to S^{-1}\slashed \partial\chi$, and the $Spin(D,D)$ invariance of the Lagrangian is manifest. It is not invariant under the full $Pin(D,D)$ because this would break the chirality of $\chi$.
In addition to the above Lagrangian, and after deriving the equations of motion, one imposes the duality relation
\be
\slashed\partial\chi=-\Kappa\slashed\partial\chi,
\ee
which are invariant under $\chi\to S^{-1}\chi,\mathbb S\to S^\dagger\mathbb SS$ for $S\in Spin^+(D,D)$. The restriction to only $Spin^+(D,D)$ (and not $Spin(D,D)$) invariance has to do with the Lorentz signature, so that timelike T-dualities would spoil the relations. Notice that the duality relation is consistent with $\Kappa^2=1$.
The equations of motion  for $\chi$ that one obtains from the above Lagrangian are $\slashed\partial(\Kappa\slashed\partial\chi)=0$. They are consistent with the duality relation and the identity $\slashed\partial^2=0$ valid on the weak constraint.
The abelian gauge symmetries acting on RR potentials are here implemented as $\delta_\lambda\chi=\slashed\partial\lambda$, and because $\delta_\lambda\slashed\partial\chi=\slashed\partial^2\lambda=0$, the invariance of the Lagrangian is obvious.

The above Lagrangian can be rewritten as a Lagrangian for the RR fields in the usual way.\footnote{To do so it is useful to notice that $\exp(\psi^mR_m{}^n\psi_n)\psi^p\ket{0}=(\exp R)_m{}^p\psi^m\ket{0}$ and $S_{[\bar\eta]}\psi^m\ket{0}=-\delta^{mn}\bar\eta_{np}\psi^p\ket{0}$.
It follows that 
\be
\begin{aligned}
S_{[G]}\psi^{m_1}\cdots\psi^{m_p}\ket{0}=-\frac{1}{\sqrt{-G}}\delta^{m_1n_1}\cdots \delta^{m_pn_p}G_{n_1q_1}\cdots G_{n_pq_p}\psi^{q_1}\cdots \psi^{q_p}\ket{0},\\
S_{[G]}^{-1}\psi^{m_1}\cdots\psi^{m_p}\ket{0}=-\sqrt{-G}G^{m_1n_1}\cdots G^{m_pn_p}\delta_{n_1q_1}\cdots \delta_{n_pq_p}\psi^{q_1}\cdots \psi^{q_p}\ket{0},
\end{aligned}
\ee
where $G$ is the determinant of the metric $G_{mn}$.
}
We recall that we are taking the spinor $\chi$ rewritten as $\ket{ \chi}=\sum_{p=0}^D\frac{1}{p!} C_{ m_1\cdots  m_p} \psi^{ m_1}\cdots \psi^{ m_p}\ket{0}$ which we associate to a polyform $ C=\sum_{p=0}^D\frac{1}{p!} C_{ m_1\cdots  m_p}d x^{ m_1}\cdots d x^{ m_p}$.
Let us now define $F_{m_1\cdots m_p}\equiv p\partial_{[m_1}C_{m_2\cdots m_p]}$ so that on the strong constraint (i.e.~taking $\tilde\partial^m=0$) we have
$
\ket{F}=\ket{\slashed\partial\chi}
=\sum_{p=1}^D\frac{1}{(p-1)!}\partial_{m_1}C_{m_2\cdots m_p}\psi^{m_1}\cdots \psi^{m_p}\ket{0}
=\sum_{p=1}^D\frac{1}{p!}F_{m_1\cdots m_p}\psi^{m_1}\cdots \psi^{m_p}\ket{0}
$
and in polyform notation $F=dC$. Using the NSNS B-field $B_{mn}$ we also define
$\ket{\hat F}\equiv \exp(-\tfrac12 B_{mn}\psi^m\psi^n)\ket{F}$
so that $\hat F=\exp(-B^{(2)})\wedge F$ where $B^{(2)}=\tfrac12 B_{mn}dx^m\wedge dx^n$. The conjugate is obtained by $\bra{\hat F}=\sum_{p=1}^D\tfrac{1}{p!}\bra{0}\psi_{m_p}\cdots\psi_{m_1}\hat F_{m_1\cdots m_p}$.
With this rewriting the RR Lagrangian becomes
\be
\begin{aligned}
\mathcal L_{RR}&=\frac14 (\slashed\partial\chi)^\dagger \mathbb S\slashed\partial\chi
=\frac14\braket{F|S_{[\mathcal H]}|F}
=\frac14\braket{\hat F|S_{[G]}^{-1}|\hat F}\\
&=\frac14 \sum_{p=1}^D\sum_{q=1}^D\frac{1}{p!q!}\braket{0|\psi_{m_p}\cdots\psi_{m_1}S_{[G]}^{-1}\psi^{n_1}\cdots \psi^{n_q}|0}\hat F_{m_1\cdots m_p}\hat F_{n_1\cdots n_q}\\
&=-\frac{\sqrt{-G}}{4}\sum_{p=1}^D|\hat F_p|^2,
\qquad\qquad
|\hat F_p|^2\equiv \frac{1}{p!}\hat F_{m_1\cdots m_p}G^{m_1n_1}\cdots G^{m_pn_p}\hat F_{n_1\cdots n_p},
\end{aligned}
\ee
where we used $\braket{0|\psi_{m_p}\cdots \psi_{m_1}\psi^{n_1}\cdots\psi^{n_q}|0}=p!\delta_{pq}\delta_{m_1}^{[n_1}\cdots \delta_{m_p}^{n_p]}$.
This is the Lagrangian in the democratic formulation, see~\cite{Hohm:2011dv}. In fact $\hat{ F}_{ m_1\cdots  m_p}$ are the RR field strengths that are commonly used in type II supergravity (they are the $\hat F$ of~\cite{Hohm:2011dv}), and they are the ones that appear in the quadratic couplings of the fermions in the Green-Schwarz formulation of the superstring (they are the $F$ of~\cite{Wulff:2013kga}).

In terms of $\ket{\hat F}$ the self duality relation reads as
$\ket{\hat F}=-S_{[G]}C\ket{\hat F}$ and in terms of $p$-forms
$\hat F^{(p)}=-(-1)^{\frac12 p(p+1)}*\hat F^{(10-p)}$.
The Hodge dual is defined as
\be
(*A)_{m_1\cdots m_p}=\frac{1}{(10-p)!}G_{m_1n_1}\cdots G_{m_pn_p}\varepsilon^{k_{p+1}\cdots k_{10}n_1\cdots n_p}A_{k_{p+1}\cdots k_{10}},
\ee
where $\epsilon^{1\cdots 10}=1$, $\epsilon_{1\cdots 10}=-1$, $\varepsilon^{m_1\cdots m_{10}}=\frac{1}{\sqrt{-G}}\epsilon^{m_1\cdots m_{10}}$, $\varepsilon_{m_1\cdots m_{10}}=\sqrt{-G}\epsilon_{m_1\cdots m_{10}}$.
We also have $**\omega^{(p)}=-(-1)^{p(10-p)} \omega^{(p)}$ where the additional minus sign is due to the Lorentz signature.

Under the abelian RR gauge transformations the spinor transforms as $\delta_{\lambda}\ket{\chi}=\slashed\partial\ket{\lambda}$ which implies $\delta_\lambda C=d\lambda$. Under $D$-dimensional diffeomorphisms $\delta_{\xi}C_{m_1\cdots m_p}=\mathcal L_\xi C_{m_1\cdots m_p}$, i.e.~it transforms with the standard Lie derivative as expected. Under the gauge transformation of the B-field with gauge parameter $\tilde \xi$ the transformation is $\delta_{\tilde \xi}C_{m_1\cdots m_p}=p(p-1)\partial_{[m_1}\tilde\xi_{m_2}C_{m_3\cdots m_p]}$, meaning $\delta_{\tilde\xi}C=d\tilde \xi\wedge C$ where $\delta_{\tilde \xi} B^{(2)}=d\tilde\xi$.
Therefore $F$ is not invariant under gauge transformations of the B-field, but $\hat F$ is invariant.
Defining $\hat A=e^{-B^{(2)}}\wedge C$ (i.e.~$C=e^{B^{(2)}}\wedge \hat A$) then $\delta_{\tilde \xi}\hat A=0$.
Notice that in terms of these new potentials $\hat F=d\hat A+H\wedge \hat A$, with $H=dB$.
If we further define $A^{(p)}=\hat A^{(p)}$ for $p\neq 4$ and $A^{(4)}=\hat A^{(4)}+\frac12 B^{(2)}\wedge \hat A^{(2)}$ then one finds
\be
\begin{aligned}
&\hat F^{(1)}=dA^{(0)},\qquad\qquad
&&\hat F^{(2)}=dA^{(1)},\\
&\hat F^{(3)}=dA^{(2)}+H\wedge A^{(0)},\qquad\qquad
&&\hat F^{(4)}=dA^{(3)}+H\wedge A^{(1)},\\
&\hat F^{(5)}=dA^{(4)}+\frac12 H\wedge A^{(2)}-\frac12 B^{(2)}\wedge dA^{(2)},\qquad\qquad&&
\end{aligned}
\ee
where on the left we have IIB and on the right IIA. These are the more familiar parametrisations of the RR field strenghts in terms of potentials~\cite{Hohm:2011dv}. 

To rewrite the above results as in~\cite{Geissbuhler:2013uka} one has to pass from a basis of Gamma matrices $\Gamma^M$ in terms of curved indices to a basis $\Gamma^A$ in terms of flat indices. This can be done by using $S_{[E]}$, the $Pin(D,D)$ representative of the generalised vielbein, as $S_{[E]}\Gamma^AS_{[E]}^{-1}=\Gamma^ME_M{}^A$. 
It is also necessary to change vacuum, since we have to go from a vacuum defined as $\Gamma_m\ket{0}=0$ to a new one such that $\Gamma_a\ket{0'}=0$. In the following we will omit the prime but always think in terms of the new vacuum. Notice also that we can still take the charge conjugation matrix $C$ and $S_{[\bar\eta]}$ to be defined as before, while we   interpret the $\psi$ as being defined in terms of flat indices.
We first define
\be
\ket{\Fsp}=e^dS_{[E]}\ket{F}.
\ee
Because we can decompose $S_{[E]}=S_{[e]}^{-1}S_{[B]}$ and we get
\be
\ket\Fsp=\sum_n\frac{e^\phi}{n!}\hat F_{m_1\cdots m_n}e_{a_1}{}^{m_1}\cdots e_{a_n}{}^{m_n}\psi^{a_1}\cdots \psi^{a_n}\ket{0}.
\ee
From the definition it follows that $\ket{\Fsp}=\slashed\partial\ket\Csp+S_E\psi^M\partial_MS_E^{-1}\ket\Csp+e^{d}\partial_Me^{-d}\psi^M\ket\Csp$ in terms of $\ket\chi=e^{-d}S_{[E]}^{-1}\ket\Csp$.
This may be rewritten (see also~\cite{Catal-Ozer:2017cqr}) using\footnote{Given the Lie algebra isomorphism betweem $O(D,D)$ and $Pin(D,D)$ one can identify the generators $J^{MN}\to \psi^{MN}$ and use them to construct group elements $E_A{}^M=\exp(\Lambda_{PQ}J^{PQ})_A{}^M$ and $S_{[E]}=\exp(\Lambda_{PQ}\psi^{PQ})$ with the same $\Lambda_{PQ}$. Therefore one has $(\partial_MEE^{-1})_{BC}=\Xi_{MPQ}(J^{PQ})_{BC}$ and $\partial_MS_{[E]}S_{[E]}^{-1}=\Xi_{MPQ}\psi^{PQ}$ with the same $\Xi_{MPQ}$. It follows that $E_A{}^M\partial_MS_{[E]}S_{[E]}^{-1}
=E_A{}^M\Xi_{MPQ}\psi^{PQ}
=\tfrac12 E_A{}^M\Xi_{MPQ}(2\delta^{[P}_B\delta^{Q]}_C)\psi^{BC}
=\tfrac12 E_A{}^M\Xi_{MPQ}(J^{PQ})_{BC}\psi^{BC}
=\tfrac12 E_A{}^M(\partial_MEE^{-1})_{BC}\psi^{BC}
=\tfrac12 \Omega_{ABC}\psi^{BC}$, where we used the matrix realisation $(J^{MN})_P{}^Q=2\delta_P^{[M}\eta^{N]Q}$. In (4.32) of~\cite{Catal-Ozer:2017cqr} only the leading term in the expansion is taken into account.} 
$E_A{}^M\partial_MS_{[E]}S_{[E]}^{-1}
=\tfrac12 \Omega_{ABC}\psi^{BC}$
as well as
$\Omega_{ABC}\psi^A\psi^B\psi^C
=\frac13 \mathcal F_{ABC}\psi^{ABC}+\Omega^B{}_{BA}\psi^A$, which is a consequence of the $C(D,D)$ relations.
To conclude 
\be
\ket\Fsp=(\psi^A\partial_A-\tfrac16 \mathcal F_{ABC}\psi^{ABC}-\tfrac12 \mathcal F_A\psi^A)\ket \Csp=\slashed\nabla\ket \Csp,\qquad
\slashed\nabla\equiv\slashed\partial-\slashed{\mathcal F}^{(3)}
-\tfrac12 \slashed{\mathcal F}^{(1)},
\ee
where $\slashed{\mathcal F}^{(n)}$ includes the $1/n!$ factor.
Notice that $\slashed\nabla^2=0$ on the strong constraint.
Now gauge transformations read as $\delta_\lambda\ket \Csp=\slashed\nabla\ket\lambda$, and the Bianchi identities as  $\slashed\nabla\ket \Fsp=0$.
The Lagrangian  of~\cite{Hohm:2011dv} is readily rewritten as 
\be
\mathcal L_{RR}
=\frac14e^{-2d}\braket{\Fsp |S_{\bar\eta}^{-1}|\Fsp}
=-\frac14e^{-2d}\braket{\bar{\Fsp}|CS_{\bar\eta}^{-1}|\Fsp}.
\ee
Now we have 
\be
\begin{aligned}
CS_{\bar\eta}^{-1}&=(\psi^0-\psi_0)(\psi^1-\psi_1)\cdots (\psi^9-\psi_9)(\psi^0\psi_0-\psi_0\psi^0)\\
&=(\psi^0-\psi_0)(\psi^0\psi_0-\psi_0\psi^0)(\psi^1-\psi_1)\cdots (\psi^9-\psi_9)\\
&=-(\psi^0+\psi_0)(\psi^1-\psi_1)\cdots (\psi^9-\psi_9)=\Psi_-,
\end{aligned}
\ee
and therefore we conclude
\be
\mathcal L_{RR}
=-\frac14e^{-2d}\braket{\bar{\Fsp}|\Psi_-|\Fsp}.
\ee
Varying the Lagrangian with respect to $\ket \Csp$ one finds the equations of motion $\slashed\nabla\Psi_-\ket \Fsp=0$.

The above Lagrangian  is different when comparing to~\cite{Geissbuhler:2013uka}, since they have $\Psi_+$ instead of $\Psi_-$, where $\Psi_\pm=(\psi_0\mp\psi^0)(\psi_1\pm\psi^1)\cdots (\psi_9\pm\psi^9)$. 
To further test the mismatch we first move to prove (A.9) of~\cite{Geissbuhler:2013uka}. 
Using $\Psi_\pm\psi^A=\mp\mathcal H^{AB}\psi_B\Psi_\pm$ one finds
\be
\begin{aligned}
\Psi_+\ket{\omega}&=\sum_n\frac{1}{n!}\omega_{a_1\cdots a_n}\Psi_+\psi^{a_1}\cdots \psi^{a_n}\ket{0}\\
& =\sum_n\frac{(-1)^n}{n!}\omega_{a_1\cdots a_n}\mathcal H^{a_1B_1}\cdots \mathcal H^{a_nB_n}\psi_{B_1}\cdots \psi_{B_n}\Psi_+\ket{0}\\
&=-\sum_n\frac{(-1)^n}{n!}\omega_{a_1\cdots a_n}\bar \eta^{a_1b_1}\cdots \bar\eta^{a_nb_n}\psi_{b_1}\cdots \psi_{b_n}\psi^0\psi^1\cdots\psi^9\ket{0}\\
&=\sum_n\frac{(-1)^n(-1)^{\frac{n(n-1)}{2}}}{n!(10-n)!}\omega_{a_1\cdots a_n}\bar \eta^{a_1b_1}\cdots \bar\eta^{a_nb_n}\epsilon_{b_1\cdots b_nb_{n+1}\cdots b_{10}}\psi^{b_{n+1}}\cdots \psi^{b_{10}}\ket{0}\\
&=\sum_n\frac{1}{n!(10-n)!}\epsilon_{b_{n+1}\cdots b_{10}}{}^{a_{1}\cdots a_n}\omega_{a_n\cdots a_1}\psi^{b_{n+1}}\cdots \psi^{b_{10}}\ket{0}.
\end{aligned}
\ee
This agrees with (A.9) of~\cite{Geissbuhler:2013uka}. 
In the first step we used the commutation relations between $\Psi_+$ and the $\psi$'s. In the second step we used $\Psi_+\ket{0}=-\psi^0\psi^1\cdots\psi^9\ket{0}$, in the third one we used (5.39) of~\cite{Hohm:2011dv} (notice that that equation is written with an $\epsilon$ with upper indices, so there is an overall minus sign when written for an epsilon with lower indices). In the last step we got a factor of $(-1)^{\frac{n(n-1)}{2}}$ by rearranging the indices of $\omega$ and a factor of $(-1)^{n(10-n)}$ by swapping the position of the $b_1\cdots b_n$ and $b_{n+1}\cdots b_D$ indices in $\epsilon$. In the last line, indices in the epsilon tensor are raised with the Minkowski $\bar\eta$. 

The computation for $\Psi_-$ is identical, since also $\Psi_-\ket{0}=-\psi^0\psi^1\cdots\psi^9\ket{0}$, but because of the different sign in the commutation relations with the $\psi$'s one gets an additional factor of $(-1)^n$
\be
\Psi_-\ket{\omega}
=\sum_n\frac{(-1)^n}{n!(10-n)!}\epsilon_{b_{n+1}\cdots b_{10}}{}^{a_{1}\cdots a_n}\omega_{a_n\cdots a_1}\psi^{b_{n+1}}\cdots \psi^{b_{10}}\ket{0}.
\ee 
Let us now compare to (A.11) of~\cite{Geissbuhler:2013uka}. First we compute
\be
\begin{aligned}
C\ket{\chi}&=\sum_n\frac{1}{n!}\chi_{a_1\cdots a_n}C\psi^{a_1}\cdots\psi^{a_n}\ket{0}\\
&=\sum_n\frac{1}{n!}\chi_{a_1\cdots a_n}\psi_{a_1}\cdots\psi_{a_n}C\ket{0}\\
&=\sum_n\frac{1}{n!}\chi_{a_1\cdots a_n}\psi_{a_1}\cdots\psi_{a_n}\psi^0\cdots\psi^9\ket{0}\\
&=\sum_n\frac{(-1)^{\frac{n(n-1)}{2}}}{n!(10-n)!}\chi_{a_1\cdots a_n}\epsilon^{a_1\cdots a_nb_{n+1}\cdots b_{10}}\psi^{b_{n+1}}\cdots\psi^{b_{10}}\ket{0},
\end{aligned}
\ee
where in the last step we used again  (5.39) of~\cite{Hohm:2011dv}.
Now  using $(C\ket{\chi})^\dagger=\bra{\chi}C^\dagger=-\bra{\chi}C$ we find
\be
\begin{aligned}
\braket{\chi|C\Psi_+|\omega}
&=-\sum_{n,m}\frac{(-1)^{\frac{n(n-1)}{2}}}{n!(10-n)!}\frac{1}{m!(10-m)!}
\chi_{a_1\cdots a_n}\epsilon^{a_1\cdots a_nb_{n+1}\cdots b_{10}}
\epsilon_{d_{m+1}\cdots d_{10}}{}^{c_1\cdots c_m}
\omega_{c_m\cdots c_1}\\
&\quad\times
\braket{0|\psi_{b_{10}}\cdots\psi_{b_{n+1}}\psi^{d_{m+1}}\cdots \psi^{d_{10}}|0}\\
&=-\sum_{n}\frac{(-1)^{\frac{n(n-1)}{2}}}{n!n!(10-n)!}
\chi_{a_1\cdots a_n}\epsilon^{a_1\cdots a_nb_{n+1}\cdots b_{10}}
\epsilon_{b_{n+1}\cdots b_{10}}{}^{c_1\cdots c_n}
\omega_{c_n\cdots c_1}\\
&=-\sum_{n}\frac{(-1)^{\frac{n(n-1)}{2}}(-1)^{n(10-n)}}{n!n!(10-n)!}
\chi_{a_1\cdots a_n}\epsilon^{b_{n+1}\cdots b_{10}a_1\cdots a_n}
\epsilon_{b_{n+1}\cdots b_{10}}{}^{c_1\cdots c_n}
\omega_{c_n\cdots c_1}\\
&=\sum_{n}\frac{(-1)^{\frac{n(n-1)}{2}}(-1)^{n(10-n)}}{n!}
\chi_{a_1\cdots a_n}\bar\eta^{a_1b_1}\cdots \bar\eta^{a_nb_n}
\omega_{b_n\cdots b_1}\\
&=\sum_{n}\frac{(-1)^{\frac{n(n-1)}{2}}(-1)^{n(10-n)}(-1)^{\frac{n(n-1)}{2}}}{n!}
\chi_{a_1\cdots a_n}\bar\eta^{a_1b_1}\cdots \bar\eta^{a_nb_n}
\omega_{b_1\cdots b_n}\\
&=\sum_{n}\frac{(-1)^{n}}{n!}
\chi_{a_1\cdots a_n}\bar\eta^{a_1b_1}\cdots \bar\eta^{a_nb_n}
\omega_{b_1\cdots b_n}.
\end{aligned}
\ee
Compared to (A.11) of~\cite{Geissbuhler:2013uka} we have an additional factor of $(-1)^{n}$.
In the computation above  we used also $\epsilon^{b_{n+1}\cdots b_{10}a_1\cdots a_n}
\epsilon_{b_{n+1}\cdots b_{10}}{}^{c_1\cdots c_n}=-n!(10-n)!\bar\eta^{[a_1|c_1}\cdots \bar\eta^{a_n]c_n}$.
The computation for $\Psi_-$ works in the same way, and because of the previous factor we have in fact
\be
\braket{\chi|C\Psi_-|\omega}
=\sum_{n}\frac{1}{n!}
\chi_{a_1\cdots a_n}\bar\eta^{a_1b_1}\cdots \bar\eta^{a_nb_n}
\omega_{b_1\cdots b_n}.
\ee
This computation is independent from the one that was giving us $\Psi_-$ in the Lagrangian, and it is a further confirmation that $\Psi_-$ rather than $\Psi_+$ should be used there, since with $\Psi_+$ there would be an additional factor of $(-1)^n$ when matching the DFT action to the one of standard supergravity.
In the IIB case this would affect the sign of the RR Lagrangian (because for $n$ odd $(-1)^n=-1$) while for IIA using $\Psi_-$ or $\Psi_+$ is inconsequential.

Let us also remark that the self-duality condition  $\Psi_-\Fsp=\Fsp$ (rather than $\Psi_+\Fsp=\Fsp$) is in fact consistent with the conventions of~\cite{Hohm:2011dv}. The left-hand-side is
\be
\begin{aligned}
&\sum_n\frac{(-1)^n}{n!(10-n)!}\epsilon_{b_{n+1}\cdots b_{10}}{}^{a_{1}\cdots a_n}\hat F_{a_n\cdots a_1}\psi^{b_{n+1}}\cdots \psi^{b_{10}}\ket{0}\\
=&\sum_n\frac{(-1)^{(10-n)}}{n!(10-n)!}\epsilon_{b_{1}\cdots b_{n}}{}^{a_{n+1}\cdots a_{10}}\hat F_{a_{10}\cdots a_{n+1}}\psi^{b_{1}}\cdots \psi^{b_{n}}\ket{0}\qquad (n\to 10-n)\\
=&\sum_n\frac{(-1)^{(10-n)}(-1)^{n(10-n)}}{n!(10-n)!}\epsilon^{a_{n+1}\cdots a_{10}}{}_{b_{1}\cdots b_{n}}\hat F_{a_{10}\cdots a_{n+1}}\psi^{b_{1}}\cdots \psi^{b_{n}}\ket{0}\qquad (\epsilon_{\cdots}{}^{\cdots}\to\epsilon^{\cdots}{}_{\cdots})\\
=&\sum_n\frac{(-1)^{(10-n)}(-1)^{n(10-n)}(-1)^{(10-n)(10-n-1)/2}}{n!(10-n)!}\epsilon^{a_{n+1}\cdots a_{10}}{}_{b_{1}\cdots b_{n}}\hat F_{a_{n+1}\cdots a_{10}}\psi^{b_{1}}\cdots \psi^{b_{n}}\ket{0}\\
& \qquad\qquad\qquad\qquad\qquad\qquad\qquad\qquad\qquad (\hat F_{a_{10}\cdots a_{n+1}}\to\hat F_{a_{n+1}\cdots a_{10}})\\
=&-\sum_n\frac{(-1)^{\frac{n(n+1)}{2}}}{n!(10-n)!}\epsilon^{a_{n+1}\cdots a_{10}}{}_{b_{1}\cdots b_{n}}\hat F_{a_{n+1}\cdots a_{10}}\psi^{b_{1}}\cdots \psi^{b_{n}}\ket{0},
\end{aligned}
\ee
which implies the self-duality relation
\be
-\frac{(-1)^{\frac{n(n+1)}{2}}}{(10-n)!}\epsilon^{a_{n+1}\cdots a_{10}}{}_{b_{1}\cdots b_{n}}\hat F_{a_{n+1}\cdots a_{10}}=\hat F_{b_{1}\cdots b_{n}},
\ee
in agreement with the conventions of~\cite{Hohm:2011dv}.

\section{DFT equations of motion}\label{app:eom}
In this appendix we continue to omit the boldface notation, since the dimensional reduction plays no role.
The action in the NSNS sector is $S_{\text{NSNS}}=\int d X e^{-2d}\mathcal R$ where we prefer to use the rewriting of~\cite{Hronek:2020skb}
\be
\mathcal R=-4\partial^A\mathcal F_A^{(-)}+2\mathcal F^A\mathcal F_A^{(-)}
-\mathcal F_{ABC}^{(-)}\mathcal F^{(-)ABC}
-\frac13 \mathcal F_{ABC}^{(--)}\mathcal F^{(--)ABC}.
\ee
Here $P_\pm=\tfrac12(\eta\pm\mathcal H)$ and $\mathcal F_A^{(\pm)}=(P_\pm)_A{}^B\mathcal F_B$, $\mathcal F_{ABC}^{(\pm)}=(P_{\mp})_A{}^D(P_{\pm})_B{}^E(P_{\pm})_C{}^F\mathcal F_{DEF}$, $\mathcal F_{ABC}^{(\pm\pm)}=(P_{\pm})_A{}^D(P_{\pm})_B{}^E(P_{\pm})_C{}^F\mathcal F_{DEF}$.
To compute the equations of motion one finds that {$\delta_ES_{\text{NSNS}}=\int dX e^{-2d} \Xi_{\text{NSNS}}^{AB}\Delta_{AB}$  where $\Delta_{AB}=-\Delta_{BA}=\delta E_A{}^ME_{BM}$ and $\delta_dS_{\text{NSNS}}=\int dX e^{-2d}\Xi_{\text{NSNS}}\delta d$ } 
where~\cite{Geissbuhler:2013uka}
\be
\begin{aligned}
\Xi^{\text{NSNS}}_{[AB]}&=4\partial_{[A}\mathcal F^{(-)}_{B]}
+(\mathcal F^C-\partial^C)\check{\mathcal F}_{C[AB]}
+\check{\mathcal F}_{CD[A}\mathcal F_{B]}{}^{CD},\\
\Xi_{\text{NSNS}}&=-2\mathcal R,
\end{aligned}
\ee
where we defined $\check{\mathcal F}_{ABC}=\check S_{ABC}{}^{A'B'C'}\mathcal F_{A'B'C'}$ and
\be
\begin{aligned}
\check S_{ABCA'B'C'}={}&\tfrac12 \mathcal H_{AA'}\eta_{BB'}\eta_{CC'}
+\tfrac12 \eta_{AA'}\mathcal H_{BB'}\eta_{CC'}
+\tfrac12 \eta_{AA'}\eta_{BB'}\mathcal H_{CC'}
-\tfrac12 \mathcal H_{AA'}\mathcal H_{BB'}\mathcal H_{CC'} \\
& -\eta_{AA'}\eta_{BB'}\eta_{CC'}\\
={}&-2\left(P^{(-)}_{AA'}P^{(-)}_{BB'}P^{(-)}_{CC'}
+P^{(+)}_{AA'}P^{(-)}_{BB'}P^{(-)}_{CC'}
+P^{(-)}_{AA'}P^{(+)}_{BB'}P^{(-)}_{CC'}
+P^{(-)}_{AA'}P^{(-)}_{BB'}P^{(+)}_{CC'}
\right).
\end{aligned}
\ee
It is convenient to write $\check S$ in the second way, which is the unique way to write it as a linear combination of products of projectors. In fact it is then easy to check that 
\be
P^{(+)}_{AA'}P^{(-)}_{BB'}\Xi_{\text{NSNS}}^{[A'B']}=2\Xi'{}^{\text{NSNS}}_{AB},\qquad
\Xi'{}^{\text{NSNS}}_{AB}\equiv 2\partial^{(+)}_A\mathcal F^{(-)}_B
+(\partial^C-\mathcal F^C)\mathcal F^{(-)}_{ABC}
-\mathcal F^{(+)}_{CDA}\mathcal F^{(-)DC}{}_B.
\ee
Notice the transposition of $CD$ indices in the last term. One also has $P^{(+)}_{AA'}P^{(+)}_{BB'}\Xi_{\text{NSNS}}^{[A'B']}=0$ and $P^{(-)}_{AA'}P^{(-)}_{BB'}\Xi_{\text{NSNS}}^{[A'B']}=2P^{(-)}_{AA'}P^{(-)}_{BB'}\mathcal Z^{A'B'}\sim 0$ where the last is a constraint of DFT. Finally, using $P^{(-)}_{AA'}P^{(+)}_{BB'}\Xi_{\text{NSNS}}^{[A'B']}=-2\Xi'{}^{\text{NSNS}}_{BA}$ we can conclude that $\Xi^{\text{NSNS}}_{[AB]}\sim 4\Xi'{}^{\text{NSNS}}_{[AB]}$ (i.e.~on the constraint of DFT).
Hence because we can project with $P^{(\pm)}$,  $\Xi^{\text{NSNS}}_{[AB]}=0$ is equivalent to $\Xi'{}^{\text{NSNS}}_{AB}=0$.

When considering the RR contribution to the action one has~\cite{Geissbuhler:2013uka}
$\delta_ES_{\text{RR}}=\int dX e^{-2d} \Xi_{\text{RR}}^{AB}\Delta_{AB}$  and $\delta_dS_{\text{RR}}=\int dX e^{-2d}\Xi_{\text{RR}}\delta d$
where $\Xi_{\text{RR}}^{[AB]}=-\frac14 \bar{\Fsp}\psi^{AB}\mathcal \Fsp$ and $\Xi_{\text{RR}}=0$ after imposing the self-duality condition.
The full equations of motion in the type II case are then
$\Xi_{[AB]}=\Xi^{\text{NSNS}}_{[AB]}+\Xi^{\text{RR}}_{[AB]}=0$ and $\Xi=\Xi^{\text{NSNS}}=0$.

\section{Other ansatze used for orbits with $H$-flux} \label{app:PDE}

In this appendix we will report on other attempts to deal with orbits with non-vanishing $H$-flux in the standard parametrisation of \eqref{eq:parU} rather than appealing to the  methodology of section \ref{s:AnsatzOrbitsH}. In particular, by preferring an ordinary Lie group $G$ action on the coordinates $y$, we will assume here that  $\rho_\mu{}^i$ of \eqref{eq:parU}  are the components of a (left-invariant) Maurer-Cartan form $\rho = g^{-1} d g = dy^\mu \rho_\mu{}^i t_i$ where $t_i \in \mathfrak{g} = \text{Lie} (G)$ are the generators of the Lie algebra and $g\in G$ is a Lie group element. In this case the components $\dr_{ij}{}^k$ defined in \eqref{eq:def-dr} are precisely the structure constants of $\mathfrak{g}$ so that we can write them in the more standard way $\dr_{ij}{}^k = f_{ij}{}^k$. Furthermore, we can use the well-known expansion of the MC form
\begin{equation}
\rho = g^{-1} dg = \sum_{N=0}^{\infty} \frac{(-)^N}{(N+1)!} \text{ad}_x^N dx , \qquad x \in \mathfrak{g} .
\end{equation}

The main difficulty with pursuing in this way is finding the most general solution to the differential equation obtained from the $Q$-flux equation of \eqref{eq:FF3}. Focusing here on the simplest homogeneous equation  ($Q = 0$) with $F = 0$ we must solve
\begin{equation} \label{eq:PDEbetaH}
 \partial_\mu \beta^{ij} + \rho_\mu{}^n \beta^{l[i} \beta^{j]m} H_{nlm} = 0 ,
\end{equation}
or equivalently  upon the $F$-flux equation 
\begin{equation} \label{eq:PDEbetaMC}
 \partial_\mu \beta^{ij} - \rho_\mu{}^n \beta^{l[i} f_{nl}{}^{j]} = 0 .
\end{equation}
Let us remark  here that  by lowering all indices in eq.~\eqref{eq:PDEbetaMC} with the Killing form and  antisymmetrising in $i,j,k$ we find that a \textit{necessary} (but not sufficient) condition is  that as a two-form $\beta$  is closed, and possibly exact. For semisimple algebras $\mathfrak{h}$ it is known that there exists  no non-trivial closed two-forms which are not exact.  An exact form such as $\beta^{ij} = \rho_\mu{}^{i} \rho_\nu{}^j \partial^{[\mu} \lambda^{\nu]}$ for some one-form $\lambda$ does not, however,  solve eq.~\eqref{eq:PDEbetaMC}. We can therefore conclude that for semisimple algebras   no non-trivial solutions to eq.~\eqref{eq:PDEbetaMC} exist.

To find a non-trivial solution of \eqref{eq:PDEbetaMC} we have considered  for $\beta^{ij}$ the following ansatze.\footnote{In the  calculations that follow it is useful to use the identity
 \begin{equation} \label{eq:identityadxdx}
[\text{ad}_x^N dx, y] = \sum_{K=0}^N (-)^K \frac{N!}{K! (N-K)!} \text{ad}_x^{N-K} \text{ad}_{dx}\text{ad}_x^K y ,
\end{equation}
which holds for any $x,y\in\text{Lie}(G)$ and which can be proved by induction and  the Jacobi identity. }
\begin{enumerate}[(i)]
\item In the $(H)$-orbit of section  \ref{s:Horbit} we found a non-trivial representative, where $\rho$ is of MC form, by considering $\beta^{ij}$ to be a constant matrix.
\item  Take now a non-constant $\beta$
\begin{equation} \label{eq:betaAnsatz1}
\begin{aligned}
\beta^{ij} &= \langle O^t R O t^i ,  t^j \rangle , \qquad  O = \sum_{N=0}^\infty a_N \text{ad}_x^N,  \\
&= \sum_{N=0}^\infty \sum_{M=0}^\infty a_N a_M (\text{ad}_x^N)^i{}_l R^{lk} (\text{ad}_x^M)^{j}{}_k , 
\end{aligned}
\end{equation}
where $R$ is an antisymmetric matrix and where $t^i = \kappa^{ij} t_j$ and $\kappa_{ij} = \langle t_i , t_j \rangle$ is the Killing form of $\mathfrak{g}$. In this case one can observe that fixing the unknown coefficients $a_N$ becomes inconsistent at order ${\cal O}(\partial_\mu x , x^1 )$. When generalising \eqref{eq:betaAnsatz1} as
\begin{equation}
\beta^{ij} = \sum_{N=0}^\infty \sum_{M=0}^\infty a_{NM} (\text{ad}_x^N)^{[i}{}_l E^{lk} (\text{ad}_x^M)^{j]}{}_k, 
\end{equation}
where $E^{ij} = S^{ij} + R^{ij}$ with $S$  a symmetric  and  $R$ an antisymmetric matrix, and $a_{NM}$ unknown coefficients, one will find that the symmetrical part does not contribute and this is effectively equivalent to the ansatz \eqref{eq:betaAnsatz1}. 
\item Instead take $\beta = O - O^t$ with $O$ a series in $\text{ad}_x$, as defined above, which is important  to resonate with the MC form. Hence
\begin{equation} \label{eq:BetaAnsatz2}
\beta^{ij} = 2  \sum_{N=0}^\infty a_N (\text{ad}_x^N)^{[i}{}_k \kappa^{k|j]} .
\end{equation}
This ansatz turns out to be empty: for every $N \in \mathbb{N}$ we find $a_{N} = 0$.
\setcounter{nameOfYourChoice}{\value{enumi}}
\end{enumerate}
We have also considered the possibility that $\rho$ is not of MC form and therefore tried to solve the PDE \eqref{eq:PDEbetaH} instead.  Assuming nevertheless an underlying Lie algebra $\mathfrak{g}$ we take $H_{ijk} = \alpha_1 f_{ij}{}^l \kappa_{lk}$ for some real constant $\alpha_1$, and  we expand $\rho$ as
\begin{equation}
 \rho_\mu{}^i = \sum_{N=0}^\infty b_N   (\text{ad}_x^N)^{l}{}_m \partial_\mu x_l \kappa^{mi}  , \qquad x\in \mathfrak{g} ,
\end{equation}
for some undetermined coefficients $b_N$. Then,
\begin{enumerate}[(i)]
\setcounter{enumi}{\value{nameOfYourChoice}}
\item Taking $\beta$ as in \eqref{eq:BetaAnsatz2} we find that the resulting recursive relations will be  empty, $a_N = 0$ for all $N$, and independent of the coefficients $b_N$.
\item Adding the inhomogeneous term $Q_i{}^{jk} = \alpha_2 f_{im}{}^n \kappa^{jm} \kappa_{kn}$ (with $\alpha_2$ a second real constant) in the PDE as $Q_i{}^{jk} = \partial_i \beta^{jk} + \beta^{l[j} \beta^{k]m} H_{ilm}$ we find for the ansatz \eqref{eq:BetaAnsatz2} the following  solution that we verified until $N=10$,
\begin{equation}
a_{2N} = 0, \qquad a_{2N-1} = \frac{4^N (4^N - 1)}{2} \frac{B_{2N}}{(2N)!} \alpha_1^{N-1} \alpha_2^N b_0^{2N-1},  
\end{equation}
and
\begin{equation}
b_{2N} =  4^{N+1} (4^{N+1} - 1) \frac{B_{2N+2}}{(2N+2)!} \alpha_1^{N} \alpha_2^N b_0^{2N+1}, \qquad b_{2N-1} = 0 ,
\end{equation}
where $B_N$ is the Bernouilli number. Given the form of $H$ and $Q$ this   should coincide with the $\lambda$-model solution of section \ref{s:QHorbit}. However, the above is clearly not workable and thus the methods of section \ref{s:AnsatzOrbitsH} are preferred.
\end{enumerate}


\bibliographystyle{nb}
\bibliography{biblio}{}

\end{document}